\documentclass[11pt,onecolumn,draftcls,journal]{IEEEtran}

\normalsize
\usepackage[USenglish]{babel}
\usepackage{multirow}
\usepackage{textcomp}
\usepackage{calc}
\usepackage{amsmath,amssymb,amsfonts}
\usepackage{bbm}
\usepackage{xcolor}
\usepackage[noadjust]{cite}
\usepackage{array}
\usepackage[tight,footnotesize]{subfigure}
\usepackage{graphicx}\DeclareGraphicsExtensions{.eps}
\usepackage{algorithm}
\usepackage[noend]{algpseudocode}

\newcommand{\vect}[1]{\underline{\boldsymbol #1}}

\newcommand{\sgn}{{\mbox{\rm sgn}}}
\newcommand{\addpr}{\mathbf{a}_{\mbox{\scriptsize pr}}}
\newcommand{\minpr}{\mathbf{m}_{\mbox{\scriptsize pr}}}
\newcommand{\addprred}{{\color{red}\mathbf{a}_{\mbox{\scriptsize pr}}}}
\newcommand{\minprred}{{\color{red}\mathbf{m}_{\mbox{\scriptsize pr}}}}
\newcommand{\xorprred}{{\color{red}\mathbf{x}_{\mbox{\scriptsize pr}}}}
\newcommand{\scuprred}{{\color{red}\text{\sc scu}_{\mbox{\scriptsize pr}}}}
\newcommand{\pxor}{p_{\text{\it x}}}
\newcommand{\padd}{p_{\text{\it a}}}
\newcommand{\pcomp}{p_{\text{\it c}}}
\newcommand{\pscu}{p_{\text{scu}}}
\newcommand{\lth}{\mathbf{lt}}
\newcommand{\lthpr}{\mathbf{lt}_{\mbox{\scriptsize pr}}}
\newcommand{\xorpr}{\mathbf{x}_{\mbox{\scriptsize pr}}}
\newcommand{\scupr}{\text{\sc scu}_{\mbox{\scriptsize pr}}}

\newcommand{\unit}{\mathbf{U}}
\newcommand{\unitpr}{\mathbf{U}_{\text{pr}}}

\newcommand{\ccol}[1]{\,\hfill#1\hfill\,}

\newcommand{\peinf}{P_{e}^{(\infty)}(\chi)}
\newcommand{\pein}{P_{e}^{(\infty)}}

\newtheorem{theo}{Theorem}
\newtheorem{prop}{Proposition}

\newtheorem{defi}{Definition}

\algrenewcommand\algorithmicforall{\hspace*{-.5\parindent}\textbf{for all}}
\algrenewcommand\algorithmicif{\hspace*{-.5\parindent}\textbf{if}}

\newlength{\EndIterLen}
\algblockdefx[Init]{Init}{EndInit}{\hspace*{-.4\parindent}\textbf{Initialization}}{}
\algblockdefx[Iter]{Iter}{EndIter}{\hspace*{-.4\parindent}\textbf{Iteration Loop}}%
{\hspace*{-.35\parindent}\textbf{End Iteration Loop}\settowidth{\EndIterLen}{\textbf{End Iteration Loop}}}
\algblockdefx[MyWhile]{MyWhile}{EndMyWhile}{\textbf{While}}%
{\textbf{End While}\settowidth{\EndIterLen}{\textbf{End Iteration Loop}}}
\algtext*{EndInit}

\IEEEoverridecommandlockouts
\title{Density Evolution and Functional Threshold for the Noisy Min-Sum Decoder \\ {\LARGE(Extended Version)}}
\author{%
\IEEEauthorblockN{Christiane L. Kameni Ngassa$^{*,\#}$, Valentin Savin$^*$, Elsa Dupraz$^\#$, David Declercq$^\#$} \\%
\IEEEauthorblockA{%
$^*$CEA-LETI, Minatec Campus, Grenoble, France \\ \{christiane.kameningassa, valentin.savin\}@cea.fr \\
$^\#$ETIS, ENSEA / CNRS UMR-8051 / Univ. Cergy-Pontoise, France \\ \{elsa.dupraz, declercq\}@ensea.fr}%
\thanks{This work was supported by the Seventh Framework Programme of the European Union, under Grant Agreement number 309129 ({\em i}-RISC project).}%
\thanks{This paper is an extended version of the paper with same title, submitted to IEEE Transactions on Communications}%
\vspace*{-5mm}}

\begin{document} 
\maketitle

\begin{abstract}
\,This paper investigates the behavior of the Min-Sum decoder running on noisy devices. The aim is to evaluate the robustness of the decoder in the presence of computation noise, {\em e.g.} due to faulty logic in the processing units, which represents a new source of errors that may occur during the decoding process. To this end, we first introduce probabilistic models for the arithmetic and logic units of the the finite-precision Min-Sum decoder, and then carry out the density evolution analysis of the noisy Min-Sum decoder. We show that in some particular cases, the noise introduced by the device can help the Min-Sum decoder to escape from fixed points attractors, and may actually result in an increased correction capacity with respect to the noiseless decoder. We also reveal the existence of a specific threshold phenomenon, referred to as functional threshold. The behavior of the noisy decoder is demonstrated in the asymptotic limit of the code-length -- by using ``noisy'' density evolution equations -- and it is also verified in the finite-length case by Monte-Carlo simulation.
\end{abstract}




\section{Introduction}
In traditional models of communication or storage systems with error correction coding, it is assumed that the operations of an error correction encoder and decoder are deterministic and that the randomness exists only in the transmission or storage channel. However, with the advent of nanoelectronics, the reliability of the forthcoming circuits and computation devices is becoming questionable. 
It is then becoming crucial to design and analyze error correcting decoders able to provide reliable error correction even if they are made of unreliable components. 

Except the pioneered works by Taylor and Kuznetsov on reliable memories \cite{taylor1968reliablestorage, taylor1968reliablecomputing, kuznetsov1973information}, later  generalized in \cite{chilappagari2006analysis, vasic2007information} to the case of  hard-decision decoders, this new paradigm of noisy decoders has merely not been addressed until recently in the coding literature. However, over the last years, the study of error correcting decoders, especially Low-Density Parity-Check (LDPC) decoders, running on noisy hardware attracted more and more interest in the coding community. In \cite{winstead2009probabilistic} and \cite{tang2012ldpc} hardware redundancy is used to develop fault-compensation techniques, able to protect the decoder against the errors induced by the noisy components of the circuit. In \cite{hussien2011class}, a class of modified Turbo and LDPC decoders has been proposed, able to deal with the noise induced by the failures of a low-power buffering memory that stores the input soft bits of the decoder. Very recently, the characterization of the effect of noisy processing on message-passing iterative LDPC decoders has been proposed. In \cite{varshney2011performance}, the concentration and convergence properties were proved for the asymptotic performance of noisy message-passing decoders, and density evolution equations were derived for the noisy Gallager-A and Belief-Propagation decoders.  In \cite{yazdi2012probabilistic, yazdi2012optimal, yazdi2013gallager}, the authors investigated the asymptotic behavior of the noisy Gallager-B decoder defined over binary and non-binary alphabets. The Min-Sum decoding under unreliable message storage has been investigated in \cite{balatsoukas2014characterization, balatsoukas2014density}. However, all these papers deal with very simple error models, which emulate the noisy implementation of the decoder, by passing each of the exchanged messages through a noisy channel.

In this work we focus on the Min-Sum  decoder, which is widely implemented in real communication systems. In order to emulate the noisy implementation of the decoder, probabilistic error models are proposed for its arithmetic components (adders and comparators). The proposed probabilistic components are used to build the noisy finite-precision decoders. We further analyze the asymptotic performance of the noisy Min-Sum decoder, and provide useful regions and target-BER-thresholds \cite{varshney2011performance} for a wide range of parameters of the proposed error models. 
 We also highlight a wide variety of more or less conventional behaviors and reveal the existence of a specific threshold phenomenon, which is referred to as {\em functional threshold}. Finally, the asymptotic results are also corroborated through finite length simulations.

The remainder of the paper is organized as follows. Section~\ref{sec:ldpc_codes_and_ms} gives a brief introduction to LDPC codes and iterative decoding. Section~\ref{sec:error_injection} presents the error models for the arithmetic components. The density evolution equations and asymptotic analysis of the noisy finite-precision Min-Sum decoding are presented in Section~\ref{sec:density_evolution} and Section~\ref{sec:asymptotic-analysis} respectively. Section~\ref{sec:finite_length_perf} provides the finite-length performance and Section~\ref{sec:noisy_ms_conslusion} concludes the paper. 


\section{LDPC Codes and the Min-Sum Algorithm}\label{sec:ldpc_codes_and_ms}

\subsection{LDPC Codes}

LDPC codes \cite{gallager1963low} are linear block codes defined by sparse parity-check matrices. They can be advantageously represented by bipartite (Tanner) graphs \cite{tanner1981recursive} and decoded by message-passing (MP) iterative algorithms. The Tanner graph of an LDPC code is a bipartite graph ${\cal H}$, whose adjacency matrix is the parity-check matrix $H$ of the code. Accordingly, ${\cal H}$ contains two types of nodes:    
\begin{itemize}
\item {\em variable-nodes}, corresponding to the columns of $H$, or equivalently to the codeword bits, and
\item {\em check-nodes}, corresponding to the rows of $H$, or equivalently to the parity equations the codeword bits are checked by.
\end{itemize} 

We consider an LDPC code defined by a Tanner graph ${\cal H}$, with $N$ variable-nodes and $M$ check-nodes. Variable-nodes are denoted by $n \in \{1, 2, ... , N\}$, and check-nodes by $m \in \{1, 2, ... , M\}$. We denote by ${\cal H}(n)$ and ${\cal H}(m)$ the {\em set of neighbor nodes} of the variable-node $n$ and of the check-node $m$, respectively. The number of elements of ${\cal H}(n)$ $\left( \mbox{or } {\cal H}(m)\right)$ is referred to as the {\em node-degree}.

The Tanner graph representation allows reformulating the {\em probabilistic decoding} initially proposed by Gallager \cite{gallager1963low} in terms of  Belief-Propagation\footnote{Also referred to as Sum-Product (SP)} (BP): an MP algorithm proposed by J. Pearl in 1982 \cite{pearl1982bp}  to perform Bayesian inference on trees, but also successfully used on general graphical models \cite{pearl1988probabilistic}. The BP decoding is known to be optimal on cycle-free graphs (in the sense that it outputs the maximum a posteriori estimates of the coded bits), but can also be successfully applied to decode linear codes defined by graphs with cycles, which is actually the case of all practical codes. However, in practical applications, the BP algorithm might be disadvantaged by its computational complexity and 
its sensitivity to the channel noise density estimation (inaccurate estimation of the channel noisy density may cause significant degradation of the BP performance).

\subsection{Min-Sum Decoding}\label{subsec:ms-decoding}

One way to deal with complexity and numerical instability issues is to simplify the computation of messages exchanged within the BP decoding.  The most complex step of the BP decoding is the computation of check-to-variable messages, which makes use of computationally intensive hyperbolic tangent functions. The Min-Sum (MS) algorithm is aimed at reducing the computational complexity of the BP, by using max-log approximations of the parity-check to coded-bit messages \cite{fossorier1999reduced, chung2000construction, eleftheriou2001reduced}. The only operations required by the MS decoding are additions, comparisons, and sign ($\pm 1$) products, which solves the complexity and numerical instability problems. The performance of the MS decoding is also known to be independent of the channel noise density estimation, for most of the usual channel models.

For the sake of simplicity, we only consider transmissions over {\em binary-input} memoryless noisy channels, and assume that the channel input alphabet is $\{-1, +1\}$, with the usual convention that $+1$ corresponds to the $0$-bit, and $-1$ corresponds to the $1$-bit.
We further consider a codeword $\vect{x} = (x_1,\dots,x_N) \in\{-1, +1\}^N$  and denote by $\vect{y} = (y_1,\dots,y_N)$ the received word. The following notation will be used throughout the paper, with respect to message passing decoders:
\begin{itemize}
\item	$\gamma_n$ is the log-likelihood ratio (LLR) value of $x_n$ according to the received $y_n$ value; it is also referred to as the {\em a priori information} of the decoder concerning the variable-node $n$;
\item	$\tilde{\gamma}_n$ is the {\em a posteriori information} (LLR value) of the decoder concerning the variable-node $n$;
\item	$\alpha_{m,n}$ is the variable-to-check message sent from variable-node $n$ to check-node $m$;
\item	$\beta_{m,n}$ is the check-to-variable message sent from check-node $m$ to variable-node $n$.
\end{itemize}

The (infinite precision) MS decoding is described in Algorithm~\ref{alg:ms}. It consists of an initialization step (in which variable-to-check messages are initialized according to the a priori information of the decoder), followed by an iteration loop, where each  iteration comprises three main steps as follows:   

\begin{itemize}
\item {\bf CN-processing} (check-node processing step): computes the check-to-variable messages $\beta_{m,n}$;
\item {\bf VN-processing} (variable-node processing step): computes the variable-to-check messages $\alpha_{m,n}$;
\item {\bf AP-update} (a posteriori information update step): computes the a posteriori information $\tilde{\gamma}_n$.
\end{itemize}

Moreover, each iteration also comprises a {\em hard decision} step, in which each transmitted bit is estimated according to the sign of the a posteriori information, and a {\em syndrome check} step, in which the syndrome of the estimated word is computed. The MS decoding stops when whether the syndrome is $+1$ (the estimated word is a codeword) or a maximum number of iterations is reached.


\begin{algorithm}[!t]
\caption{Min-Sum (MS) decoding }\label{alg:ms}
\begin{algorithmic}[0]
\State{Input:}  $\vect{y} = (y_1,\dots,y_N)\in{\cal Y}^N$ (${\cal Y}$ is the channel output alphabet)\Comment{received word}
\State{Output:} $\hat{\vect{x}} = (\hat{x}_1,\dots,\hat{x}_N)\in\{-1, +1\}^N$ \Comment{estimated codeword}  

\Init 
   \ForAll{$n=1,\dots,N$} 
          $\gamma_{n} = \displaystyle\log\frac{\Pr(x_n=+1\mid y_n)}{\Pr(x_n=-1\mid y_n)}$;   
   \EndFor
   \ForAll{$n=1,\dots,N$ and $m\in{\cal H}(n)$} 
          $\alpha_{m,n} = \gamma_n$;
   \EndFor
\EndInit
\Iter 
   \ForAll{$m=1,\dots,M$ and $n\in{\cal H}(m)$} \
       \Comment{{\bf CN-processing}}
       \State 
       $\beta_{m,n}=\displaystyle\left(\prod_{{n'}\in{\cal H}(m)\setminus n} 
       \!\!\!\!\!\sgn(\alpha_{m,n'})\right)\left(\min_{{n'}\in{\cal H}
       (m)\setminus n}\!\!\!\!\!|\alpha_{m,n'}|\right)$;
   \EndFor
   
   \ForAll{$n=1,\dots,N$ and $m\in{\cal H}(n)$} \
       \Comment{{\bf VN-processing}}       
		\State $\alpha_{m,n} = \displaystyle \gamma_n + \sum_{{m'}
		        \in{\cal H}(n)\setminus m} \beta_{m',n}$; 
   \EndFor
	
	\ForAll{$n=1,\dots,N$}\ 
	    \Comment{{\bf AP-update}}        
        \State $\tilde{\gamma}_{n}=\displaystyle \gamma_n + \sum_{{m}
		        \in{\cal H}(n)} \beta_{m,n}$;
	\EndFor
	
	\ForAll{$\{v_n\}_{n=1,\dots,N}$} $\hat{x}_n = \sgn(\tilde{\gamma}_{n})$; \ 
	    \Comment{hard decision}
	\EndFor
	
	\If{$\hat{\vect{x}}$ is a codeword} exit the iteration loop \Comment{syndrome check}
	\EndIf
\EndIter
%
%
\end{algorithmic}
\end{algorithm}

The a priori information (LLR) of the decoder is defined by
$\displaystyle\gamma_n = \log\frac{\Pr(x_n=+1\mid y_n)}{\Pr(x_n=-1\mid y_n)}$, and
for the two following channel models (predominantly used in this work), it can be computed as follows:
\begin{itemize}
\item For the Binary Symmetric Channel (BSC), $\vect{y}\in \{-1,+1\}^N$ is obtained by flipping each entry of $\vect{x}$ with some probability $\varepsilon$, referred to as the channel's crossover probability. Consequently:
\begin{equation}
\gamma_n = \log\left(\frac{1-\varepsilon}{\varepsilon}\right)y_n
\end{equation}
\item For the Binary-Input Additive White Gaussian Noise (BI-AWGN) channel,  $\vect{y}\in \mathbb{R}^N$ is obtained by $y_n = x_n + z_n$, where $z_n$ is the white Gaussian noise with variance $\sigma^2$. It follows that:   
\begin{equation}
\gamma_n = \frac{2}{\sigma^2}y_n
\end{equation}
\end{itemize}

\noindent{\bf Remark:} It can be easily seen that if the a priori information vector $\vect{\gamma}=(\gamma_1,\dots,\gamma_N)$ is multiplied by a constant value, this value will factor out from all the processing steps in Algorithm~\ref{alg:ms} (throughout the decoding iterations), and therefore it will not affect in any way the decoding process. It follows that for both the BSC and BI-AWGN channel models, one can simply define the a priori information of the decoder by $\gamma_n = y_n$, $\forall n = 1,\dots, N$.

\section{Error Injection and Probabilistic Models for Noisy Computing}\label{sec:error_injection}

\subsection{Noisy Message-Passing decoders}
The model for noisy MP decoders proposed in \cite{varshney2011performance} incorporates two different sources of noise: {\em computation noise} due to noisy logic in the processing units, and {\em message-passing noise} due to noisy wires (or noisy memories) used to exchange messages between neighbor nodes. 
\begin{itemize}
\item The computation noise is modeled as a random variable, which the variable-node or the check-node processing depends on. 
Put differently, an outgoing message from a (variable or check) node depends not only on the incoming messages to that node (including the a priori information for the variable-node processing), but also on the realization of a random variable which is assumed to be independent of the incoming messages. 
\item The message-passing noise is simply modeled as a noisy channel. Hence, transmitting a message over a noisy wire is emulated by passing that message through the corresponding noisy channel.
\end{itemize}

However, in \cite{varshney2011performance} it has been noted that {\em ``there is no essential loss of generality by combining computation noise and message-passing noise into a single form of noise''} (see also \cite[Lemma 3.1]{dobrushin1977lower}). Consequently, the approach adopted has been to merge noisy computation into message-passing noise, and to emulate noisy decoders by passing the exchanged messages through different noisy channel models. Thus, the noisy Gallager-A decoder has been emulated by passing the exchanged messages over independent and identical BSC wires, while the noisy BP decoder has been emulated by corrupting the exchanged messages with bounded and symmetrically distributed additive noise ({\em e.g.} uniform noise or truncated Gaussian noise).

\medskip The approach we follow in this work differs from the one in \cite{varshney2011performance} in that the computation noise is modeled at the lower level of {\em arithmetic and logic operations} that compose the variable-node and check-node processing units. This finer-grained noise modeling is aimed at determining the level of noise that can be tolerated in each type of operation. As the main focus of this work is on computation noise, we shall consider that messages are exchanged between neighbor nodes through error-free wires (or memories). However, we note that {\em this work can readily be extended to include different error models for the message-passing noise} (as defined in \cite{varshney2011performance}). Alternatively, we may assume that the message-passing noise is merged into the computation noise, in the sense that adding noise in wires would modify the probabilistic model of the noisy logic or arithmetic operations.

\subsection{Error Injection Models}
We only consider the case of finite-precision operations, meaning that the inputs (operands) and the output of the operator are assumed to be bounded integer numbers. We simulate a noisy operator by injecting errors into the output of the noiseless one. In the following, ${\cal V} \subset {\mathbb Z}$ denotes a finite set consisting of all the possible outputs of the noiseless operator.

\begin{defi}
An {\em error injection model} on ${\cal V}$, denoted by $({\cal E}, p_{{\cal E}}, \imath \mid {\cal V})$, is given by:
\begin{itemize}
\item A finite {\em error set} ${\cal E}\subset {\mathbb Z}$ together with a probability mass function  $p_{{\cal E}}:{\cal E} \rightarrow [0, 1]$, referred to as the {\em error distribution};
\item A function $\imath : {\cal V} \times {\cal E} \rightarrow {\cal V}$, referred to as the {\em error injection function}.
\end{itemize} 
\end{defi}

For a given set of inputs, the output of the noisy operator is the random variable defined by $\imath(v, e)$, where $v\in{\cal V}$ is the corresponding output of the noiseless operator, and $e$ is drawn randomly from ${\cal E}$ according to the probability distribution $p_{{\cal E}}$.

The error injection probability is defined by
\begin{equation}\label{eq:err-injection-proba}
p_0 = \frac{1}{|{\cal V}|} \sum_{v}\sum_{e} \bar{\delta}_{\imath(v,e)}^v p_{\cal E}(e),
\end{equation}
where $\bar{\delta}_{\imath(v,e)}^v = 0$ if $v = \imath(v,e)$, and $\bar{\delta}_{\imath(v,e)}^v = 1 $ if $v \neq \imath(v,e)$. In other word, $p_0 = \Pr(v \neq \imath(v, e))$, assuming that $v$ is drawn uniformly from ${\cal V}$ and $e$ is drawn from ${\cal E}$ according to $p_{{\cal E}}$.

 The above definition makes some implicit assumptions which are discussed below.
\begin{itemize}
\item The set of possible outputs of the noisy operator is the same as the set of  possible outputs of the noiseless operator (${\cal V}$). This is justified by the fact that, in most common cases, ${\cal V}$ is the set of all (signed or unsigned) integers that can be represented by a given number of bits. Thus, error injection will usually alter the bit values, but not the number of bits.
\item The injected error does not depend on the output value of the noiseless operator and, consequently, neither on the given set of inputs. In other words, the injected error is independent on the data processed by the noiseless operator. The validity of this assumption does actually depend on the size of the circuit implementing the operator. Indeed, this assumption tends to hold fairly well for large circuits \cite{i-RISC-D2.1}, but becomes more tenuous as the circuit size decreases. 
\end{itemize}

Obviously, it would be possible to define more general error injection models, in which the injected error would depend on the data (currently and/or previously) processed by the operator. Such an error injection model would certainly be more realistic, but it would also make it very difficult to analytically characterize the behavior of noisy MP decoders. As a side effect, the decoding error probability would be dependent on the transmitted codeword, which would prevent the use of the {\em density evolution} technique for the analysis of the asymptotic decoding performance (since the density evolution technique relies on the all-zero codeword assumption). 

\medskip However, the fact that the error injection model is data independent does not guarantee that the decoding error probability is independent of the transmitted codeword. In order for this to happen, the error injection model must also satisfy a {\em symmetry condition} that can be stated as follows. 

\begin{defi} 
An error injection model $({\cal E}, p_{{\cal E}}, \imath \mid {\cal V})$ is said to be {\em symmetric} if ${\cal V}$ is symmetric around the origin (meaning that $v\in{\cal V} \Leftrightarrow -v\in{\cal V}$, but $0$ does not necessarily belong to ${\cal V}$), and 
the following equality holds
\begin{equation}
\sum_{\{e \mid\, \imath(v,e) = w\}} p_{\cal E}(e) \ \ = \sum_{\{e \mid\, \imath(-v,e) = -w\}} p_{\cal E}(e), \ \ \forall v, w\in {\cal V}
\end{equation}
\end{defi}

The meaning of the symmetry condition is as follows. Let $V$ be a random variable on ${\cal V}$. Let $\phi^{(\imath)}_{V}$ and $\phi^{(\imath)}_{-V}$ denote the probability mass functions of the random variables obtained by injecting errors in the output of $V$ and $-V$, respectively.  Then the above symmetry condition is satisfied if and only if  for any $V$ the following equality holds
\begin{equation}
\phi^{(\imath)}_{V}(w) = \phi^{(\imath)}_{-V}(-w), \ \forall w\in{\cal V} 
\end{equation}
A particular case in which the symmetry condition is fulfilled is when $\imath(-v, e) = -\imath(v, e)$, for all $v\in {\cal V}$ and $e\in {\cal E}$. In this case, the error injection model is said to be {\em highly symmetric}.

\medskip Messages exchanged within message-passing decoders are generally in {\em belief-format}, meaning that the sign of the message indicates the bit estimate and the magnitude of the message the confidence level. As a consequence, errors occurring on the sign of the exchanged messages are expected to be more harmful than those occurring on their magnitude.  
This motivates the following definition, which will be used in the following section (see also the discussion in Section~\ref{subsec:unit-sign-prez}). 

\begin{defi}
An error injection model $({\cal E}, p_{{\cal E}}, \imath \mid {\cal V})$ is said to be {\em sign-preserving} if for any $v\in{\cal V}$ and $e\in{\cal E}$,  $v$ and $\imath(v, e)$ are either both non-negative ($\geq 0$) or both non-positive ($\leq 0$). 
\end{defi}

\subsection{Bitwise-XOR Error Injection}\label{subsec:bitwise-xor-injection}

We focus now on the two main symmetric error injection models that will be used in this work. Both models are based on a bitwise {\sc xor} operation between the noiseless output $v$ and the error $e$. The two models differ in the definition of the error set ${\cal E}$, which is chosen such that the bitwise {\sc xor} operation may or may not affect the sign of the noiseless output. In the first case, the bitwise {\sc xor} error injection model is said to be {\em full-depth}, while in the second it is said to be {\em sign-preserving}. These error injection models are rigorously defined below.

\medskip In the following, we fix $\theta \geq 2$ and set ${\cal V} = \{-\Theta, \dots, -1, 0, +1, \dots, +\Theta\}$, where $\Theta = 2^{\theta-1}-1 \geq 1$.\break We also fix a {\em signed number binary representation}, which can be any of the {\em sign-magnitude}, {\em one's complement}, or {\em two's complement} representation. There are exactly $2^\theta$ signed numbers that can be represented by $\theta$ bits in any of the above formats, one of which does not belong to ${\cal V}$ (note that ${\cal V}$ contains only $2\Theta + 1 = 2^\theta-1$ elements for symmetry reasons!). We denote this element by $\zeta$. Hence:
\begin{itemize}
\item In sign-magnitude format, $\zeta = -0$, with binary representation $10\cdots0$;
\item In one's complement format, $\zeta = -0$, with binary representation $11\cdots1$;
\item In two's complement format, $\zeta = -(\Theta+1)$, with binary representation $10\cdots0$.
\end{itemize}
For any $u,v\in {\cal V}$, we denote by $u\wedge v$ the bitwise {\sc xor} operation between  $u$ and $v$. From the above discussion, it follows that $u\wedge v \in {\cal V}\cup \{\zeta\}$.

\subsubsection{Full-depth error injection} For this error model, the error set is  ${\cal E} = {\cal V}$. 
The error injection probability is denoted by $p_0$, and all the possible error values $e\neq 0$ are assumed to occur with the same probability (for symmetry reasons). It follows that the error distribution function is given by $p_{\cal E}(0) = 1-p_0$ and $p_{\cal E}(e) = \frac{p_0}{2\Theta}$, $\forall e\neq 0$.  Finally, the error injection function is defined by:
\begin{equation}
\imath(v, e) = \left\{\begin{array}{cl}
  v\wedge e, & \mbox{if }   v\wedge e \in {\cal V}\\
  e, & \mbox{if }   v\wedge e = \zeta\end{array}\right.
\end{equation}
  
  \subsubsection{Sign-preserving error injection} For this error model, the error set is  ${\cal E} = \{0, +1,\dots, +\Theta\}$.
The error injection probability is denoted by $p_0$, and all the possible error values $e\neq 0$ are assumed to occur with the same probability (for symmetry reasons). It follows that the error distribution function is given by $p_{\cal E}(0) = 1-p_0$ and $p_{\cal E}(e) = \frac{p_0}{\Theta}$, $\forall e\neq 0$.  Finally, the error injection function is defined by:
\begin{equation}
\imath(v, e) = \left\{\begin{array}{cl}
  v\wedge e, & \mbox{if }  v \neq 0 \mbox{ and } v\wedge e \in {\cal V}\\
  \pm e, & \mbox{if } v = 0 \\
  0, & \mbox{if }   v\wedge e = \zeta\end{array}\right.
\end{equation}
  In the above definition, $\imath(0, e)$ is randomly set  to either $-e$ or $+e$, with equal probability (this is due once again to symmetry reasons). Note also that the last two conditions, namely  $v = 0$ and $v\wedge e = \zeta$, cannot hold simultaneously (since $e\neq \zeta$).
  
  Finally, we note that both of the above models are highly symmetric, if one of the sign-magnitude or the one's complement representation is used. In case that the two's complement representation is used, they are both symmetric, but not highly symmetric.  
  
  An example of sign-preserving bitwise-{\sc xor} error injection is given in Table~\ref{tab:sign-preserve-bitwise-xor}. The number of bits is $\theta = 5$ and two's complement binary representation is used. The sign bit of the error is not displayed, as it is equal to zero for any $e\in {\cal E}$.  The positions of $1$'s in the binary representation of $e$ correspond to the positions of the erroneous bits in the noisy output. 
 
\begin{table}[t!]
\begin{center}
\caption{Example of sign-preserving bitwise-xor error injection}     
\label{tab:sign-preserve-bitwise-xor}
\begin{tabular}{|@{\,}r@{\,}|c@{\,}||*{5}{@{\,}m{12.5mm}@{\,}|}}
\hline
    & integer & \multicolumn{5}{c|}{$2$'s complement binary representation}\\
\hline
noiseless output: $v$ & $-11$ &\ccol{$1$}&\ccol{$0$}&\ccol{$1$}&\ccol{$0$}&\ccol{$1$}\\
\hline
error: $e$ &  $6$  &          &\ccol{$0$}&\ccol{$1$}&\ccol{$1$}&\ccol{$0$}\\
\hline
noisy output: $\imath(v, e)$ & $-13$  & \ccol{$1$} & \ccol{$0$} & \ccol{$\textcolor{red}0$} & \ccol{$\textcolor{red}1$} & \ccol{$1$} \\
\hline
\multicolumn{2}{|@{\,}r@{\,}||@{\,}}{bit position} &  \ccol{$\theta\!=\!5$} &  \ccol{$4$} & \ccol{$3$} &  \ccol{$2$} &   \ccol{$1$} \\
 \hline
\end{tabular}
\end{center}
\end{table}

\medskip\noindent{\bf Remark:} It is also possible to define a {\em variable depth error injection} model, in which errors are injected in only the $\lambda$ least significant bits, with $\lambda\leq \theta$. Hence, $\lambda = \theta$ corresponds to the above full-depth model, while $\lambda = \theta-1$ corresponds to the sign-preserving model. However, for  $\lambda < \theta-1$ such a model is {\bf not} symmetric, if the the two's complement representation is used. 
  
 \subsection{Output-Switching Error Injection}\label{subsec:output-swithcing}
 A particular case is represented by error injection on binary output. Assuming that  ${\cal V} = \{0, 1\}$,  
 the {\em bit-flipping} error injection model is defined as follows. The error set is  ${\cal E} = \{0, 1\}$,  with error distribution function given by $p_{\cal E}(0) = 1-p_0$ and $p_{\cal E}(1) = p_0$, where $p_0$ is the error injection probability, and the error injection function is given by $\imath(v, e) = v\wedge e$. Put differently, the error injection model flips the value of a bit in ${\cal V}$ with probability $p_0$. 
 
 Clearly, the above error injection model can be applied on any set ${\cal V}$ with two elements, by switching one value to another with probability $p_0$. In this case, we shall refer to this error injection model as {\em output-switching}, rather than bit-flipping.
 
 Moreover, if one takes  ${\cal V} = \{-1, +1\}$ (with the usual $0{,}1$ to $\pm 1$ conversion), it can be easily verified that this error injection model is highly symmetric.

 \subsection{Probabilistic models for noisy adders, comparators and XOR-gates}\label{subsec:proba-models-operators}
 In this section we describe the probabilistic models for noisy adders, comparators and xor-gates, built upon the above error injection models. These probabilistic models will be used in the next section, in order to emulate the noisy implementation of the quantized (finite-precision) MS decoder.
 
 \subsubsection{Noisy adder model} We consider a $\theta$-bit adder, with $\theta \geq 2$. The inputs and the output of the adder are assumed to be in ${\cal V} = \{-\Theta, \dots, -1, 0, +1, \dots, +\Theta\}$, where $\Theta = 2^{\theta-1}-1$. 
 
We denote by $\mathbf{s}_{\cal V} : \mathbb{Z}\rightarrow {\cal V}$, the $\theta$-bit saturation map, defined by: 
\begin{equation}\label{eq:saturation-map}
\mathbf{s}_{\cal V}(v) = \left\{\begin{array}{ll}
-\Theta, & \mbox{if } v < -\Theta \\
\ \ v,  & \mbox{if } v \in {\cal V}\\
+\Theta,  & \mbox{if } v > +\Theta \end{array}\right.
\end{equation}
 
 For inputs $(x, y)\in{\cal V}$, the output of the noiseless adder is defined as $s_{\cal V}(x + y)$. Hence, for a given error injection model $({\cal E}, p_{{\cal E}}, \imath \mid {\cal V})$, the output of the noisy adder is given by:
 \begin{equation}\label{eq:noisy-add-def}
 \addpr(x, y) =   \imath\left(\mathbf{s}_{\cal V} (x+y), e\right),
 \end{equation}
 where $e$ is drawn randomly from ${\cal E}$ according to the probability distribution $p_{{\cal E}}$. The {\em error probability of the noisy adder}, assuming uniformly distributed inputs, is equal to the error injection probability (parameter $p_0$ defined in (\ref{eq:err-injection-proba})), and will be denoted in the sequel by $\padd$.

\subsubsection{Noisy comparator model} Let $\lth$ denote the noiseless {\em less than} operator, defined by $\lth(x, y) = 1$ if $x < y$, and $\lth(x, y) = 0$ otherwise.
 The {\em noisy less than} operator, denoted by $\lthpr$, is defined by injecting errors on the output of the noiseless one, according to the bit-flipping model defined in  Section~\ref{subsec:output-swithcing}. In other words, the output of the noiseless $\lth$ operator is flipped with some probability value, which will be denoted in the sequel by $\pcomp$. 
 
 Finally, the {\em noisy minimum} operator is defined by:
 \begin{equation}\label{eq:noisy-comp-def}
 \minpr(x, y) = \left\{\begin{array}{rl}
 x, & \mbox{if } \lthpr(x, y) = 1\\
 y, & \mbox{if } \lthpr(x, y) = 0
 \end{array}\right.
 \end{equation}
 
 \subsubsection{Noisy XOR model} The noisy {\sc xor} operator, denoted by $\xorpr$ is defined by flipping the output of the noiseless operator with some probability value, which will be denoted in the sequel by $\pxor$ (according to the bit-flipping error injection model in Section~\ref{subsec:output-swithcing}). It follows that:
 \begin{equation}\label{eq:noisy-xor-def}
 \xorpr(x, y) = \left\{\begin{array}{rl}
 x\wedge y, & \mbox{with probability } 1-\pxor\\
 \overline{x\wedge y}, & \mbox{with probability } \pxor
 \end{array}\right.
 \end{equation}

\medskip\noindent{\bf Assumption:} We further assume that the inputs and the output of the {\sc xor} operator may take values in either $\{0, 1\}$ or $\{-1, +1\}$ (using the usual $0{,}1$ to $\pm 1$ conversion). This assumption will be implicitly made throughout the paper.

\medskip\noindent{\bf Remark:} As a general rule, we shall refer to a noisy operator according to its underlying error injection model. For instance, a sign-preserving (resp. full-depth or sign-preserving bitwise-{\sc xor}ed) noisy adder, is a noisy adder whose underlying error injection model is sign-preserving (resp. one of the bitwise-{\sc xor} error injection models defined in Section~\ref{subsec:bitwise-xor-injection}). We shall also say that a noisy operator is {\em (highly) symmetric} if its underlying error injection model is so.
 
\subsection{Nested Operators}
As it can be observed from Algorithm~\ref{alg:ms}, several arithmetic/logic operations must be nested\footnote{For instance, $(d_n-1)$ additions -- where $d_n$ denotes the degree of the variable-node $n$ -- are required in order to compute each $\alpha_{m,n}$ message. Similarly, each  $\beta_{m,n}$ message requires $(d_m-2)$ {\sc xor} operations and $(d_m-2)$ comparisons.} in order to compute the exchanged messages. Since all these operations (additions, comparisons, {\sc xor}) are commutative, the way they are nested does not have any impact on the infinite-precision MS decoding. However, this is no longer true for finite-precision decoding, especially in case of noisy operations. Therefore, one needs an assumption about how operators extend from two to more inputs. 

\medskip Our assumption is the following. For $n \geq 2$ inputs, we randomly pick any two inputs and apply the operator on this pair. Then we replace the pair by the obtained output, and repeat the above procedure until there is only one output (and no more inputs) left.


The formal definition goes as follows. Let $\Omega\subset {\mathbb Z}$ and $\omega: \Omega\times \Omega\rightarrow \Omega$ be a noiseless or noisy operator with two operands. Let $\{x_i\}_{i=1:n} \subset \Omega$ be an unordered set of $n$ operands. We define:
$$\omega\left(\{x_i\}_{i=1:n}\right) = \omega(\cdots(\omega(x_{\pi(1)}, x_{\pi(2)}), \cdots), x_{\pi(n)}),$$
where $\pi$ is a random permutation of $1,\dots,n$.

\section{Noisy Min-Sum Decoding}\label{sec:noisy_ms_decoding}

\subsection{Finite-Precision Min-Sum Decoder}\label{subsec:finite-prec-dec-notation}
We consider a finite-precision MS decoder, in which the a priori information ($\gamma_n$) and the exchanged messages ($\alpha_{m,n}$ and $\beta_{m,n}$) are quantized on $q$ bits. The a posteriori information ($\tilde{\gamma}_n$) is quantized on $\tilde{q}$ bits, with $\tilde{q}>q$ (usually $\tilde{q}=q+1$, or $\tilde{q}=q+2$). We further denote:
\begin{itemize} 
\item ${\cal M} = \{-Q,\dots,-1,0,+1,\dots,Q\}$, where $Q=2^{q-1}-1$, the alphabet of both the a priori information and the exchanged messages;
\item $\widetilde{\cal M} = \{-\widetilde{Q},\dots,-1,0,+1,\dots,\widetilde{Q}\}$, where $\widetilde{Q}=2^{\tilde{q}-1}-1$, the alphabet of the a posteriori information;
\item $\mathbf{q} : {\cal Y} \rightarrow {\cal M}$, a quantization map, where ${\cal Y}$ denotes the channel output alphabet;
\item $\mathbf{s}_{\cal M} : \mathbb{Z}\rightarrow {\cal M}$, the $q$-bit saturation map (defined in a similar manner as in (\ref{eq:saturation-map}));
\item $\mathbf{s}_{\widetilde{\cal M}} : \mathbb{Z}\rightarrow \widetilde{\cal M}$, the $\tilde{q}$-bit saturation map 
\end{itemize}

\medskip\noindent{\bf Remark:} The quantization map $\mathbf{q}$ determines the $q$-bit quantization of the decoder soft input. Since $\mathbf{q}$ is defined on the channel input ({\em i.e.} $y_n$ values), it must also encompass the computation of the corresponding LLR values, whenever is necessary (see also the Remark at the end of Section~\ref{subsec:ms-decoding}).

Saturation maps $\mathbf{s}_{\cal M}$ and  $\mathbf{s}_{\widetilde{\cal M}}$ define the finite-precision saturation of the exchanged messages and of the a posteriori information, respectively.

\subsection{Noisy Min-Sum Decoder}

The noisy (finite-precision) MS decoding is presented in Algorithm~\ref{alg:noisy_ms}.
We assume that $\tilde{q}$-bit adders are used to compute both $\alpha_{m,n}$ messages in the {\bf VN-processing} step, and $\tilde{\gamma}_n$ values in the {\bf AP-update} processing step. This is usually the case in practical implementations\footnote{In practical implementation, the $\tilde{\gamma}_n$ is computed first, and then $\alpha_{m,n}$ is obtained from $\tilde{\gamma}_n$ by subtracting the incoming $\beta_{m,n}$ message}, and allows us to use the same type of adder in both processing steps. This assumption explains as well the $q$-bit saturation of $\alpha_{m,n}$ messages in the {\bf VN-processing} step. Note also that the saturation of $\tilde{\gamma}_n$ values is actually done within the adder (see Equation~(\ref{eq:noisy-add-def})).

Finally, we note that the {\em hard decision}  and the {\em syndrome check} steps in Algorithm~\ref{alg:noisy_ms} are assumed to be {\em noiseless}. We note however that the  syndrome check step is optional, and if missing, the decoder stops when the maximum number of iterations is reached. 


\begin{algorithm}[!t]
\caption{Noisy Min-Sum (Noisy-MS) decoding }\label{alg:noisy_ms}
\begin{algorithmic}[0]
\State{Input:}  $\vect{y} = (y_1,\dots,y_N)\in{\cal Y}^N$ (${\cal Y}$ is the channel output alphabet)\Comment{received word}
\State{Output:} $\hat{\vect{x}} = (\hat{x}_1,\dots,\hat{x}_N)\in\{-1, +1\}^N$ \Comment{estimated codeword}  

\Init 
   \ForAll{$n=1,\dots,N$} 
          $\gamma_{n} = \mathbf{q}(y_n)$;   
   \EndFor
   \ForAll{$n=1,\dots,N$ and $m\in{\cal H}(n)$} 
          $\alpha_{m,n} = \gamma_n$;
   \EndFor
\EndInit
\Iter 
   \ForAll{$m=1,\dots,M$ and $n\in{\cal H}(m)$} \
       \Comment{{\bf CN-processing}}
       \State $\beta_{m,n}=\displaystyle
		\xorprred\left( \{\sgn(\alpha_{m,n'})\}_{{n'}\in{\cal H}(m)\setminus n} \right) 
		\minprred\left( \{|\alpha_{m,n'}|\}_{{n'}\in{\cal H}(m)\setminus n} \right)$;
   \EndFor
   
   \ForAll{$n=1,\dots,N$ and $m\in{\cal H}(n)$} \
       \Comment{{\bf VN-processing}}       
		\State  $\begin{array}{r@{\ }c@{\ }l}
                \alpha_{m,n}&  = & \addprred\left(\{\gamma_n\} \cup 
                \{\beta_{m',n}\}_{{m'}
		        \in{\cal H}(n)\setminus m}\right)\text{;} \\
                \alpha_{m,n}& = &\mathbf{s}_{\cal M}
                \left(\alpha_{m,n} \right)\text{;}
                \end{array}$
   \EndFor
	
	\ForAll{$n=1,\dots,N$}\ 
	    \Comment{{\bf AP-update}}        
        \State $\tilde{\gamma}_n = \addprred\left(\{\gamma_n\} \cup 
        \{\beta_{m,n}\}_{{m} \in{\cal H}(n)}\right)\text{;}$
	\EndFor
	
	\ForAll{$\{v_n\}_{n=1,\dots,N}$} $\hat{x}_n = \sgn(\tilde{\gamma}_{n})$; \ 
	    \Comment{hard decision}
	\EndFor
	
	\If{$\hat{\vect{x}}$ is a codeword} exit the iteration loop \Comment{syndrome check}
	\EndIf
\EndIter
%
%
\end{algorithmic}
\end{algorithm}

\subsection{Sign-Preserving Properties}\label{subsec:unit-sign-prez}

Let $\unit$ denote any of the VN-processing or CN-processing units of the noiseless MS decoder. We denote by $\unitpr$ the corresponding unit of the noisy MS decoder. We say that $\unitpr$ is {\em sign-preserving} if for any incoming messages and any noise realization, the outgoing message is of the same sign as the message obtained when the same incoming messages are supplied to $\unit$.  

\medskip Clearly, $\text{CN}_{\text{pr}}$ is sign-preserving if and only if the {\sc xor}-operator is noiseless ($\pxor = 0$). In case that the noisy {\sc xor}-operator severely degrades the decoder performance, it is possible to increase its reliability by using classical fault-tolerant techniques (as for instance modular redundancy, or multi-voltage design by increasing the supply voltage of the corresponding {\sc xor}-gate). The price to pay, when compared to the size or the energy consumption of the whole circuit, would be reasonable. 

\medskip Concerning the VN-processing, it is worth noting that the $\text{VN}_{\text{pr}}$ is {\bf not} sign-preserving, even if the noisy adder is. This is due to the fact that multiple adders must be nested in order to complete the VN-processing. However, a sign-preserving adder might have several benefits. First, the error probability of the sign of variable-node messages would be lowered, which would certainly help the decoder. Second, if the noisy adder is sign-preserving and all the variable-node incoming messages have the same sign, then the $\text{VN}_{\text{pr}}$  does preserve the sign of the outgoing message. Put differently, in case that all the incoming messages agree on the same hard decision, the noisy VN-processing may change the confidence level, but cannot change the decision. This may be particularly useful, especially during the last decoding iterations.

\medskip Finally, the motivation behind the sign-preserving noisy adder model is to investigate its possible benefits on the decoder performance.  If the benefits are worth it ({\em e.g.} one can ensure a target performance of the decoder), the sign-bit of the adder could be protected by using classical fault-tolerant solutions.

\section{Density Evolution}\label{sec:density_evolution}

\subsection{Concentration and Convergence Properties}

First, we note that our definition of {\em symmetry} is slightly more general than the one used in \cite{varshney2011performance}. Indeed, even if all the error injection models used within the noisy MS decoder are {\em symmetric}, the noisy MS decoder does not necessarily verify the \textsl{\textbf{symmetry}} property from \cite{varshney2011performance}. However, this property is verified in case of {\em highly symmetric} fault injection\footnote{According to the probabilistic models introduced in Section~\ref{subsec:proba-models-operators}, the noisy comparator and the noisy {\sc xor}-operator are highly symmetric, but the noisy adder does not necessarily be so!}. Nevertheless, the concentration and convergence properties proved in \cite{varshney2011performance} for \textsl{\textbf{symmetric}} noisy message-passing decoders, can easily be generalized to our definition of  {\em symmetry}. 

We summarize below the most important results; the proof relies essentially on the same arguments as in \cite{varshney2011performance}. We consider an {\em ensemble} of LDPC codes, with length $N$ and fixed degree distribution polynomials \cite{richardson2001capacity}. We choose a random code $\mathbf{C}$ from this ensemble and assume that a random codeword $\vect{x}\in\{-1, +1\}^N$ is sent over a binary-input memoryless symmetric channel. We fix some number of decoding iterations $\ell > 0$, and denote by $E^{(\ell)}_{\mathbf{C}}(\vect{x})$ the expected fraction of incorrect {\em messages}\footnote{Here, {\em ``messages''} may have any one of the three following meanings: ``variable-node messages'', or ``check-node messages'', or ``a posteriori information values''.}  at iteration $\ell$. 

\begin{theo}
Assume that all the error injection models used within the MS decoder are symmetric. Then,  the following properties hold:

\begin{enumerate}
\item{}[{\em Conditional Independence of Error}] For any decoding iteration $\ell > 0$, the
expected fraction of incorrect messages $E^{(\ell)}_{\mathbf{C}}(\vect{x})$ does not depend on $\vect{x}$. Therefore, we may define $E^{(\ell)}_{\mathbf{C}} := E^{(\ell)}_{\mathbf{C}}(\vect{x})$.

\item{} [{\em Cycle-Free Case}] If the graph of $\mathbf{C}$  contains no cycles of length $2\ell$ or less, the expected fraction of incorrect messages $E^{(\ell)}_{\mathbf{C}}$ does not depend on the code $\mathbf{C}$ or the code-length $N$, but only on the degree distribution polynomials; in this case, it will be further denoted by $E^{(\ell)}_{\infty}(\vect{x})$.

\item{} [{\em Concentration Around the Cycle-Free Case}] For any $\delta > 0$, the probability that $E^{(\ell)}_{\mathbf{C}}$ lies outside the interval $\left(E^{(\ell)}_{\infty}(\vect{x})-\delta, E^{(\ell)}_{\infty}(\vect{x})+\delta\right)$ converges to zero exponentially fast in $N$.
\end{enumerate}

\end{theo}

\subsection{Density Evolution Equations}

In this section we derive density evolution equations for the noisy finite-precision MS decoding for a regular $(d_v,d_c)$ LDPC code. The study can be easily generalized to irregular LDPC codes, simply by averaging according to the degree distribution polynomials.  

The objective of the density evolution technique is to recursively compute the probability mass functions of exchanged messages, through the iterative decoding process. This is done under the independence assumption of exchanged messages, holding in the asymptotic limit of the code length, in which case the decoding performance converges to the cycle-free case.  Due to the symmetry of the decoder, the analysis can be  further simplified by assuming that the all-zero codeword is transmitted through the channel. We note that our analysis applies to any memoryless symmetric channel. 

\medskip Let $\ell > 0$ denote the decoding iteration. Superscript ${(\ell)}$ will be used to indicate the messages and the a posteriori information computed at iteration $\ell$.  To indicate the value of a message on a randomly selected edge, we drop the variable and check node indexes from the notation (and we proceed in a similar manner for the a priori and a posteriori information). The corresponding probability mass functions are denoted as follows.
$$\begin{array}{r@{\ }c@{\ }ll}
C(z) & = & \Pr(\gamma = z), & \forall z\in{\cal M} \\
\widetilde{C}^{(\ell)}(\tilde{z})& = &\Pr(\tilde{\gamma}^{(\ell)}= \tilde{z}), &\forall \tilde{z}\in{\cal \widetilde{M}}\\
A^{(\ell)}(z)& = &\Pr\left(\alpha^{(\ell)} = z\right),& \forall z\in {\cal M}\\
B^{(\ell)}(z)& = &\Pr\left(\beta^{(\ell)} = z\right), &\forall z\in {\cal M}
\end{array}$$

\subsubsection{Expression of the input probability mass function $C$} \label{subsubsec:express-C}
The probability mass function $C$ depends only on the channel and the quantization map $\mathbf{q} : {\cal Y} \rightarrow {\cal M}$, where ${\cal Y}$ denotes the channel output alphabet (Section \ref{subsec:finite-prec-dec-notation}). We also note that for $\ell = 0$, we have $A^{(0)} = C$.

We give below the expression of $C$ for the BSC and the BI-AWGN channel models (see Section~\ref{subsec:ms-decoding}). For the BSC, the channel output alphabet is ${\cal Y} = \{-1, +1\}$, while for the BI-AWGN channel, ${\cal Y} = {\mathbb R}$.

Let $\mu$ be a positive number, such that $\mu \leq Q$. The quantization map $\mathbf{q}_{\mu}$ is defined as follows:
\begin{equation} \label{eq:quant-map}
\mathbf{q}_{\mu} : {\cal Y} \rightarrow {\cal M}, \ \ \mathbf{q}_{\mu}(y) = \mathbf{s}_{\cal M}([\mu{\cdot}y]),
\end{equation}
where $[\mu{\cdot}y]$ denotes the nearest integer to $\mu{\cdot}y$, and $\mathbf{s}_{\cal M}$ is the saturation map (Section~\ref{subsec:finite-prec-dec-notation}). For the BSC, we will further assume that $\mu$ is an integer.  It follows that  $\mathbf{q}_{\mu}(y) = \mu{\cdot}y$, $\forall y \in {\cal Y} = \{-1, +1\}$. 

Considering the all-zero ($+1$) codeword assumption, the probability mass function $C$ can be computed as follows.
\begin{itemize}
\item For the BSC with crossover probability $\varepsilon$:
\begin{equation}
  C(z) = \left\{\begin{array}{cl} 
         1-\varepsilon, & \mbox{if } z =  \mu \\
          \varepsilon,  & \mbox{if } z = -\mu \\
          0, & \mbox{otherwise}
         \end{array}\right.
\end{equation}

\item For the BI-AWGN channel with noise variance $\sigma^2$:
\begin{equation}\label{eq:awgn_c}
  C(z) = \left\{\begin{array}{cl} 
          1-q\left(\frac{-Q+0.5-\mu}{\mu\sigma}\right), & \mbox{if } z =  -Q \\
          q\left(\frac{z-0.5-\mu}{\mu\sigma}\right) - q\left(\frac{z+0.5-\mu}{\mu\sigma}\right),  
          & \mbox{if } -Q < z < +Q  \\
          q\left(\frac{Q-0.5-\mu}{\mu\sigma}\right), & \mbox{if } z = +Q
         \end{array}\right.
\end{equation}
where $\displaystyle q(x) = \frac{1}{\sqrt{2\pi}}\int_{x}^{+\infty}\mbox{exp}\left(-\frac{u^2}{2}\right) du$ is the  tail probability of the standard normal distribution (also known as the {\em $Q$-function}).
\end{itemize}

\subsubsection{Expression of $B^{(\ell)}$ as a function of $A^{(\ell-1)}$} \

In the sequel, we make the convention that $ \Pr(\sgn(0) = 1)= \Pr(\sgn(0) = -1)=1/2$.
The following notation will be used: 

\begin{itemize}
\item $A_{[x,y]}     = \displaystyle\sum_{z=x}^y A(z)$, for $x\leq y\in{\cal M}$
\item $A_{[0^{+},y]} = \displaystyle\frac{1}{2}A(0) + \sum_{z=1}^y A(z)$, for $y\in{\cal M}, \ y > 0$
\item $A_{[x, 0^{-}]} = \displaystyle\frac{1}{2}A(0) + \sum_{z=x}^{-1} A(z)$, for $x\in{\cal M}, \ x < 0$
\end{itemize}


For the sake of simplicity, we drop the iteration index, thus $B := B^{(\ell)}$ and  $A := A^{(\ell-1)}$. We proceed by recursion on $i=2,\dots,d_c-1$, where $d_c$ denotes the check-node degree. 

\noindent Let $\beta_1 :=  \alpha_1$, and for $i=2,\dots,d_c-1$ define:
$$\beta_i = \xorprred(\sgn(\beta_{i-1}),\sgn(\alpha_i))\minprred(|\beta_{i-1}|, |\alpha_i|)$$

\noindent Let also $B_{i-1}$ and $B_i$ denote the probability mass functions of $\beta_{i-1}$ and $\beta_i$, respectively (hence, $B_1 = A$).
 
\medskip \noindent First of all, for $z = 0$, we have:\\
  $B_i(0) = \Pr(\beta_i = 0) =  A(0)B_{i-1}(0) + \left[ B_{i-1}(0)(1-A(0)) + A(0)(1-B_{i-1}(0)) \right](1-p_c)$. 
  
\medskip \noindent  For $z\neq 0$, we proceed in several steps as follows:

\medskip\noindent\begin{minipage}{.495\linewidth}
For $z > 0$:  

\medskip\resizebox{\linewidth}{!}{$\begin{array}{@{}r@{\ }c@{\ }l}
  F^\prime_i(z) &\stackrel{\mbox{\tiny def}}{=}& \Pr(\beta_i \geq z \mid \pxor=0)\\ &  = &
  \left[{B_{i-1}}_{[0^+,z-1]}A_{[z,Q-1]} + A_{[0^+,z-1]}{B_{i-1}}_{[z,Q-1]}\right]p_c \\                      
  & + & \left[{B_{i-1}}_{[1-z,0^-]}A_{[-Q,-z]} + A_{[1-z,0^-]}{B_{i-1}}_{[-Q,-z]}\right]p_c \\
           & + &   {B_{i-1}}_{[z,Q-1]}A_{[z,Q-1]}  + {B_{i-1}}_{[-Q,-z]}A_{[-Q,-z]}
\end{array}$}

\medskip$\begin{array}{@{}r@{\ }c@{\ }l}
        F_i(z) & \stackrel{\mbox{\tiny def}}{=}& \Pr(\beta_i \geq z)   \\
        & = & (1-\pxor).F^\prime_i(z)+\pxor.G^\prime_i(-z)\\
        B_i(z) & = & \Pr(\beta_i = z) = F_i(z) - F_i(z+1)
  \end{array}$ 
\end{minipage}\hfill\begin{minipage}{.495\linewidth}
For $z < 0$:  

\medskip\resizebox{\linewidth}{!}{$\begin{array}{@{}r@{\ }c@{\ }l}
  G^\prime_i(z) &\stackrel{\mbox{\tiny def}}{=}& \Pr(\beta_i \leq z \mid \pxor=0) \\ &  = &
      \left[{B_{i-1}}_{[0^+,-z-1]}A_{[-Q,z]} + A_{[0^+,-z-1]}{B_{i-1}}_{[-Q,z]}\right]p_c \\
  &+ & \left[{B_{i-1}}_{[-z,Q-1]}A_{[z+1,0^{-}]} + A_{[-z,Q-1]}{B_{i-1}}_{[z+1,0^{-}]}\right]p_c  \\
  & + & {B_{i-1}}_{[-z,Q-1]}A_{[-Q,z]} + A_{[-z,Q-1]}{B_{i-1}}_{[-Q,z]}
\end{array}$} 

\medskip$\begin{array}{@{}r@{\ }c@{\ }l}
  G_i(z) & \stackrel{\mbox{\tiny def}}{=}&\Pr(\beta_i \geq z)\\
 & =&(1-\pxor).G^\prime_i(z)+\pxor.F^\prime_i(-z)\\
 B_i(z) & = & \Pr(\beta_i = z) = G_i(z) - G_i(z+1)
  \end{array}$ 
\end{minipage}

\medskip \noindent Finally, we have that $B = B_{d_c-1}$.
  
\subsubsection{Expression of $A^{(\ell)}$ as a function of $B^{(\ell)}$ and $C$}
We derive at the same time the {\em expression of $\widetilde{C}^{(\ell)}$ as a function of $B^{(\ell)}$ and $C$}. 

\medskip For simplicity, we drop the iteration index, so $A:=  A^{(\ell)}$, $B:=B^{(\ell)}$, and $\widetilde{C}:=\widetilde{C}^{(\ell)}$. We denote by $\left({\cal E}, p_{\cal E}, \imath \mid \widetilde{\cal M}\right)$ the error injection model used to define the noisy adder.
We decompose each noisy addition into three steps (noiseless infinite-precision addition, saturation, and error injection), and proceed by recursion on  $i=0,1,\dots,d_v$, where $d_v$ denotes the variable-node degree:

\begin{itemize}
\item For $i = 0$:\\
$\begin{array}{r@{\ }c@{\ }lr@{\ }c@{\ }l}
 \Omega_0 & \stackrel{\mbox{\tiny def}}{=} & 
 \gamma\in {\cal M}\subseteq \widetilde{\cal M}, \hspace*{6.3mm}& 
 \widetilde{C}_{0}(\tilde{z}) & \stackrel{\mbox{\tiny def}}{=} & 
 \Pr(\Omega_0 = \tilde{z})
 = \left\{\begin{array}{cl} C(\tilde{z}), &\mbox{if } \tilde{z}\in{\cal M}\\
 0, & \mbox{if } \tilde{z}\in\widetilde{\cal M}\setminus{\cal M} 
 \end{array}\right.
 \end{array}$ 
 
\item For $i = 1,\dots, d_v$:\\
$\begin{array}{r@{\ }c@{\ }lr@{\ }c@{\ }l}
  \omega_i & \stackrel{\mbox{\tiny def}}{=} & 
  \Omega_{i-1} + \beta_{m_i,n}  \in \mathbb{Z},  &
  c_{i}(w) & \stackrel{\mbox{\tiny def}}{=} & 
  \Pr(\omega_i = w) = \sum_u \widetilde{C}_{i-1}(u) B(w-u),  \forall w \in \mathbb{Z} \\[3mm]
  \tilde{\omega}_i & \stackrel{\mbox{\tiny def}}{=} & 
  \mathbf{s}_{\widetilde{\cal M}}(\omega_i) \in \widetilde{\cal M}, &  
  \tilde{c}_{i}(\tilde{w}) & \stackrel{\mbox{\tiny def}}{=} & 
  \Pr(\tilde{\omega}_i = \tilde{w}) = 
  \renewcommand{\arraystretch}{1.2}
  \left\{\begin{array}{l@{\ }l}
      c_i(\tilde{w}), & \mbox{if } \tilde{w} \in \widetilde{\cal M} \setminus\{\pm \widetilde{Q}\} \\
      \sum_{w \leq -\widetilde{Q}} c_i(w), & \mbox{if } \tilde{w} = -\widetilde{Q}\\
      \sum_{w \geq +\widetilde{Q}} c_i(w), & \mbox{if } \tilde{w} = +\widetilde{Q}
  \end{array}\right.\\[10mm]
  \Omega_i & \stackrel{\mbox{\tiny def}}{=} & 
  \imath(\tilde{\omega}_i, e) \in \widetilde{\cal M}, &
  \widetilde{C}_{i}(\tilde{z}) & \stackrel{\mbox{\tiny def}}{=} &
  \Pr(\Omega_i = \tilde{z}) = 
  \sum_{\tilde{\omega}} \sum_{e} {\delta}_{\imath(\tilde{\omega}, e)}^{\tilde{z}}
  p_{\cal E}(e) \tilde{c}_{i}(\tilde{\omega}),  
  \forall  \tilde{z} \in \widetilde{\cal M} \\
  & & & \multicolumn{3}{l}{\mbox{where } {\delta}_{x}^y = 1 \mbox{ if } x = y, 
  \mbox{ and } {\delta}_{x}^y = 0 \mbox{ if } x \neq y. }
\end{array}$ 
\end{itemize}
\renewcommand{\arraystretch}{1}

\medskip\noindent 
Note that in the definition of $\Omega_i$ above, $e$ denotes an error drown from the error set ${\cal E}$ according to the error probability distribution $p_{\cal E}$.

\medskip\noindent Finally, we have:
\begin{itemize}
\item ${A} = \mathbf{s}_{\cal M}\left( \widetilde{C}_{d_v-1}\right)$
\item $\widetilde{C} = \widetilde{C}_{d_v}$ 
\end{itemize}  
In the first equation above, applying the saturation operator $\mathbf{s}_{\cal M}$  on the probability mass function $\widetilde{C}_{d_v-1}$  means that all the probability weights corresponding to values $\tilde{w}$  outside ${\cal M}$ must be accumulated to the probability of the corresponding boundary value of ${\cal M}$ (that is, either $-Q$ or $+Q$, according to whether  $\tilde{w}< -Q$ or $\tilde{w}< +Q$).

\medskip\noindent {\bf Remark:} If the noisy adder is defined by one of the bitwise-{\sc xor} error injection models (Section \ref{subsec:bitwise-xor-injection}), then the third equation from the above recursion (expression of $\widetilde{C}_{i}$ as a function of $\tilde{c}_{i}$) may be rewritten as follows:
\begin{itemize}
 \item Sign-preserving bitwise-{\sc xor}ed noisy adder
 \renewcommand{\arraystretch}{1.5}
 \begin{equation}\label{eq:ct-sign-preserve}
 \widetilde{C}_{i}(\tilde{z}) =   \left\{\begin{array}{ll}
  \displaystyle (1-\padd)\tilde{c}_{i}(\tilde{z}) + 
 \frac{1}{\widetilde{Q}}\padd\left( \tilde{c}_{i\,{[\leq\, 0^{-}]}}
 - \tilde{c}_{i}(z) \right), & \mbox{if } \tilde{z} < 0 \\
  \displaystyle (1-\padd)\tilde{c}_{i}(0) + 
 \frac{1}{\widetilde{Q}}\padd\left(1 - \tilde{c}_{i}(0) \right), &
 \mbox{if } \tilde{z} = 0 \\
  \displaystyle (1-\padd)\tilde{c}_{i}(\tilde{z}) + 
 \frac{1}{\widetilde{Q}}\padd\left( \tilde{c}_{i\,{[\geq\, 0^{+}]}} 
 - \tilde{c}_{i}(z) \right), & \mbox{if } \tilde{z} > 0 
 \end{array}\right.
 \end{equation}
 where $\tilde{c}_{i\,{[\leq\, 0^{-}]}} = \sum_{\tilde{\omega} < 0} \tilde{c}_{i}(\tilde{\omega}) + \frac{1}{2}\tilde{c}_{i}(0)$, and $\tilde{c}_{i\,{[\geq\, 0^{+}]}} = \frac{1}{2}\tilde{c}_{i}(0) + \sum_{\tilde{\omega} > 0} \tilde{c}_{i}(\tilde{\omega})$.
 \renewcommand{\arraystretch}{1}
 \item Full-depth bitwise-{\sc xor}ed noisy adder
 \begin{equation}\label{eq:ct-full-depth}
 \widetilde{C}_{i}(\tilde{z}) = (1-\padd)\tilde{c}_{i}(\tilde{z}) + \frac{1}{2\widetilde{Q}}\padd 
 \left(1 - \tilde{c}_{i}(\tilde{z})\right)
 \end{equation}
\end{itemize}

\medskip Finally, we note that the density evolution equations for the noiseless finite-precision  MS decoder can be obtained by setting $\padd = \pcomp = \pxor=0$.

\subsection{Error Probability and Useful and Target-BER regions} \label{subsec:err_proba_and_useful_reg}

\subsubsection{Decoding Error Probability}
The error probability at decoding iteration $\ell$, is defined by: 
\begin{equation}\label{eq:p_iter}
P_e^{(\ell)} = \displaystyle\sum_{\tilde{z}=-\widetilde{Q}}^{-1}\widetilde{C}^{(\ell)}(\tilde{z})+\frac{\widetilde{C}^{(\ell)}(0)}{2}
\end{equation}

\begin{prop}\label{prop:lower-bounds}
The error probability at decoding iteration $\ell$ is lower-bounded as follows:
\begin{itemize}
\item[(a)] For the sign-preserving bitwise-{\sc xor}ed noisy adder: $P_e^{(\ell)} \geq \displaystyle \frac{1}{2\widetilde{Q}}\padd$.
\item[(b)] For the full-depth bitwise-{\sc xor}ed noisy adder: $P_e^{(\ell)} \geq \displaystyle \frac{1}{2}\padd + \frac{1}{4\widetilde{Q}}\padd$.
\end{itemize}
\end{prop}
\noindent{\em Proof.} {\em (a)} Using $\widetilde{C} = \widetilde{C}_{d_v}$ and equations (\ref{eq:p_iter}) and (\ref{eq:ct-sign-preserve}), it follows that
 $P_e^{(\ell)} = (1-\padd) \tilde{c}_{d_v\,{[\leq\, 0^{-}]}} + \frac{1}{2\widetilde{Q}}\padd\left(1-2\tilde{c}_{d_v\,{[\leq\, 0^{-}]}}\right) + 
 \padd\left(\tilde{c}_{d_v\,{[\leq\, 0^{-}]}} - \frac{1}{2}\tilde{c}_{d_v}(0)\right) \geq (1-\padd) \tilde{c}_{d_v\,{[\leq\, 0^{-}]}} + \frac{1}{2\widetilde{Q}}\padd\left(1-2\tilde{c}_{d_v\,{[\leq\, 0^{-}]}}\right) \geq \frac{1}{2\widetilde{Q}}\padd$, since the function 
 $(1-\padd) x + \frac{1}{2\widetilde{Q}}\padd\left(1-2x\right)$ is an increasing function of $x\in[0,1]$.
 
 \noindent {\em (b)} Equations (\ref{eq:p_iter}) and (\ref{eq:ct-full-depth}) imply that $P_e^{(\ell)} = \frac{1}{2}\padd + (1-\padd)\tilde{c}_{d_v\,{[\leq\, 0^{-}]}} + \frac{1}{4\widetilde{Q}}\padd\left(1 - 2\tilde{c}_{d_v\,{[\leq\, 0^{-}]}}\right)\geq \frac{1}{2}\padd + \frac{1}{4\widetilde{Q}}\padd$\\
  $\mbox{ }$ \hfill$\square$

\medskip Note that the above lower bounds are actually inferred from the error injection in the {\em last (the $d_v$-th) addition}  performed when computing the a posteriori information value. Therefore, these lower bounds are not expected to be tight. However, if the channel error probability is small enough, the sign-preserving lower bound proves to be tight in the asymptotic limit of $\ell$ (this will be discussed in more details in Section~\ref{sec:asymptotic-analysis}). Note also that by protecting the sign of the noisy adder, the bound is lowered by a factor of roughly $\widetilde{Q}$, which represents an exponential improvement with respect to the number of bits of the adder.


\medskip In the asymptotic limit of the code-length, $P_e^{(\ell)}$ gives the probability of the hard bit estimates being in error at decoding iteration $\ell$. For the (noiseless, infinite-precision) BP decoder, the error probability is usually a decreasing function of $\ell$. This is no longer true for the noiseless, infinite-precision MS decoder,  for which the error probability may increase with $\ell$.  However, both decoders exhibit a {\em threshold phenomenon}, separating the region where error probability goes to zero (as the number of decoding iterations goes to infinity), from that where it is bounded above zero \cite{richardson2001capacity}.

Things get more complicated for the noisy (finite-precision) MS decoder. First, the error probability have a more unpredictable behavior. It does not always converge and it may become periodic\footnote{In fact, for both BSC and BI-AWGN  channels, the only cases we observed, in which the sequence $\left(P_e^{(\ell)}\right)_{\ell > 0}$ does not converge, are those cases in which this sequence becomes periodic for $\ell$ large enough.} when the number of iterations goes to infinity. Second, the error probability is always bounded above zero (Proposition~\ref{prop:lower-bounds}), since there is a non-zero probability of fault injection at any decoding iteration. Hence, a decoding threshold, similar to the noiseless case, cannot longer be defined.

\medskip 
Following \cite{varshney2011performance}, we define below the notions of useful decoder and target error rate threshold.  We consider a channel model depending on a channel parameter $\chi$, such that the channel is degraded by increasing $\chi$ (for example, the crossover probability for the BSC, or the noise variance for the BI-AWGN channel). We will use subscript $\chi$ to indicate a quantity that depends on $\chi$. Hence, in order to account that $P_{e}^{(\ell)}$ depends also on the value of the channel parameter, it will be denoted in the following by $P_{e,\chi}^{(\ell)}$.

\subsubsection{Useful Region}
The first step is to evaluate the channel and hardware parameters yielding a final probability of error (in the asymptotic limit of the number of iterations) less than the {\em input error probability}. The latter probability is given by 
 $P_{e,\chi}^{(0)} = \sum_{z=-Q}^{-1} C(z) + \frac{1}{2} C(0)$, where $C$ is the probability mass function of the quantized a priori information of the decoder (see Section~\ref{subsubsec:express-C}).

Following  \cite{varshney2011performance}, the decoder is said to be {\em useful} if $\left(P_{e,\chi}^{(\ell)}\right)_{\ell > 0}$ is convergent, and:
\begin{equation}
P_{e,\chi}^{(\infty)} \stackrel{\text{def}}{=} \lim_{\ell \rightarrow \infty} P_{e,\chi}^{(\ell)} < P_{e,\chi}^{(0)}
\end{equation}
 
The ensemble of the parameters that satisfy this condition constitutes the {\em useful region} of the decoder.

\subsubsection{Target Error Rate Threshold}
For noiseless-decoders, the decoding threshold is defined as the supremum channel noise, such that the error probability converges to zero as the number of decoding iterations goes to infinity. However, for noisy decoders this error probability does not converge to zero, and an  alternative definition of the decoding threshold has been introduced in \cite{varshney2011performance}. Accordingly, for a target bit-error rate $\eta$, the $\eta$-threshold is defined\footnote{In \cite{varshney2011performance}, the $\eta$-threshold is defined by $\chi^{\ast} (\eta)=\sup \left\{\chi \mid P_{e,\chi}^{(\infty)} \mbox{ exists and }  P_{e,\chi}^{(\infty)} < \eta \right\}$, and consequently, there might exist a channel parameter value $\chi' < \chi^{\ast} (\eta)$, for which $P_{e,\chi'}^{(\infty)}$ does not exist. In order to avoid this happening, our definition is slightly different from the one in \cite{varshney2011performance}.} by: 
\begin{equation}\label{eq:eta_threshold}
\chi^{\ast} (\eta)=\sup \left\{\chi \mid P_{e,\chi'}^{(\infty)} \mbox{ exists and }  P_{e,\chi'}^{(\infty)} < \eta, \ \forall \chi'\in[0, \chi] \right\} 
\end{equation}

\subsection{Functional Threshold}\label{subsec:funct_threshold}
Although the $\eta$-threshold definition allows determining the maximum channel noise for which the bit  error probability can be reduced below a target value, there is not significant change in the behavior of the decoder when the channel noise parameter $\lambda$ increases beyond the value of $\chi^{\ast} (\eta)$.  
In this section, a new threshold definition is introduced in order to identify the channel and hardware parameters yielding to a sharp change in the decoder behavior, similar to the change that occurs around the threshold of the noiseless decoder.   
This threshold will be referred to as the {\em functional threshold}.
The aim is to detect a sharp increase (e.g. discontinuity) in the error probability of the noisy decoder, when $\lambda$ goes beyond this functional threshold value. The threshold definition we propose make use of the Lipschitz constant of the function $\chi \mapsto \peinf$ in order to detect a sharp change of $\peinf$ with respect to $\chi$. The definition of the Lipschitz constant is first restated for the sake of clarity.

\begin{defi}
 Let $f:I \rightarrow \mathbb{R}$ be a function defined on an interval $I\subseteq \mathbb{R}$.
The {\em Lipschitz constant} of $f$ in $I$ is defined as
\begin{equation}
L(f, I) = \sup_{x\neq y \in I} \frac{|f(x)-f(y)|}{|x-y|} \in \mathbb{R}_{+} \cup \{+\infty\}
\end{equation} 
For $a\in I$ and $\delta > 0$, let $I_a(\delta) = I \cap (a-\delta, a+\delta)$. 
The {\em (local) Lipschitz constant} of $f$ in $a\in I$ is defined by:
\begin{equation}
L(f, a) = \inf_{\delta > 0} L(f, I_a(\delta)) \in \mathbb{R}_{+} \cup \{+\infty\}
\end{equation} 
\end{defi}

Note that if $a$ is a discontinuity point of $f$, then $L(f, a) = +\infty$.
On the opposite, if $f$ is differentiable in $a$, then the Lipschitz constant in $a$ corresponds to the absolute value of the derivative.
Furthermore, if $L(f, I) < +\infty$, then $f$ is  uniformly continuous on $I$ and almost everywhere differentiable. 
In this case, $f$ is said to be {\em Lipschitz continuous} on $I$.

The functional threshold is then defined as follows.
\begin{defi}\label{def:ft}
 For given hardware parameters and a channel parameter $\chi$, the decoder is said to be {\em functional} if
\begin{description}
\item[$(a)$] The function $x \mapsto\pein(x)$ is defined on $[0,\chi]$
\item[$(b)$] $\pein$ is Lipschitz continuous on $[0, \chi]$
\item[$(c)$] $L\left(\pein, x\right)$ is an increasing function of $x\in [0, \chi]$
\end{description}

Then, the functional threshold $\bar{\chi}$ is defined as:
\begin{equation}
 \bar{\chi} = \sup \{ \chi \mid \mbox{conditions } (a), (b) \mbox{ and } (c) \mbox{ are satisfied}\} 
\end{equation}
\end{defi}

The use of the Lipschitz constant allows a rigorous definition of the functional threshold, while avoiding the use of the derivative (which would require $\pein(\lambda)$ to be a piecewise differentiable function of $\lambda$).  As it will be further illustrated in Section~\ref{sec:asymptotic-analysis}, the functional threshold corresponds to a transition between two modes.
The first mode corresponds to the channel parameters leading to a low level of error probability, \emph{i.e.}, for which the decoder can correct most of the errors from the channel.
In the second mode, the channel parameters lead to a much higher error probability level.
If $L\left(\pein, \bar{\chi}\right) = +\infty$, then $\bar{\chi}$ is a discontinuity point of $\pein$ and the transition between the two levels is sharp.
If $L\left(\pein, \bar{\chi}\right) < +\infty$, then $\bar{\chi}$ is an inflection point of $\pein$ and the transition is smooth. 
With the Lipschitz constant, one can characterize the transition in both cases.
However, the second case corresponds to a degenerated one, in which the hardware noise is too high and leads to a non-standard asymptotic behavior of the decoder.
That is why a set of admissible hardware noise parameters is defined as follows.
\begin{defi}
 The set of \emph{admissible hardware parameters} is the set of hardware noise parameters $(p_a,p_c,p_x)$ for which $L\left(\pein, \bar{\chi}\right) = +\infty$.
\end{defi}
In the following, as each threshold definition helps at illustrating different effects, one or the other definition will be used, depending on the context.

\section{Asymptotic analysis of the noisy Min-sum decoder}\label{sec:asymptotic-analysis}

In this section, the density evolution equations derived previously are used to analyze the asymptotic performance ({\em i.e.} in the asymptotic limit of both the code length and number of iterations) of the noisy MS decoder.

Unless specified otherwise, the following parameters are used throughout this section:

\medskip\noindent{\bf Code parameters:}
\begin{itemize}
\item We consider the ensemble of regular LDPC codes with variable-node degree $d_v=3$ and check-node degree $d_c=6$
\end{itemize}

\noindent{\bf Quantization parameters:} 
\begin{itemize}
\item The a priori information and exchanged messages are quantized on $q = 4$ bits; hence, $Q = 7$ and ${\cal M} = \{-7,\dots, +7\}$.
\item The a posteriori information is quantized on $\tilde{q} = 5$ bits; hence, $\widetilde{Q} = 15$ and $\widetilde{\cal M} = \{-15,\dots, +15\}$.
\end{itemize}

\medskip\noindent We analyze the decoding performance depending on:
\begin{itemize}
\item The quantization map $\mathbf{q}_{\mu} : {\cal Y} \rightarrow {\cal M}$, defined in Equation~(\ref{eq:quant-map}). The factor $\mu$ will be referred to as the {\em channel-output scale factor}, or simply the {\em channel scale factor}.
\item The parameters of the noisy adder,  comparator, and {\sc xor}-operator, defined respectively in Equations (\ref{eq:noisy-add-def}), (\ref{eq:noisy-comp-def}), and (\ref{eq:noisy-xor-def}).
\end{itemize}

\subsection{Numerical results for the BSC}\label{subsec:num_results_bsc}

For the BSC, the channel output alphabet is ${\cal Y} = \{-1, +1\}$ and the quantization map is defined by $\mathbf{q}_{\mu}(-1) = -\mu$ and  $\mathbf{q}_{\mu}(+1) = +\mu$, with $\mu\in\{1,\dots,Q\}$.

The infinite-precision MS decoder (Algorithm~\ref{alg:ms}), is known to be independent of the scale factor $\mu$. This is because $\mu$ factors out from all the processing steps in Algorithm~\ref{alg:ms}, and therefore does not affect in any way the decoding process. This is no longer true for the finite precision decoder (due to saturation effects), and we will show in this section that, even in the noiseless case, the scale factor $\mu$ can significantly impact the performance of the finite precision MS decoder. 

We start by analyzing the performance of the MS decoder with quantization map $\mathbf{q}_{1}$, and then we will analyze its performance with an optimized quantization map $\mathbf{q}_{\mu}$.

\subsubsection{Min-Sum decoder with quantization map $\mathbf{q}_{1}$}\label{subsec:ms_q1}

The case $\mu=1$ leads to an ``unconventional'' behavior, 
 as in some particular cases the noise introduced by the device can help the MS decoder to escape from fixed points attractors, and may actually result in an increased correction capacity with respect to the noiseless decoder. This behavior will be discussed in more details in this section.
 
\begin{figure}[!thb]
\centering
\includegraphics[width=90mm]{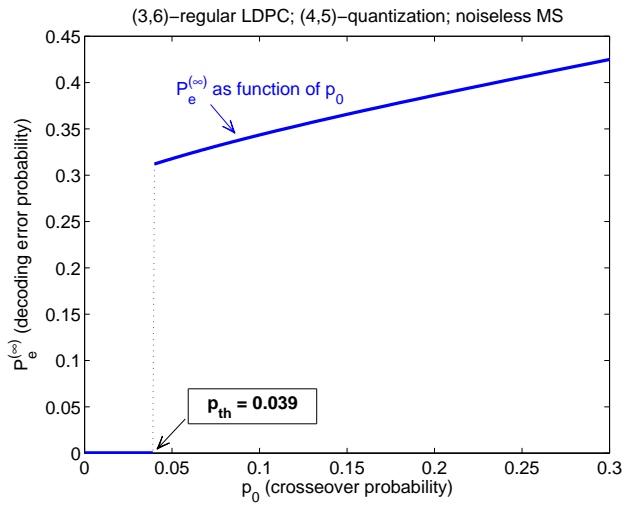}
\caption{Asymptotic error probability $P_{e}^{(\infty)}$ of the noiseless MS decoder as a function of $p_0$}
\label{fig:pinf_noiselessMS}
\end{figure}

We start with the noiseless decoder case. Figure~\ref{fig:pinf_noiselessMS} shows the asymptotic error probability $P_{e}^{(\infty)}$ as a function of $p_0$. It can be seen that $P_{e}^{(\infty)}$ decreases slightly with $p_0$, until $p_0$ reaches a threshold value $p_{\text{th}} = 0.039$, where $P_{e}^{(\infty)}$ drops to zero. This is the {\em classical} threshold phenomenon mentioned in Section~\ref{subsec:err_proba_and_useful_reg}: for $p_0 > p_{\text{th}}$, the decoding error probability is bounded far above zero ($P_{e}^{(\infty)} > 0.31$), while for $p_0 < p_{\text{th}}$, one has  $P_{e}^{(\infty)} = 0$.

\begin{figure}[!thb]
\centering
\subfigure[$P_{e}^{(\ell)}$ plotted in linear scale]{\includegraphics[width=80mm]{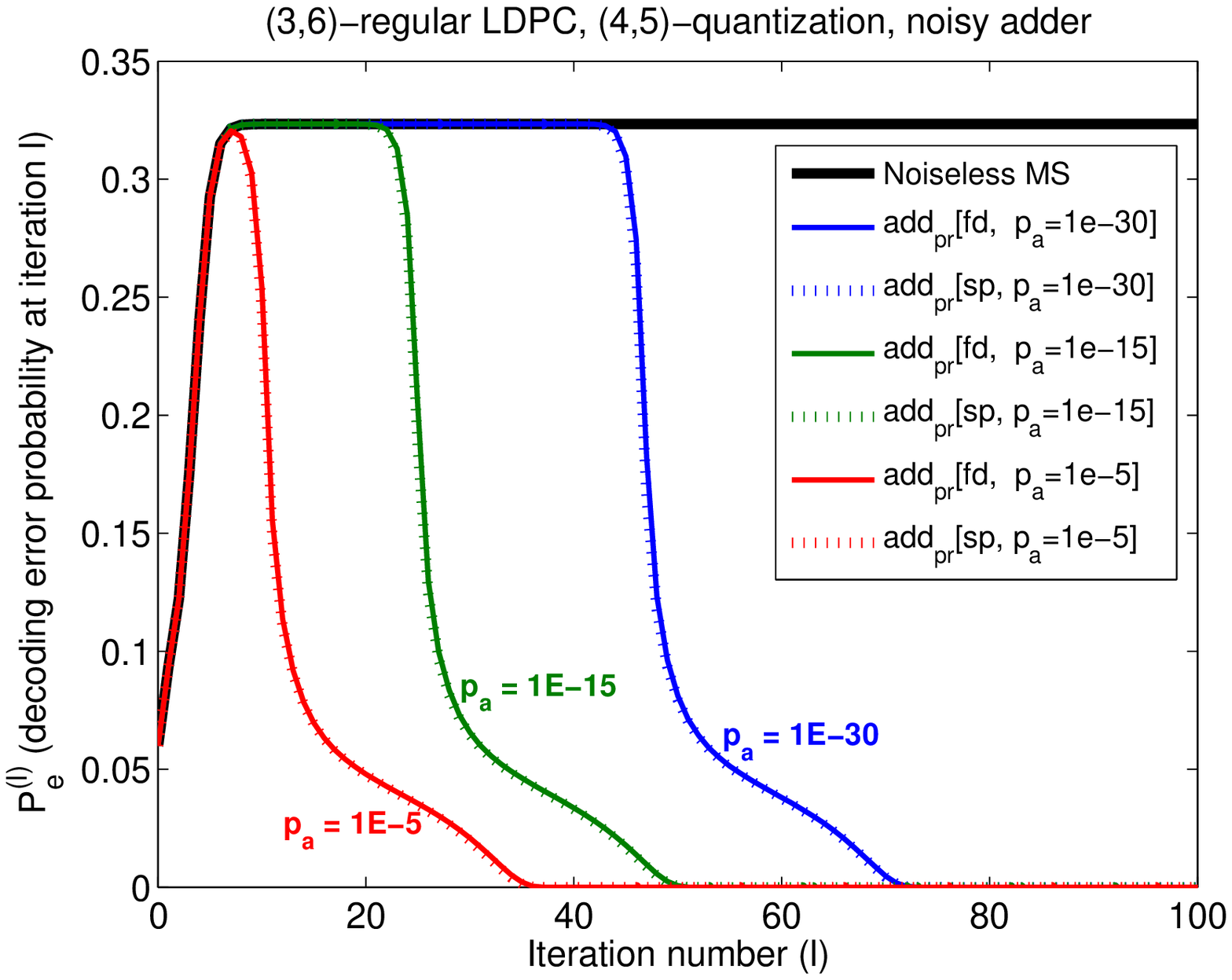}\label{subfig:piter_noisy_adder_lin}}
\subfigure[$P_{e}^{(\ell)}$ plotted in logarithmic scale]{\includegraphics[width=80mm]{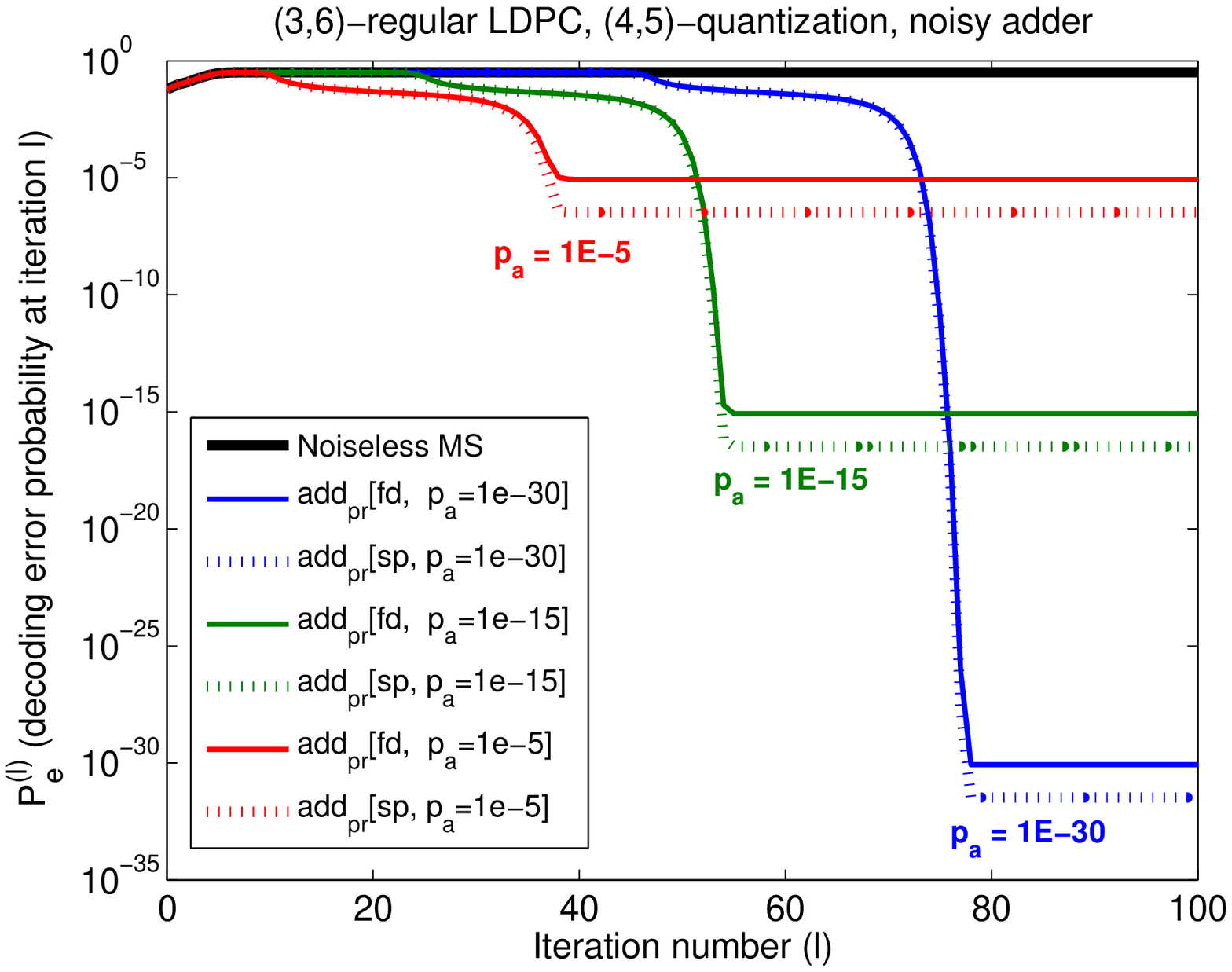}\label{subfig:piter_noisy_adder_log}}
\caption{Effect of the noisy adder on the asymptotic performance of the MS decoder  ($p_0 = 0.06$)}
\label{fig:piter_noisy_adder}
\end{figure}

 Now, we consider a $p_0$ value slightly greater than the threshold of the noiseless decoder, and investigate the effect of the noisy adder on the decoder performance. Let us fix $p_0 = 0.06$. 
Figure~\ref{subfig:piter_noisy_adder_lin} shows the decoding error probability at iteration $\ell$, for different parameters $p_a \in\{10^{-30}, 10^{-15}, 10^{-5}\}$ of the noisy adder. For each $p_a$ value, there are two superimposed curves, corresponding to the full-depth (``fd'', solid curve) and sign-preserving (``sp'', dashed curve) error models of the noisy adder. 

\noindent The error probability of the noiseless decoder is also plotted (solid black curve): it can be seen that it increases rapidly from the initial value $P_{e}^{(0)} = p_0$ and closely approaches the limit value $P_{e}^{(\infty)} = 0.323$ after a few number of iterations.
When the adder is noisy, the error probability increases during the first decoding iterations, and behaves similarly as in the noiseless case. It may approach the limit value from the noiseless case, but starts decreasing after some number of decoding iterations. However, it remains bounded above zero, according to the lower bounds from Proposition~\ref{prop:lower-bounds}. This can be seen in Figure~\ref{subfig:piter_noisy_adder_log}, where $P_{e}^{(\ell)}$ plotted in logarithmic scale.  The asymptotic values $P_{e}^{(\infty)}$ and the corresponding lower-bounds values from Proposition~\ref{prop:lower-bounds} are shown in Table~\ref{tab:pinf_noisy_ader}. It can be seen that these bounds are tight, especially in the sign-preserving case.

\begin{table}[!htb]
\caption{Asymptotic error probability of the MS decoding with noisy adder ($p_0 = 0.06$)}
\label{tab:pinf_noisy_ader}
\centering
\begin{tabular}{|c|c|c|c|c|}
\hline
\multicolumn{2}{|r|}{$p_a$} & $10^{-30}$ & $10^{-15}$ & $10^{-5}$ \\ 
\hline\hline
full  & $P_{e}^{(\infty)}$ & $8.500\times 10^{-31}$ & $8.500\times 10^{-16}$ & $8.507\times 10^{-6}$ \\
\cline{2-5} 
depth & lower-bound        & $5.167\times 10^{-31}$ & $5.167\times 10^{-16}$ & $5.167\times 10^{-6}$ \\
\hline\hline
sign      & $P_{e}^{(\infty)}$ & $3.333\times 10^{-32}$ & $3.333\times 10^{-17}$ & $3.333\times 10^{-7}$  \\
\cline{2-5}
preserving & lower-bound        & $3.333\times 10^{-32}$ & $3.333\times 10^{-17}$ &  $3.333\times 10^{-7}$ \\
\hline
\end{tabular}
\end{table} 

\begin{figure}[!htb]
\centering
\subfigure[Iteration $\ell = 0$]{\includegraphics[width=.33\linewidth]{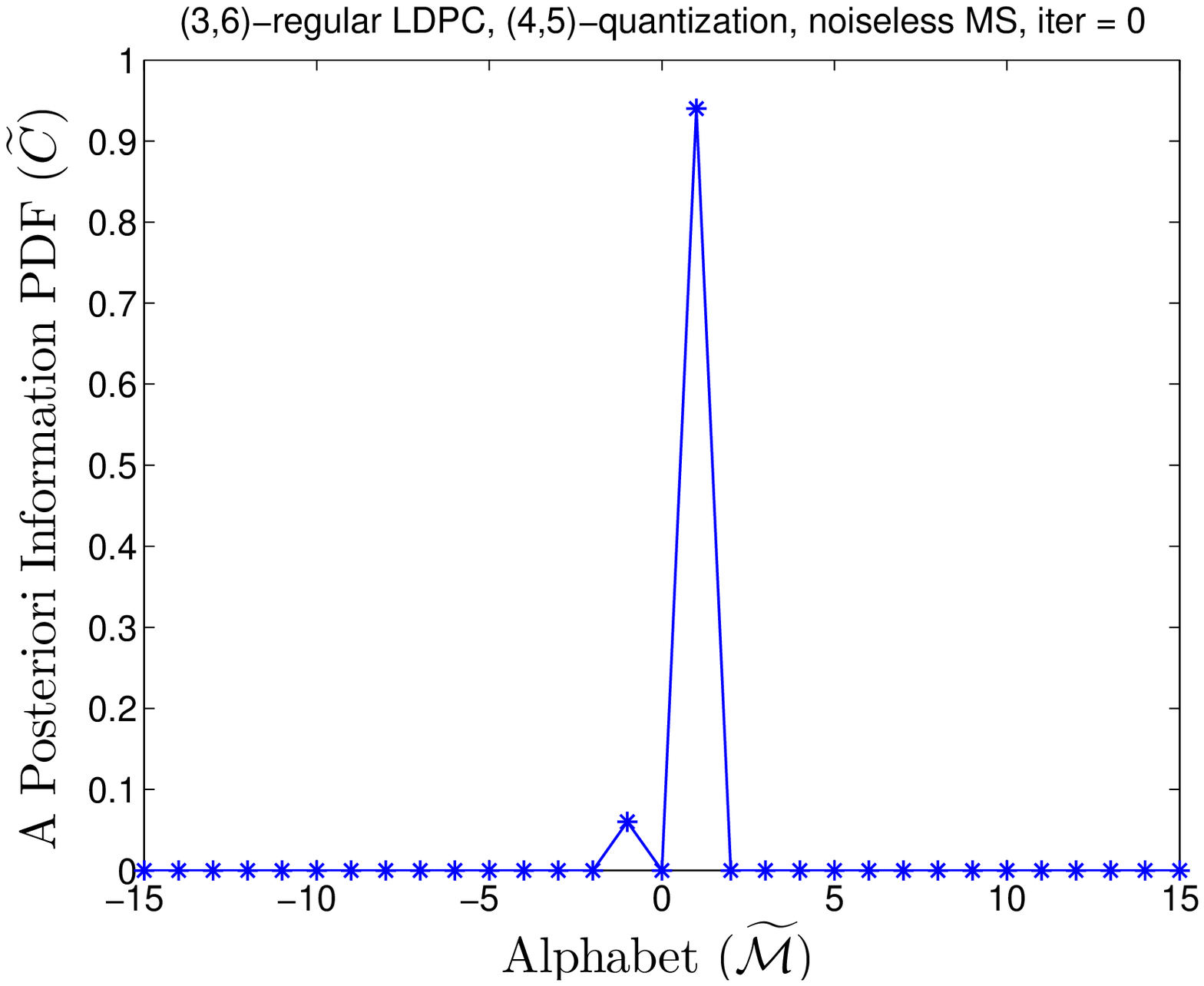}\label{subfig:ct_noiselessMS_i0}}%
\subfigure[Iteration $\ell = 5$]{\includegraphics[width=.33\linewidth]{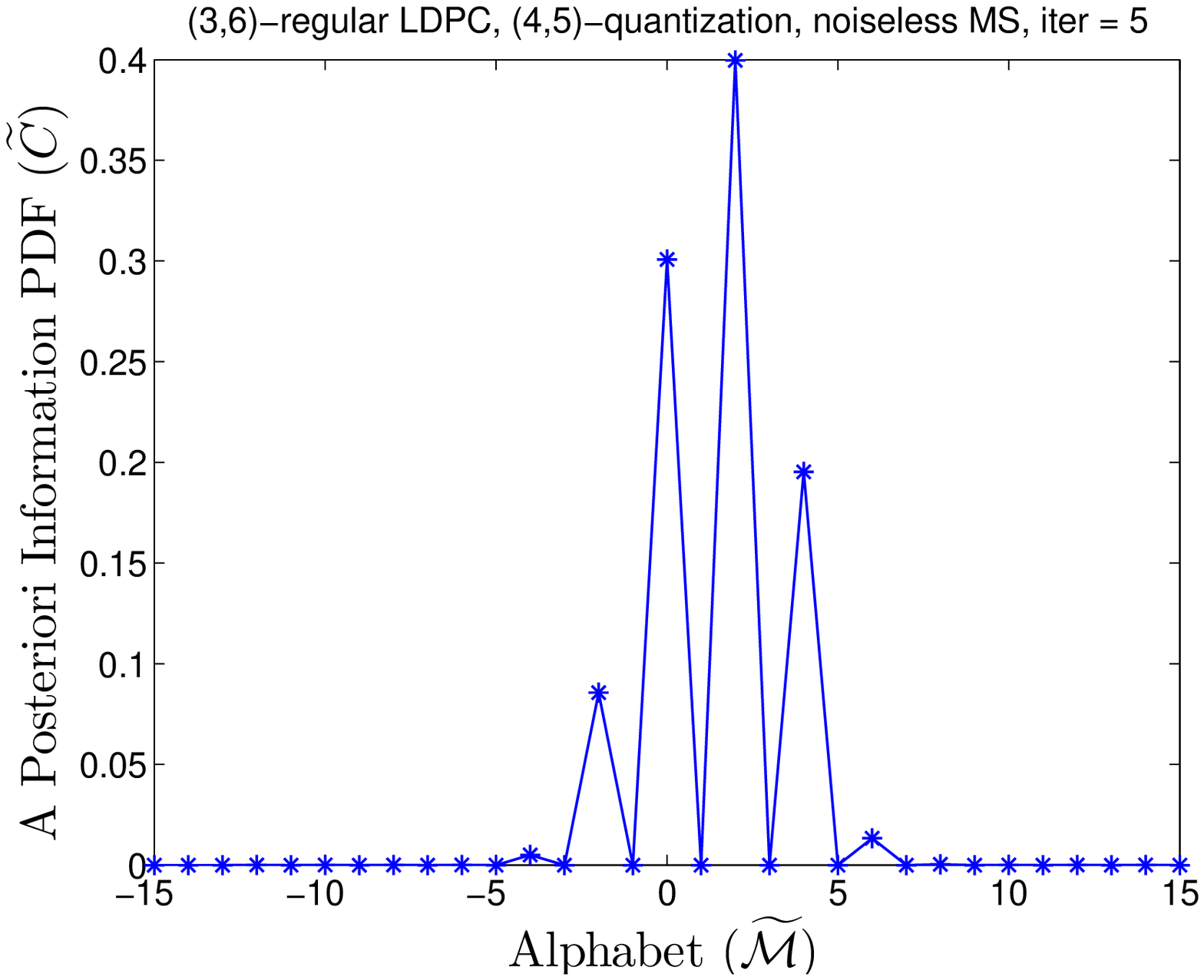}\label{subfig:ct_noiselessMS_i5}}%
\subfigure[Iteration $\ell = 20$]{\includegraphics[width=.33\linewidth]{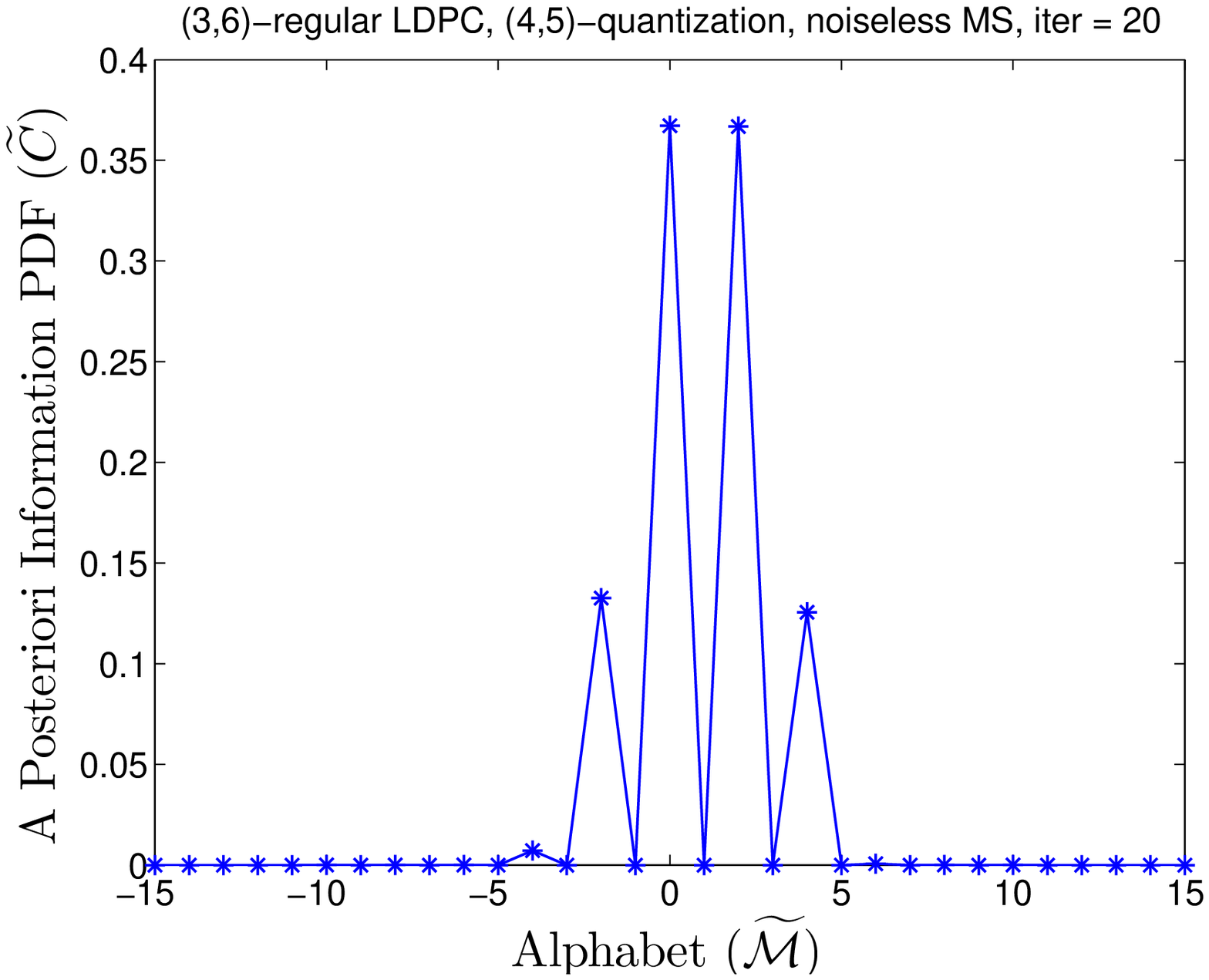}\label{subfig:ct_noiselessMS_i20}}%
\caption{Probability mass function of the a posteriori information $\widetilde{C}^{(\ell)}$ (noiseless MS decoder)}
\label{fig:ct_noiselessMS}
\end{figure} 

\begin{figure}[!htb]
\centering
\subfigure[Iteration $\ell = 20$]{\includegraphics[width=.33\linewidth]{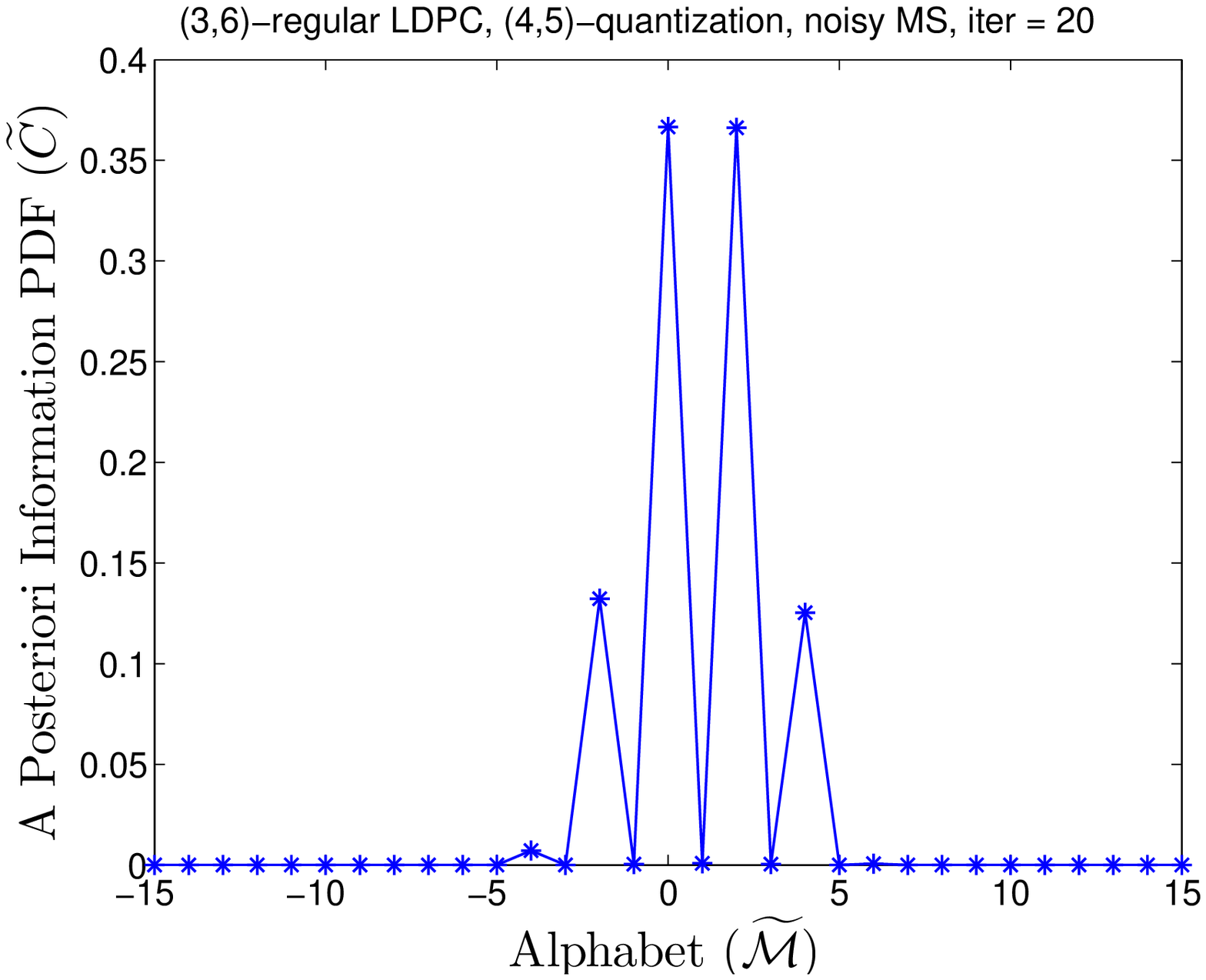}\label{subfig:ct_noisy_adder_i20}}%
\subfigure[Iteration $\ell = 23$]{\includegraphics[width=.33\linewidth]{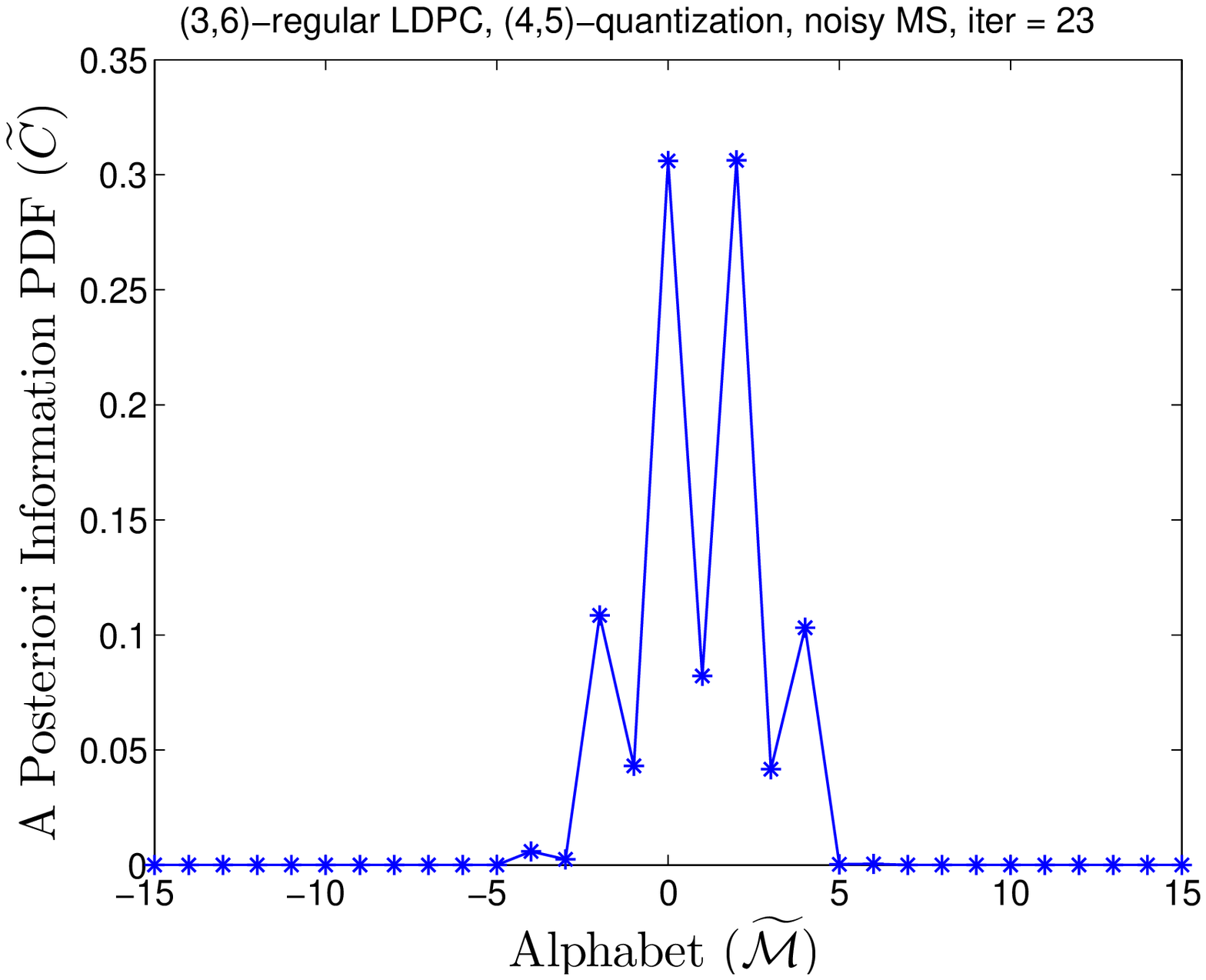}\label{subfig:ct_noisy_adder_i23}}%
\subfigure[Iteration $\ell = 30$]{\includegraphics[width=.33\linewidth]{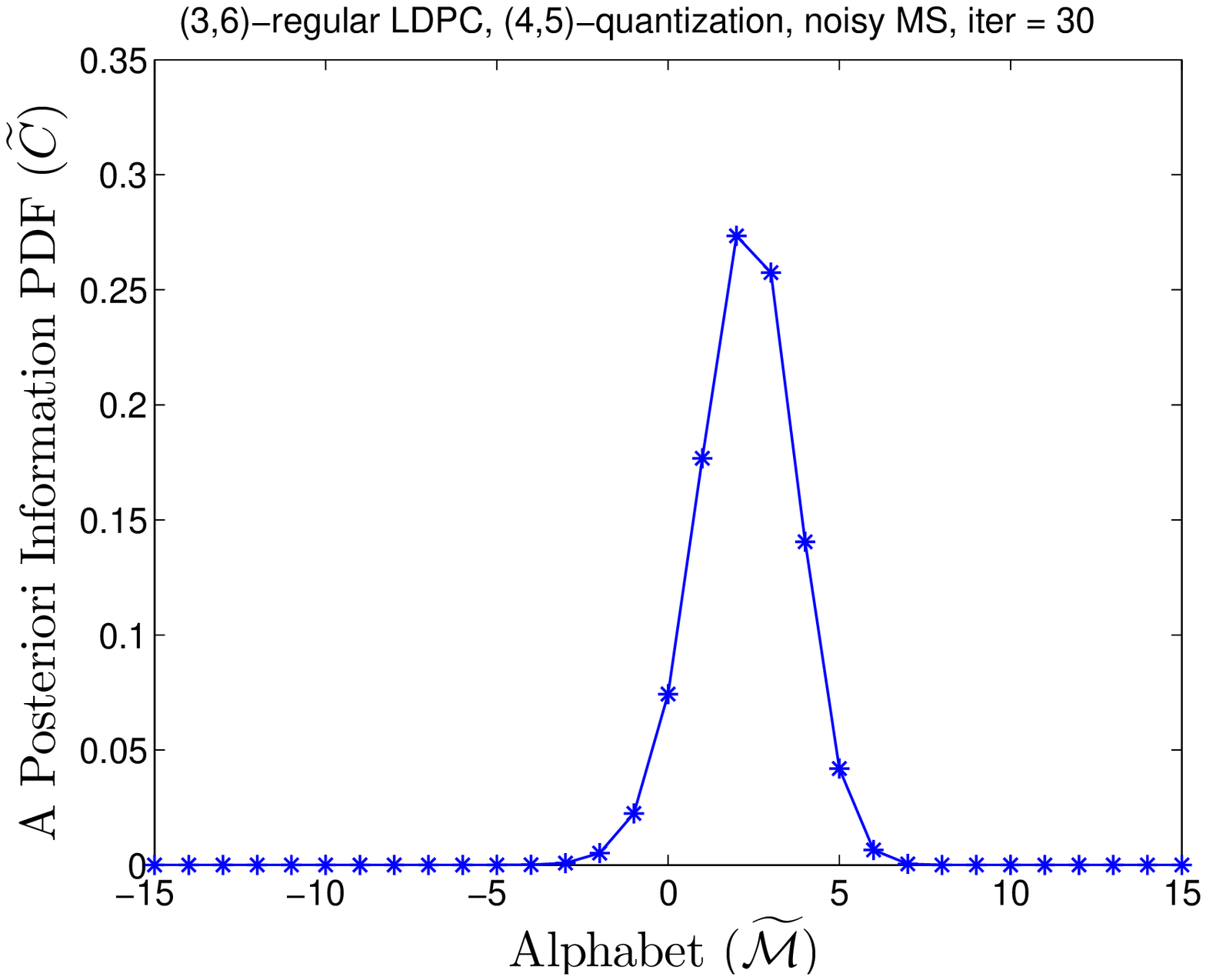}\label{subfig:ct_noisy_adder_i30}}%
\caption{Probability mass function of the a posteriori information $\widetilde{C}^{(\ell)}$ (MS decoder with full-depth noisy adder, $p_a = 10^{-15}$)}
\label{fig:ct_noisy_adder}
\end{figure}

\medskip The above behavior of the MS decoder is explained by the fact that the noise present in the adder helps the MS decoder to escape from fixed points attractors.  Figure~\ref{fig:ct_noiselessMS} illustrates the evolution of the probability mass function $\widetilde{C}^{(\ell)}$ for the noiseless decoder. At iteration $\ell = 0$, $\widetilde{C}^{(0)}$ is supported in $\pm 1$, with $\widetilde{C}^{(0)}(-1) = p_0$ and $\widetilde{C}^{(0)}(+1) = 1-p_0$. It evolves during the iterative decoding, and reaches a fixed point of the density evolution for $\ell = 20$. Note that since all variable-nodes are of degree $d_v = 3$, it can be easily seen that, for $\ell \geq 1$, $\widetilde{C}^{(\ell)}$ is supported only on even values. These ``gaps'' in the probability mass function seem lead to favorable conditions for the occurrence of density-evolution fixed-points.

Figure~\ref{fig:ct_noisy_adder} illustrates the evolution of the probability mass function $\widetilde{C}^{(\ell)}$ when the full-depth noisy adder with $p_a = 10^{-15}$ is used within the MS decoder. At iteration $\ell = 20$,  $\widetilde{C}^{(\ell)}$ is virtually the same as in the noiseless case. However, the noisy adder allows the decoder to escape from this fixed-point, as it can be seen for iterations $\ell = 23$ and $\ell = 30$. For $\ell > 30$, the  $\widetilde{C}^{(\ell)}$ moves further on the right, until the corresponding error probability $P_e^{(\ell)}$ reaches the limit value  $P_e^{(\infty)} = 8.5\times 10^{-16}$. 

It is worth noting that neither the noisy comparator nor the {\sc xor}-operator can help the decoder to escape from fixed-point distributions, as they do not allow ``filling the gaps'' in the support of $\widetilde{C}^{(\ell)}$.

\medskip We focus now on the useful region of the noisy MS decoder. We assume that only the adder is noisy, while the comparator and the {\sc xor}-operator are noiseless.

\begin{figure}[!thb]
\centering
\includegraphics[width=90mm]{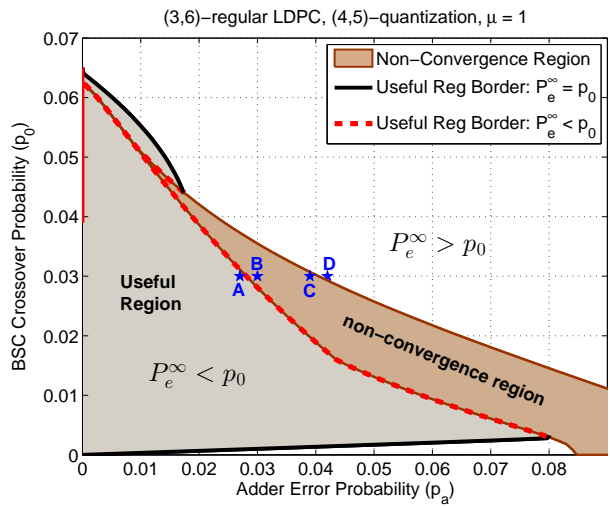}
\vspace*{-3mm}\caption{Useful and non-convergence regions of the MS decoder with sign-preserving noisy adder}
\label{fig:ureg_add_sp_mu1_pt}
\end{figure}  

\begin{figure}[!thb]
\centering
\subfigure[Point $A(p_0 = 0.03, p_a = 0.027)$]{\includegraphics[width=.4\linewidth]{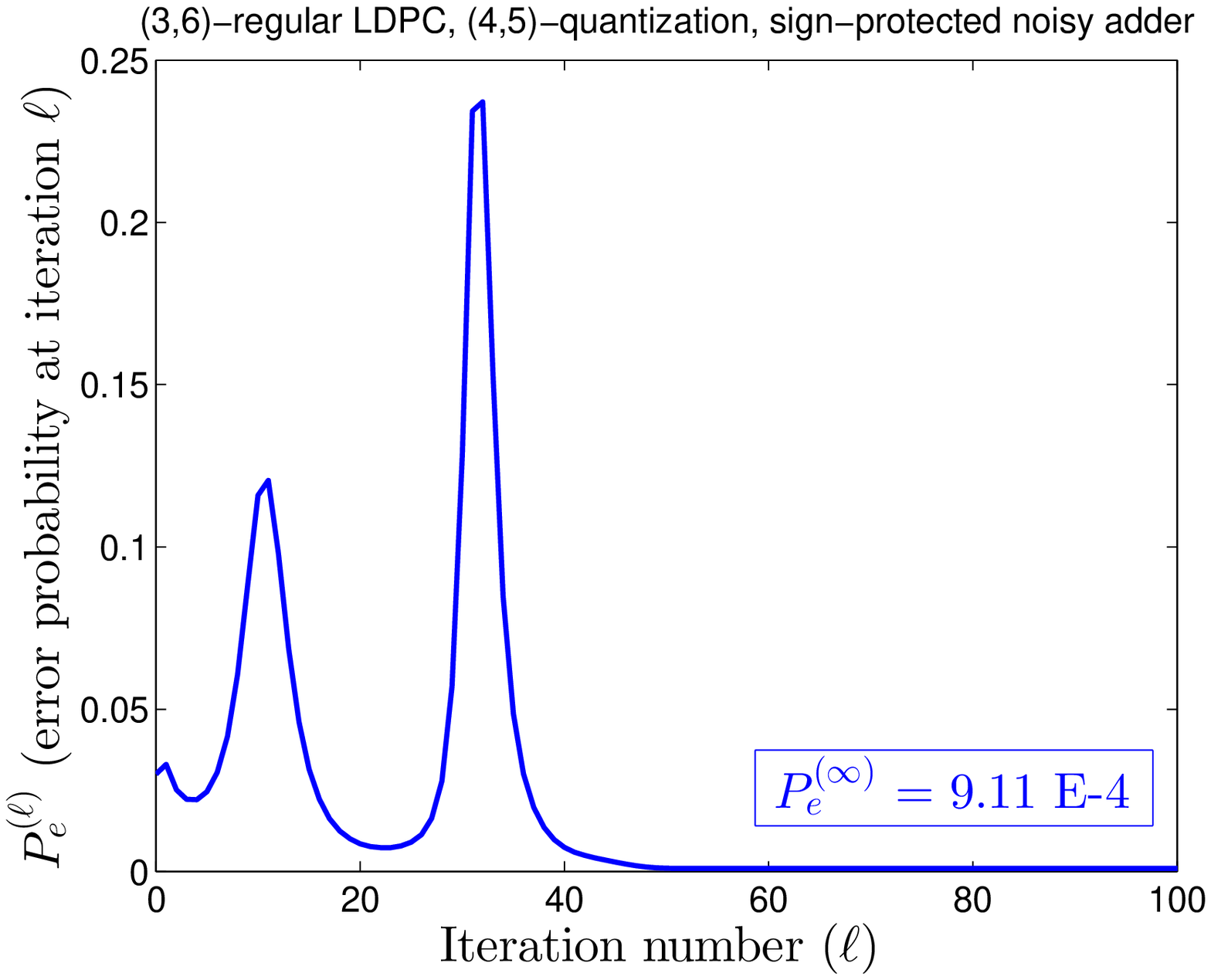}\label{subfig:piter_noisy_adder_ptA}}\ \ 
\subfigure[Point $B(p_0 = 0.03, p_a = 0.03)$]{\includegraphics[width=.4\linewidth]{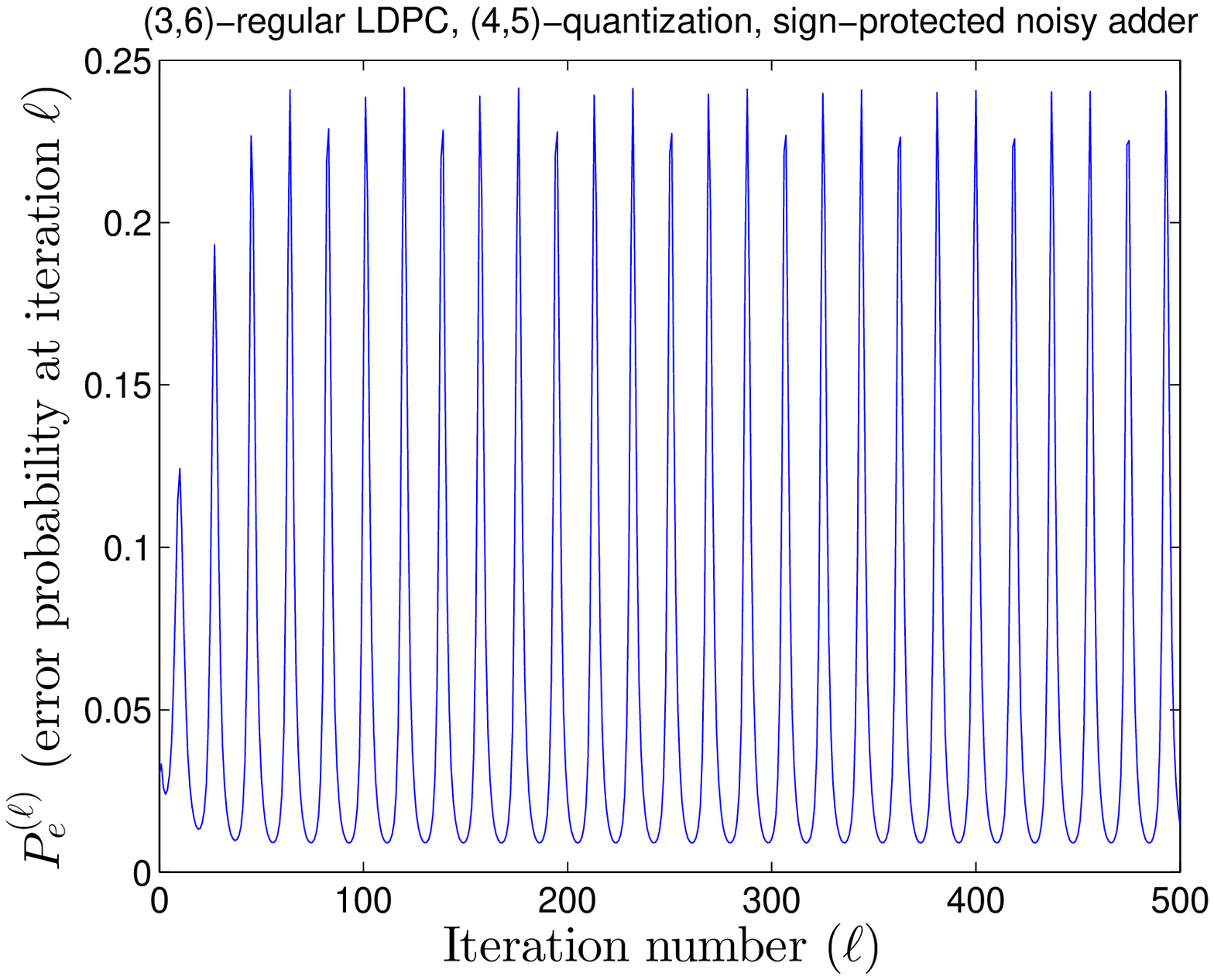}\label{subfig:piter_noisy_adder_ptB}}%

\subfigure[Point $C(p_0 = 0.03, p_a = 0.039)$]{\includegraphics[width=.4\linewidth]{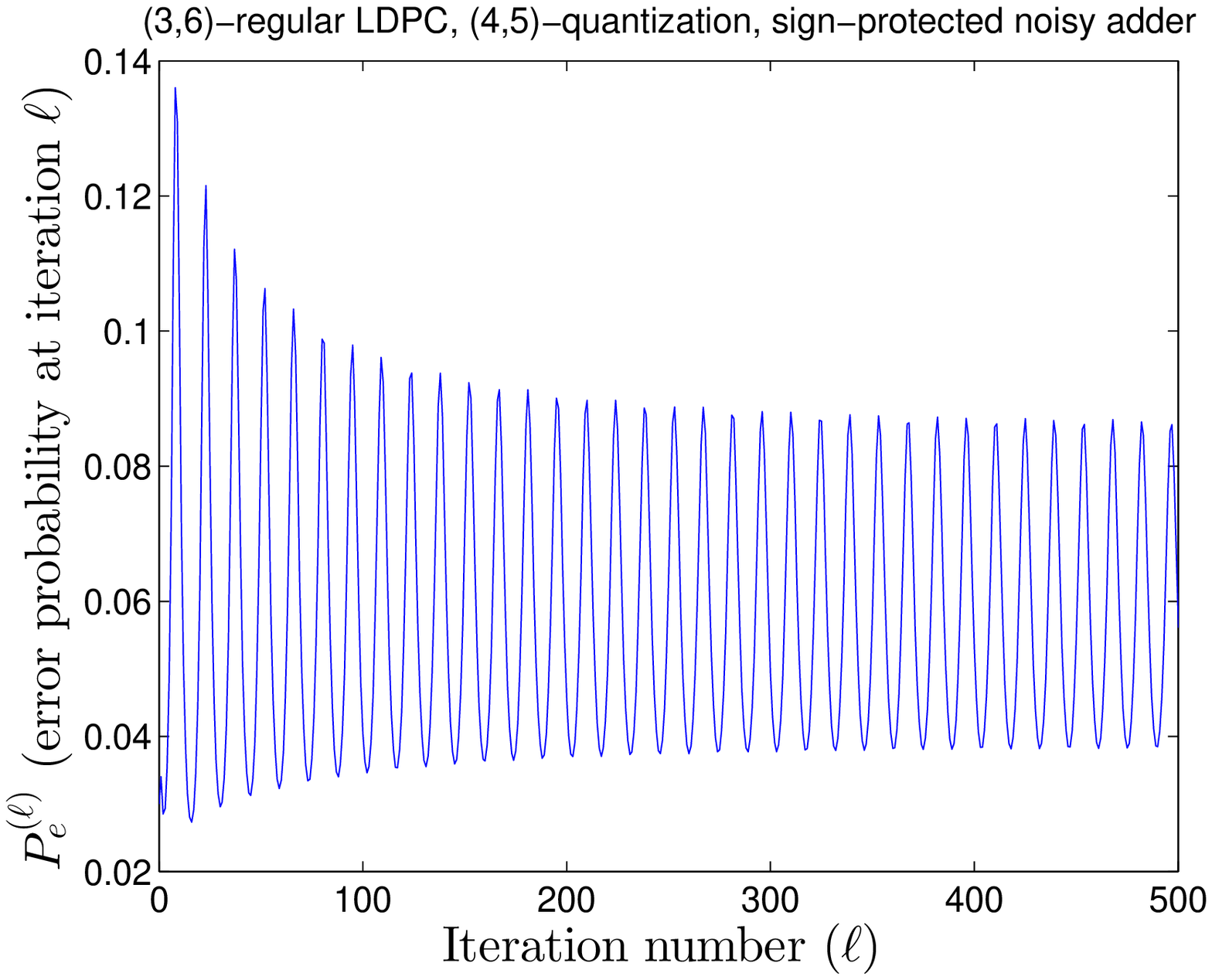}\label{subfig:piter_noisy_adder_ptC}}\ \ 
\subfigure[Point $D(p_0 = 0.03, p_a = 0.042)$]{\includegraphics[width=.4\linewidth]{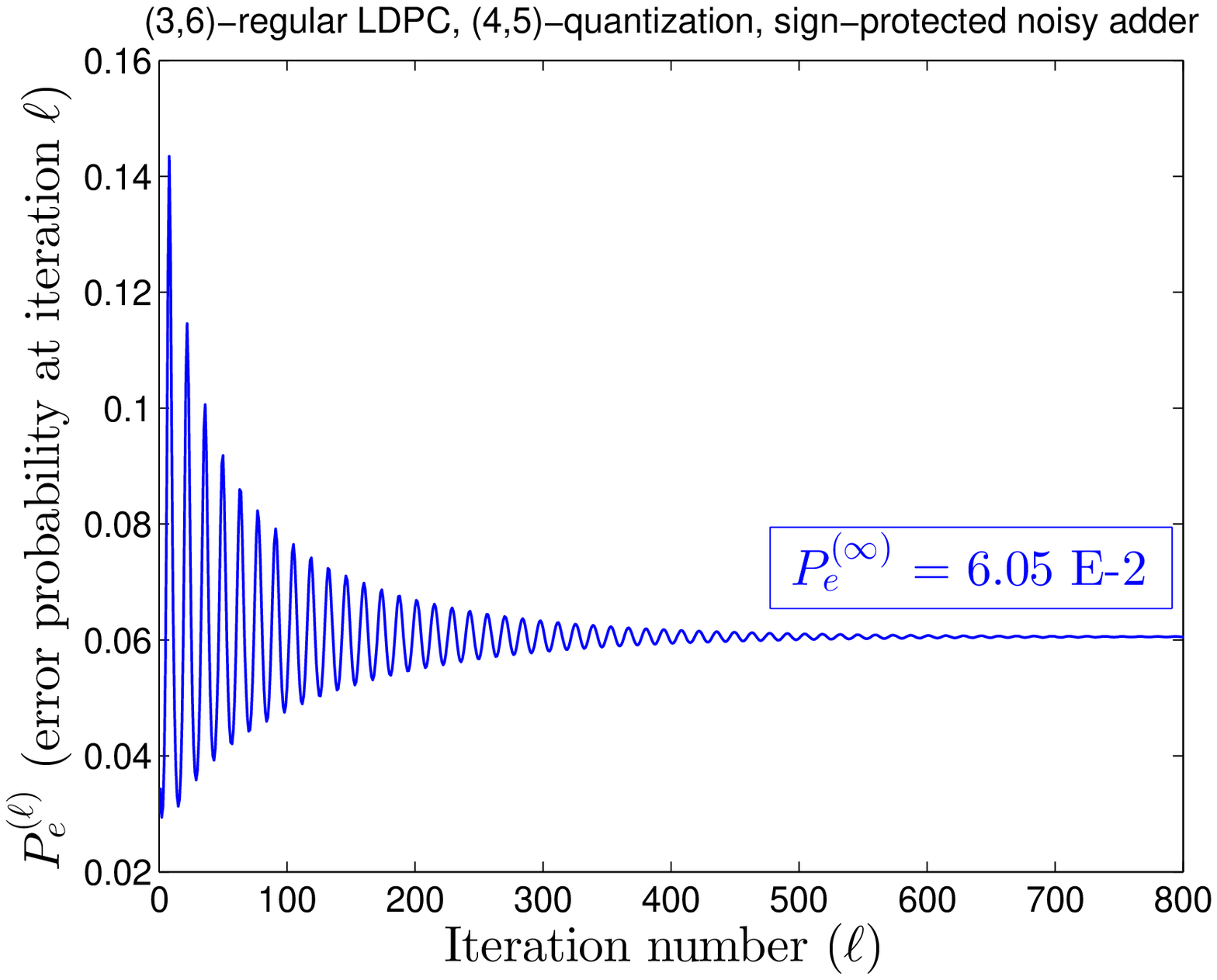}\label{subfig:piter_noisy_adder_ptD}}%
\vspace*{-2mm}\caption{Decoding error probability $P_{e}^{(\ell)}$ of the noisy MS decoder, for $p_0 = 0.03$ and sign-preserving noisy adder with various $p_a$ values}
\label{fig:piter_noisy_adder_pts}
\end{figure} 

The useful region for the sign-preserving noisy adder model is shown in Figure~\ref{fig:ureg_add_sp_mu1_pt}. The 
useful region is shaded in gray and delimited by either a solid black curve or a dashed red curve. Although one would expect that $P_e^{(\infty)} = p_0$ on the border of the useful region, this equality only holds on the solid black border. On the dashed red border, one has $P_e^{(\infty)} < p_0$. The reason why the useful region does not extend beyond the dashed red border is that for points located on the other side of this border the sequence $(P_e^{(\ell)})_{\ell > 0}$ is periodic, and hence it does not converge! The region shaded in brown in Figure~\ref{fig:ureg_add_sp_mu1_pt} is the {\em non-convergence region} of the decoder. Note that the non-convergence region gradually narrows in the upper part, and there is a small portion of the useful region delimited by the non-convergence region on the left and the black border on the right. Finally, we note that points with $p_a = 0$ (noiseless decoder) and $p_0 > 0.039$ (threshold of the noiseless decoder) -- represented by the solid red line superimposed on the vertical axis in Figure~\ref{fig:ureg_add_sp_mu1_pt} -- are excluded from the useful region.  Indeed, for such points $P_e^{(\infty)} > p_0$; however, for $p_a$ greater than but close to zero, we have $P_e^{(\infty)}\approx \frac{p_a}{2\widetilde{Q}}$ (see Figure \ref{fig:piter_noisy_adder} and related discussion). 

We exemplify the decoder behavior on four points located on one side and the other of the left and right boundaries of the non-convergence region. These points are indicated in Figure~\ref{fig:ureg_add_sp_mu1_pt} by $A, B, C$, and $D$. For all the four points $p_0 = 0.03$, while $p_a = 0.027, 0.03, 0.039$, and $0.042$, respectively.  
The error probability $(P_e^{(\ell)})_{\ell > 0}$ is plotted for each one of these points in Figure~\ref{fig:piter_noisy_adder_pts}. The point $A$ belongs to the useful region, and it can be seen from Figure~\ref{subfig:piter_noisy_adder_ptA} that $(P_e^{(\ell)})_{\ell > 0}$ converges to $P_e^{(\infty)} = 9.11\times 10^{-4} < p_0$. For the point $B$, located just on the other side of the dashed red border of the useful region,  $(P_e^{(\ell)})_{\ell > 0}$ exhibits a periodic behavior (although we only plotted the first $500$ iterations, we verified the periodic behavior on the first $5\times 10^4$ iterations). Crossing the non-convergence region from left to the right, the amplitude between the inferior and superior limits of $(P_e^{(\ell)})_{\ell > 0}$ decreases (point C), until it reaches again a convergent behavior (point D).  Note that $D$ is outside the useful region, as $(P_e^{(\ell)})_{\ell > 0}$ converges to $P_e^{(\infty)} = 0.0605 > p_0$. 

The non-convergence region gradually narrows in the upper part, and for $0\leq p_a < 0.01$ it takes the form of a {\em discontinuity line}: $P_e^{(\infty)}$ takes values close to $10^{-4}$ just below this line, and values greater than $0.05$ above this line. 

Note that points $(p_a, p_0)$  with $p_0 < \frac{p_a}{2\widetilde{Q}} = \frac{p_a}{30}$ cannot belong to the useful region, since from Proposition~\ref{prop:lower-bounds} we have $P_e^{(\infty)}\geq \frac{p_a}{2\widetilde{Q}} > p_0$. Moreover, we note that the bottom border of the useful region (solid black curve) is virtually identical to, but slightly above, the line defined by $p_0 = \frac{p_a}{2\widetilde{Q}}$.

\subsubsection{Optimization of the quantization map}
\begin{figure}[!bht]
\centering
\includegraphics[width=90mm]{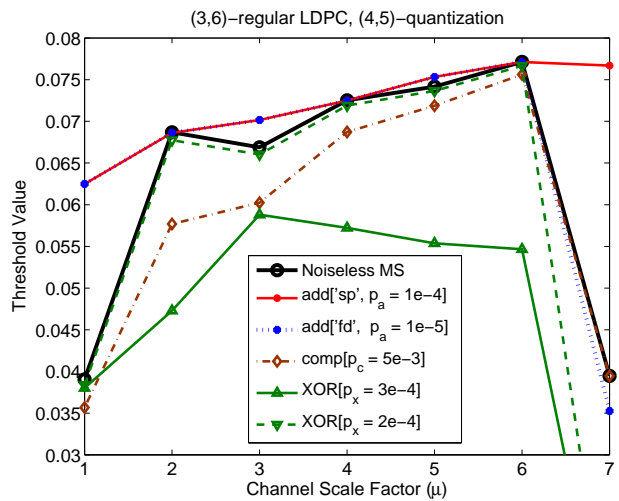}
\caption[Threshold values of noiseless and noisy MS decoders with various channel scale factors]{Threshold values of noiseless and noisy MS decoders with various channel scale factors (for noisy decoders, threshold values correspond to a target error probability $\eta = 10^{-5}$)}
\label{fig:csf_optimisation}
\end{figure}

In this section we show that the decoder performance can be significantly improved by using an appropriate choice of the channel scale factor $\mu$. Figure~\ref{fig:csf_optimisation} shows the threshold values for the noiseless and several noisy decoders with channel scale factors $\mu\in\{1,2,\dots,7\}$. For the noisy decoders, the threshold values are computed for a target error probability $\eta = 10^{-5}$ (see Equation~(\ref{eq:eta_threshold})). 

The solid black curve in Figure~\ref{fig:csf_optimisation} correspond to the noiseless decoder. The solid red curve and the dotted blue curve correspond to the MS decoder with sign-preserving noisy adder and full-depth noisy adder, respectively. The adder error probability is $p_a = 10^{-4}$ for the sign-preserving noisy adder, and $p_a = 10^{-5}$  for the full-depth adder\footnote{Note that according to Proposition~\ref{prop:lower-bounds}, a necessary condition to achieve a target error probability $P_{e}^{(\infty)} \leq \eta = 10^{-5}$ is $p_a\leq 2\widetilde{Q}\eta = 3\times 10^{-4}$ for the signed-preserving adder, and $p_a\leq 2\eta\frac{2\widetilde{Q}+1}{2\widetilde{Q}} = 2.07\times 10^{-5}$ for the full-depth adder.}. The two curves are superimposed for $1\leq \mu \leq 6$, and differ only for $\mu=7$. The corresponding threshold values are equal to those obtained in the noiseless case for $\mu\in\{2, 4, 6\}$.  For $\mu\in\{1, 3, 5\}$, the   MS decoders with noisy-adders exhibit better thresholds than the noiseless decoder. This is due to the fact that the
messages alphabet ${\cal M}$ is underused by the noiseless decoder, since all the exchanged messages are necessarily odd (recall that all variable-nodes are of degree $d_v=3$). For the MS decoders with noisy adders, the noise present in the adders leads to a more efficient use of the  messages alphabet, which allows the decoder 
to escape from fixed-point attractors and hence results in better thresholds (Section \ref{subsec:ms_q1}).

Figure~\ref{fig:csf_optimisation} also shows a curve corresponding to the MS decoder with a noisy comparator having $\pcomp = 0.005$, and two curves for the MS decoder with noisy {\sc xor}-operators, having respectively  $\pxor = 2\times 10^{-4}$ and  $\pxor = 3\times 10^{-4}$. 

Concerning the noisy {\sc xor}-operator, it can be seen that the threshold values corresponding to $\pxor = 2\times 10^{-4}$  are very close to those obtained in the noiseless case, except for $\mu = 7$ (the same holds for values $\pxor < 2\times 10^{-4}$). However, a significant degradation of the threshold can be observed when slightly increasing the {\sc xor} error probability to $\pxor = 3\times 10^{-4}$. Moreover, although not shown in the figure, it is worth mentioning that for $\pxor \geq 5\times 10^{-4}$, the target error probability $\eta = 10^{-5}$ can no longer be reached (thus, all threshold values are equal to zero).

Finally, we note that except for the noisy {\sc xor}-operator with $\pxor = 3\times 10^{-4}$, the best choice of the channel scale factor is $\mu=6$. For the  noisy {\sc xor}-operator with $\pxor = 3\times 10^{-4}$, the best choice of the channel scale factor is $\mu=3$. This is rather surprising, as in this case the messages alphabet is underused by the decoder: all the exchanged messages are odd, and the fact that the {\sc xor}-operator is noisy does not change their parity.

\medskip\noindent{\bf Assumption:} In the following sections, we will investigate the impact of the noisy adder, comparator and  {\sc xor}-operator on the   MS decoder performance, assuming that the channel scale factor is $\mu=6$.

\subsubsection{Study of the impact of the noisy adder (quantization  map $\mathbf{q}_6$)}

\begin{figure}[!thb]
\centering
\subfigure[Sign-preserving noisy adder]{\includegraphics[width=.49\linewidth]{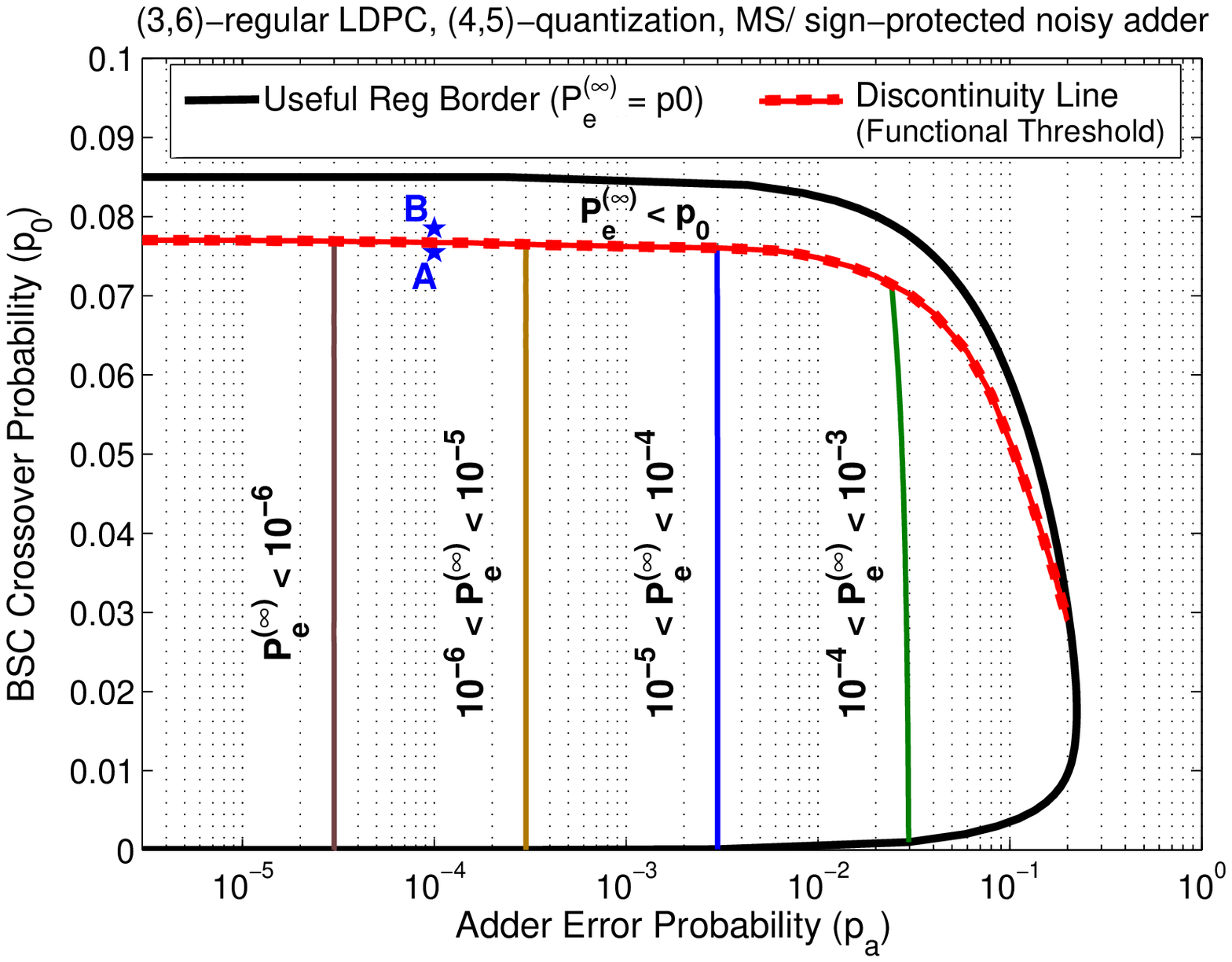}\label{subfig:bsc_45_eta_regions_add_sp}}%
\subfigure[Full-depth noisy adder]{\includegraphics[width=.49\linewidth]{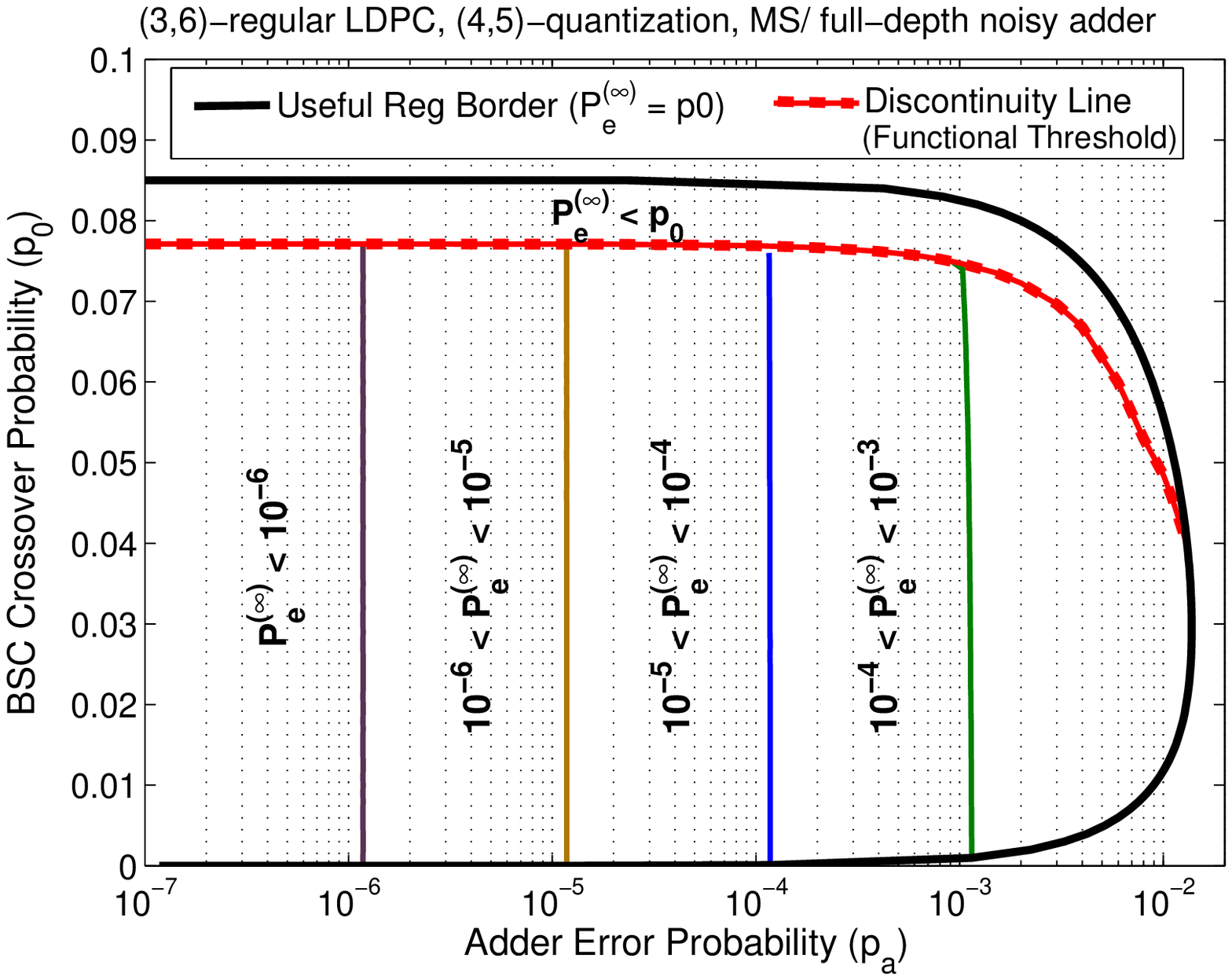}\label{subfig:bsc_45_eta_regions_add_fd}}%
\caption{Useful and $\eta$-threshold regions of the MS decoder with noisy adder}
\label{fig:bsc_45_eta_regions_add}
\end{figure}

In order to evaluate the impact of the noisy adder on the MS decoder performance, the useful region and the $\eta$-threshold regions have been computed, assuming that only the adders within the VN-processing step are noisy ($p_a > 0$), while the CN-processing step is noiseless ($\pxor = p_c = 0$).  This regions are represented in Figure~\ref{fig:bsc_45_eta_regions_add}, for both sign-preserving and full-depth noisy adder models.

The useful region is delimited by the solid black curve. The vertical lines delimit the $\eta$-threshold regions, for $\eta = 10^{-3}, 10^{-4}, 10^{-5}, 10^{-6}$ (from right to the left).

\begin{figure}[!thb]
\centering
\subfigure[Point $A(p_0 = 0.0770, p_a = 10^{-4})$]{\includegraphics[width=.49\linewidth]{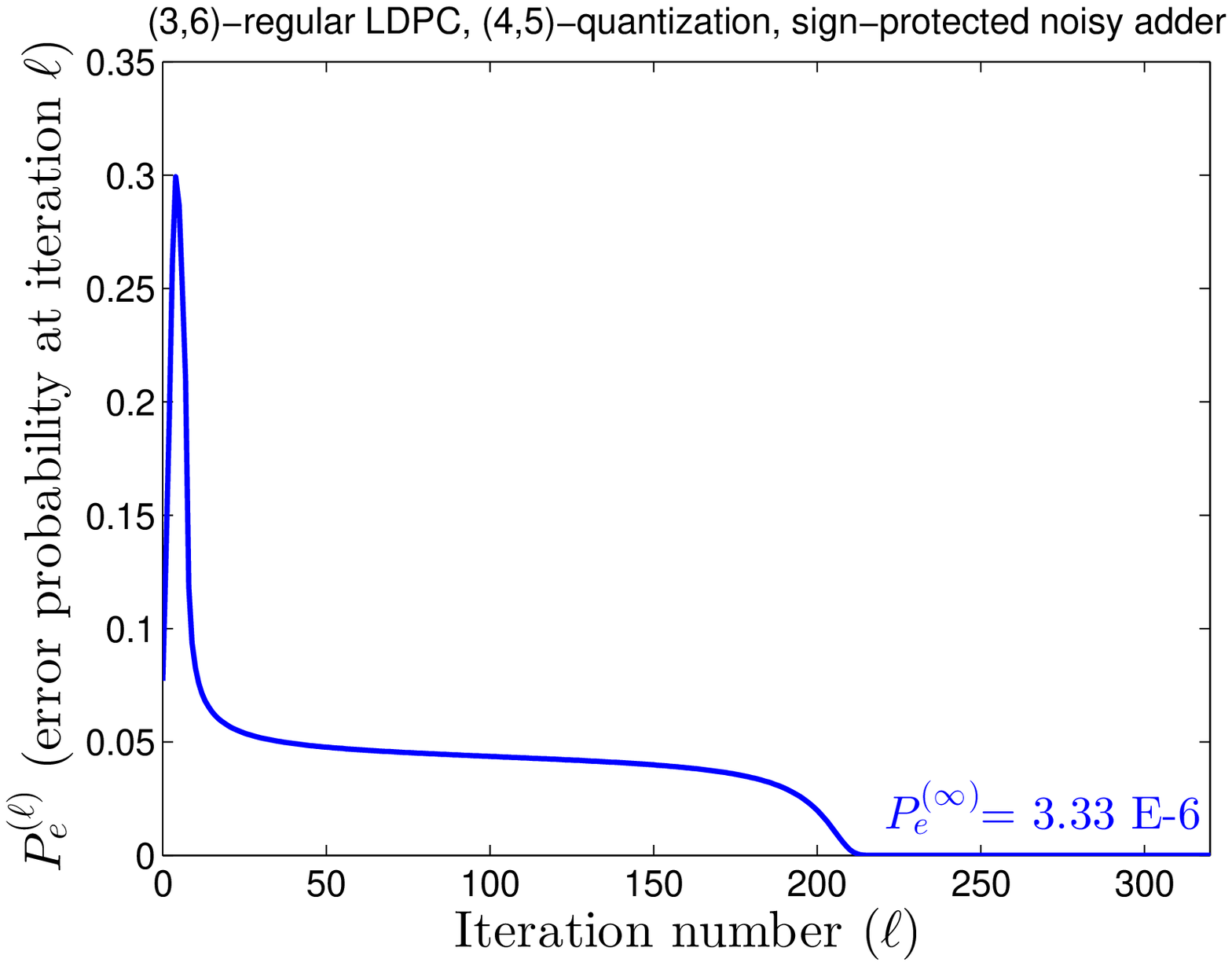}\label{subfig:bsc_45_plateau_ptA}}%
\subfigure[Point $B(p_0 = 0.0772, p_a = 10^{-4})$]{\includegraphics[width=.49\linewidth]{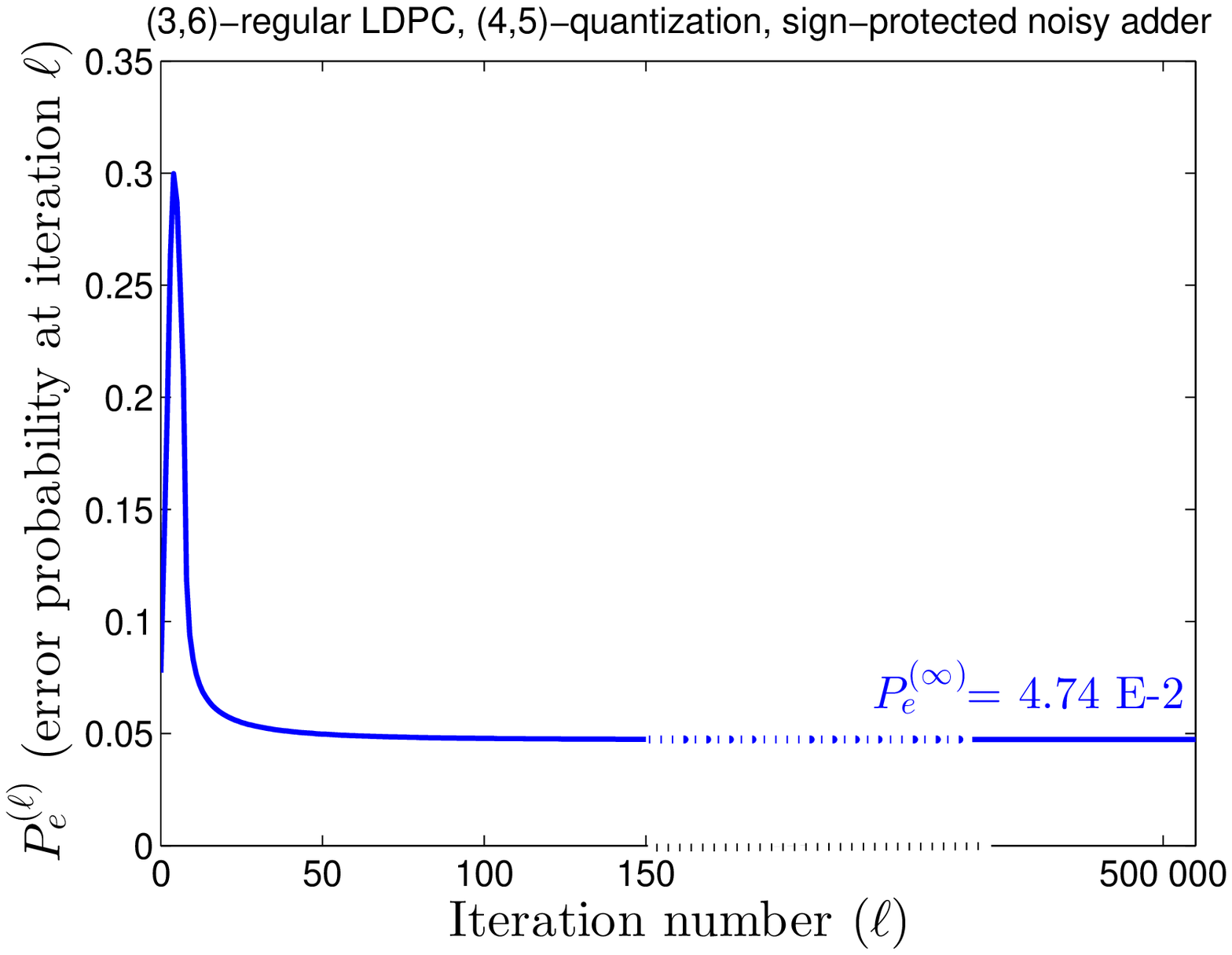}\label{subfig:bsc_45_plateau_ptB}}%
\caption[Illustration of the early plateau phenomenon]{Illustration of the early plateau phenomenon (points A and B from Figure~\ref{subfig:bsc_45_eta_regions_add_sp})}
\label{fig:bsc_45_plateau}
\end{figure}

Note that unlike the case $\mu = 1$ (Section~\ref{subsec:ms_q1}), there is no non-convergence region when the channel scale factor is set to $\mu=6$. Hence, the border of the useful region corresponds to points $(p_a, p_0)$ for which $P_{e}^{(\infty)} = p_0$. However, it can be observed that there is still a {\em discontinuity line} (dashed red curve) inside the useful region. This discontinuity line does not hide a periodic (non-convergent) behavior, but it is due to the occurrence of an {\em early plateau phenomenon} in the convergence of  $(P_{e}^{(\ell)})_\ell$. This phenomenon is illustrated in Figure~\ref{fig:bsc_45_plateau}, where the error probability $(P_{e}^{(\ell)})_\ell$ is plotted as a function of the iteration number $\ell$, for the two points A and B from Figure~\ref{subfig:bsc_45_eta_regions_add_sp}. For point A, it can be observed that the error probability $P_{e}^{(\ell)}$ reaches a first plateau for $\ell \approx 50$, then drops to $3.33\times 10^{-6}$ for $\ell \geq 250$.  For point B, $P_{e}^{(\ell)}$ behaves in a similar manner during the first iterations, but it does not decrease below the plateau value as $\ell$ goes to infinity. Although we have no analytic proof of this fact, it was numerically verified for $\ell \leq 5\times 10^{5}$.

\begin{figure}[!htb]
\centering
\subfigure[$p_a = 0$ (noiseless decoder )]{\includegraphics[width=.32\linewidth]{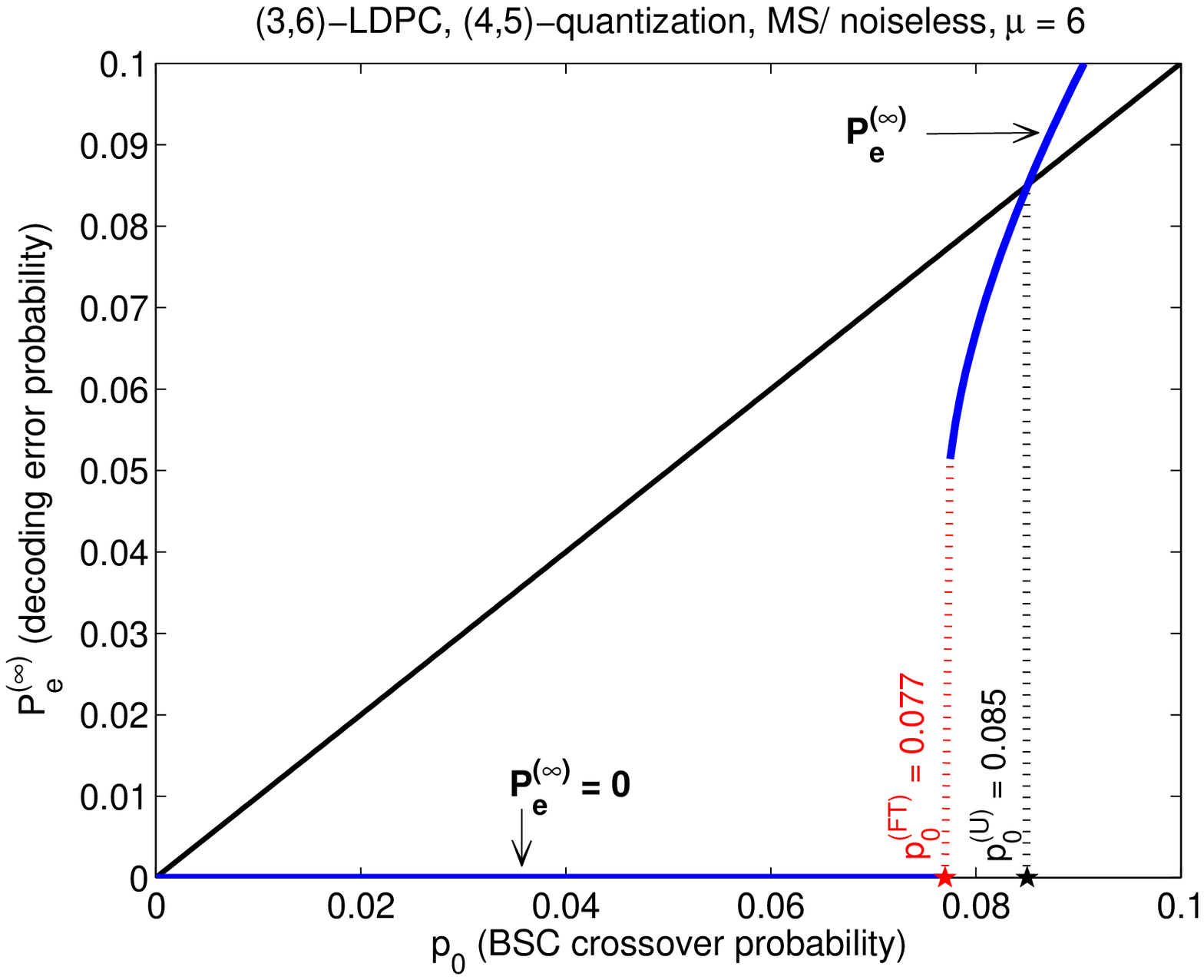}\label{subfig:pinf_noiselessMS_mu6}}%
\subfigure[$p_a = 10^{-4}$]{\includegraphics[width=.33\linewidth]{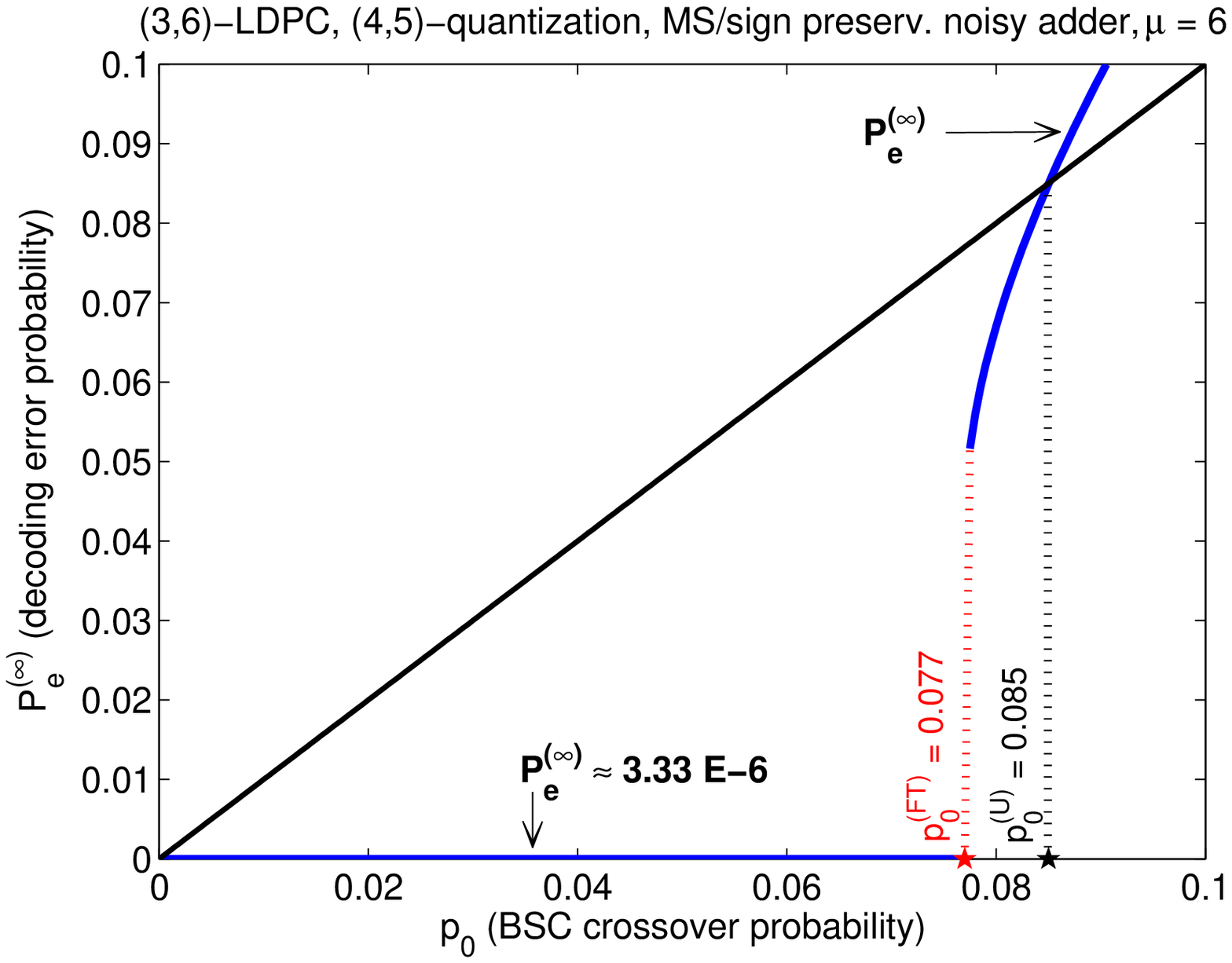}\label{subfig:pinf_noisy_adder_mu6}}%
\subfigure[$p_a = 0.05$]{\includegraphics[width=.32\linewidth]{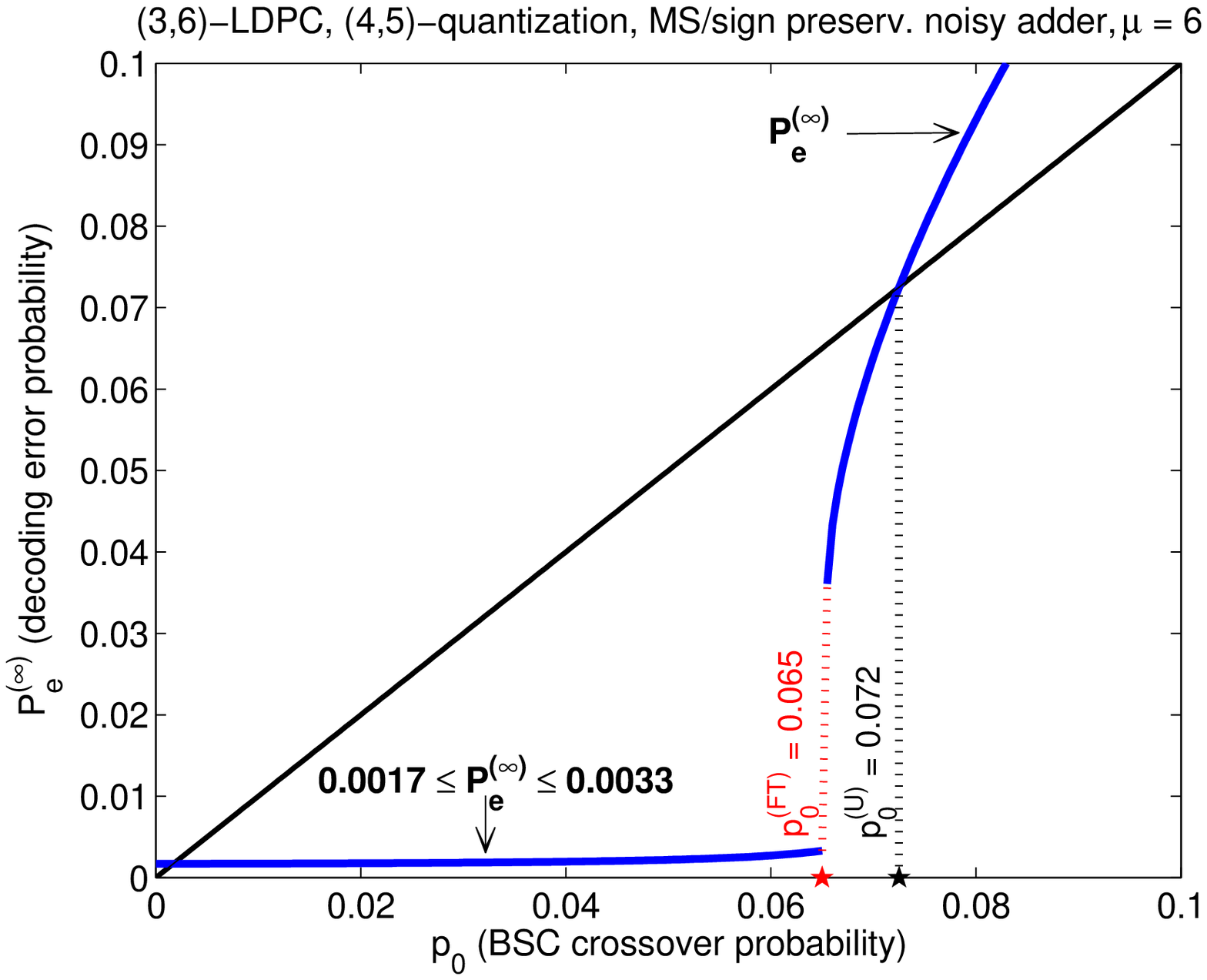}\label{subfig:pinf_noisy_adder_mu6_sec}}%
\caption{Asymptotic error probability $P_e^{(\infty)}$ as a function of $p_0$; noiseless and noisy MS decoder with sign-preserving noisy adder}
\label{fig:pinf_mu6}
\end{figure} 

In Figure~\ref{fig:pinf_mu6}, we plotted the asymptotic error probability $P_e^{(\infty)}$ as a function of $p_0$, for the noiseless decoder ($p_a = 0$), and for the sign-preserving noisy adder with error probability values  $p_a = 10^{-4}$ and $p_a = 0.05$. In each plot we have also represented two points $p_0^{(\text{U})}$ and   $p_0^{(\text{FT})}$, corresponding respectively to the values of $p_0$ on the upper-border of the useful region, and on the discontinuity line. Hence, $p_0^{(\text{FT})}$ coincides with the classical threshold of the MS decoder in the noiseless case, and it is equal to the functional threshold defined in Section~\ref{subsec:funct_threshold} in case of noisy decoders. In the following, the  
sub-region of the useful region located below the discontinuity line will be referred to as the \textbf{\em functional region}. Within this region, if the adder error probability is small enough, it can be observed that:

\noindent\parbox[t]{6mm}{$(a)$}\parbox[t]{\linewidth-6mm}{For the sign-preserving adder: $P_{e}^{(\infty)} \approx \frac{p_a}{30}$, for $p_a \lessapprox 3\times 10^{-2}$, which corresponds to the value given by the lower-bound ($\frac{1}{2\widetilde{Q}}p_a = \frac{1}{30} p_a$) from Proposition~\ref{prop:lower-bounds}.} 

\noindent\parbox[t]{6mm}{$(b)$}\parbox[t]{\linewidth-6mm}{For the full-depth adder: $P_{e}^{(\infty)} \approx 1.17 p_a$, for $p_a \lessapprox 10^{-3}$, which is about twice higher than the value given by the lower-bound ($\frac{1}{2}\padd + \frac{1}{4\widetilde{Q}}\padd = 0.52 p_a$) from Proposition~\ref{prop:lower-bounds}.} 

Finally, we note that by protecting the sign of the noisy adder, the useful region is expanded by a factor of roughly $2\widetilde{Q}$, representing an exponential improvement with respect to the number of bits of the adder (see also the discussion following the proof of Proposition ~\ref{prop:lower-bounds}).

\subsubsection{Study of the impact of the noisy XOR-operator (quantization  map $\mathbf{q}_6$)}

\begin{figure}[!htb]
\centering
\includegraphics[width=.5\linewidth]{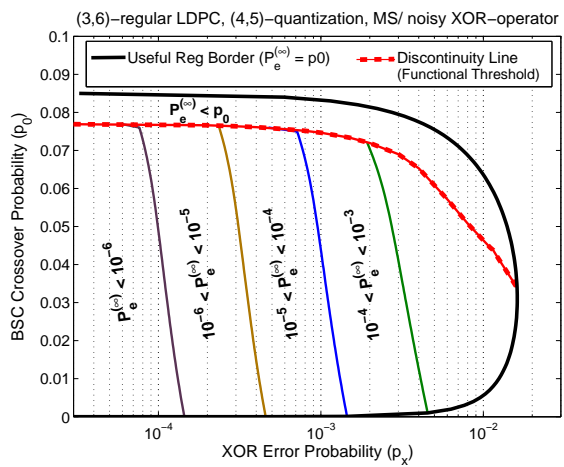}
\caption{Useful and $\eta$-threshold regions of the MS decoder with noisy {\sc xor}-operator}
\label{fig:bsc_45_eta_regions_xor}
\end{figure}

The useful region and the $\eta$-threshold regions of the decoder, assuming that only the {\sc xor}-operator used within the CN-processing step is noisy, are plotted in Fig. \ref{fig:bsc_45_eta_regions_xor}. Similar to the noisy-adder case, a discontinuity (functional threshold) line can be observed inside the useful region, which delimits the {\em functional region} of the decoder.  

Comparing the $\eta$-threshold regions from Figure~\ref{fig:bsc_45_eta_regions_add} and Figure~\ref{fig:bsc_45_eta_regions_xor}, it can be observed that in order to achieve a target error probability $P_e^{(\infty)} \leq 10^{-6}$, the error probability parameters of the noisy adder and of the noisy {\sc xor}-operator must satisfy:  
\begin{itemize}
\item $\padd < 1.17 \times 10^{-6}$, for the full-depth noisy-adder;
\item $\padd < 3 \times 10^{-5}$, for the sign-preserving noisy-adder;
\item $\pxor < 7 \times 10^{-5}$, for the noisy {\sc xor}-operator. \\ (moreover, values of $\pxor$ up to $1.4 \times 10^{-4}$ are tolerable if $p_0$ is sufficiently small) 
\end{itemize}
The most stringent requirement concerns the error probability of the  full-depth noisy-adder, thus we may consider that it has the most negative impact on the decoder performance. On the other hand, the less stringent requirement concerns the error probability of the noisy {\sc xor}-operator.   

Finally, it is worth noting that in practical cases the value of $\pxor$ should be significantly lower than the value of $p_a$ (given the high number of elementary gates contained in the adder). Moreover, since the {\sc xor}-operators used to compute the signs of CN messages represent only a small part of the decoder, this part of the circuit could be made reliable by using classical fault-tolerant methods, with a limited impact on the overall decoder design. 

\subsubsection{Study of the impact of the noisy comparator (quantization  map $\mathbf{q}_6$)}
This section investigates the case when comparators used within the CN-processing step are noisy ($\pcomp > 0$), but $\padd = \pxor = 0$. Contrary to the previous cases, this case exhibits a ``classical'' threshold phenomenon, similar to the noiseless case: for a given $\pcomp > 0$, there exists a $p_0$-threshold value, denoted by $p_0^{\text{(TH)}}$, such that $P_e^{(\infty)} = 0$ for any $p_0 < p_0^{\text{(TH)}}$. 

\begin{figure}[!thb]
\centering
\includegraphics[width=.5\linewidth]{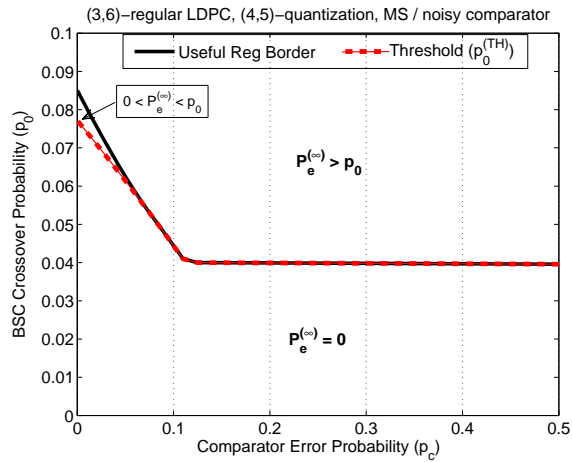}
\caption{Useful region and threshold curve of the MS decoder with noisy comparator}
\label{fig:bsc_45_eta_regions_comp}
\end{figure}

The threshold value $ p_0^{\text{(TH)}}$ is plotted as a function of $\pcomp$ in Figure~\ref{fig:bsc_45_eta_regions_comp}. The functional region of the decoder is located below the threshold curve, and $P_e^{(\infty)} = 0$ for any point within this region. In particular, it can be seen that $P_e^{(\infty)} = 0$ for any $p_0 \lessapprox 0.039$ and any  $p_c > 0$.
Although such a threshold phenomenon might seem surprising for a noisy decoder, it can be easily explained. 
The idea behind is that in this case the crossover probability of the channel is small enough, so that in the CN-processing step only the sign of check-to-variable messages is important, but not their amplitudes. In other words a decoder that only computes (reliably) the signs of check-node messages and randomly chooses their amplitudes, would be able to perfectly decode the received word.   

Finally, we note that the useful region of the decoder extends slightly above the threshold curve: for $p_c$ close to $0$, there exists a small region above the threshold curve, within which $0 < P_e^{(\infty)} <p_0$.

\subsection{Numerical results for the BI-AWGN channel}

For the BI-AWGN, the channel output is given by $y = x + z$, where $x\in\{\pm 1\}$ is the channel input and $z$ is the additive white Gaussian noise with variance $\sigma^2$. Threshold values and useful regions of the decoder will be described in terms of Signal to Noise Ratio (SNR), defined by $\text{SNR} = -10\log_{10}(\sigma^2)$.

\medskip For a given channel scale factor $\mu$, the quantization map $\mathbf{q}_{\mu}$ is defined by $\mathbf{q}_{\mu}(y) = \mathbf{s}_{\cal M}([\mu{\cdot}y])$, where $[\mu{\cdot}y]$ denotes the nearest integer to $\mu{\cdot}y$, and $\mathbf{s}_{\cal M}$ is the saturation map (see also Equation~(\ref{eq:quant-map})).

Similar to the BSC case, the choice of the channel scale factor $\mu$ may significantly impact the decoder performance. Hence, we start first by optimizing the channel scale factor value, and then we investigate the impact of the different noisy components on the decoder performance. 

\medskip \noindent{\bf Remark:} For the BI-AWGN channel we denote by $p_0 \stackrel{\text{def}}{=} P_e^{(0)}$ the error probability at iteration $0$, which is, by definition, the probability of the {\em a priori information} $\gamma = \mathbf{q}_{\mu}(y)$ being in error. Hence, $p_0 =  \sum_{z=-Q}^{-1}C(z)+\frac{1}{2}C(0)$. Using Equation~(\ref{eq:awgn_c}) it follows that:
\begin{equation}
p_0 = 1 - \frac{1}{2}\left[q\left(\frac{-0.5-\mu}{\mu\sigma}\right) +  q\left(\frac{0.5-\mu}{\mu\sigma}\right) \right]
\end{equation}

\subsubsection{Optimization of the quantization map}

\begin{figure}[!bht]
\centering
\includegraphics[width=90mm]{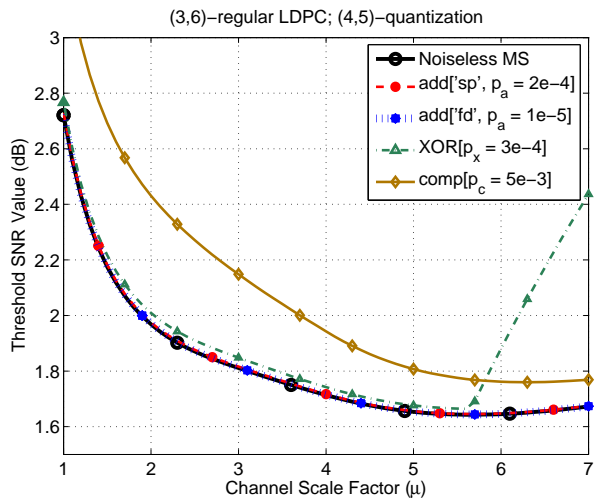}
\caption[Threshold SNR values of noiseless and noisy decoders with various channel scale factors]{Threshold SNR values of noiseless and noisy decoders with various channel scale factors (for noisy decoders, threshold values correspond to a target error probability $\eta = 10^{-5}$)}
\label{fig:awgn_csf_optimisation}
\end{figure}

The goal of this section is to provide an optimal choice of the channel scale factor $\mu$. Figure~\ref{fig:awgn_csf_optimisation} shows the threshold SNR values for the noiseless and several noisy decoders for channel scale factors $\mu$ varying within the interval $[1,7]$. For the noisy decoders, the threshold values are computed for a target error probability $\eta = 10^{-5}$ (see Equation~(\ref{eq:eta_threshold})). 

The solid black curve in Figure~\ref{fig:awgn_csf_optimisation} correspond to the noiseless decoder. The dashed red curve and the dotted blue curve correspond to the MS decoder with sign-preserving noisy adder and full-depth noisy adder, respectively. The adder error probability is $p_a = 2\times 10^{-4}$ for the sign-preserving noisy adder, and $p_a = 10^{-5}$  for the full-depth adder\footnote{Note that according to Proposition~\ref{prop:lower-bounds}, a necessary condition to achieve a target error probability $P_{e}^{(\infty)} \leq \eta = 10^{-5}$ is $p_a\leq 2\widetilde{Q}\eta = 3\times 10^{-4}$ for the signed-preserving adder, and $p_a\leq 2\eta\frac{2\widetilde{Q}+1}{2\widetilde{Q}} = 2.07\times 10^{-5}$ for the full-depth adder.}. These three curves are virtually indistinguishable.

Figure~\ref{fig:awgn_csf_optimisation} also shows two curves corresponding respectively to the MS decoder with a noisy {\sc xor}-operator ($\pxor = 2\times 10^{-4}$) and to the MS decoder with a noisy comparator ($\pcomp = 0.005$). Finally, we note that in all cases the best choice of the channel scale factor is $\mu\approx 5.5$. 

\medskip \noindent{\bf Assumption:} In the following sections, we will investigate the impact of the noisy adder, comparator and  {\sc xor}-operator on the   MS decoder performance, assuming that the channel scale factor is $\mu=5.5$.

\subsubsection{Study of the impact of the noisy adder}

\begin{figure}[!b]
\centering
\subfigure[Sign-preserving noisy adder]{\includegraphics[width=.49\linewidth]{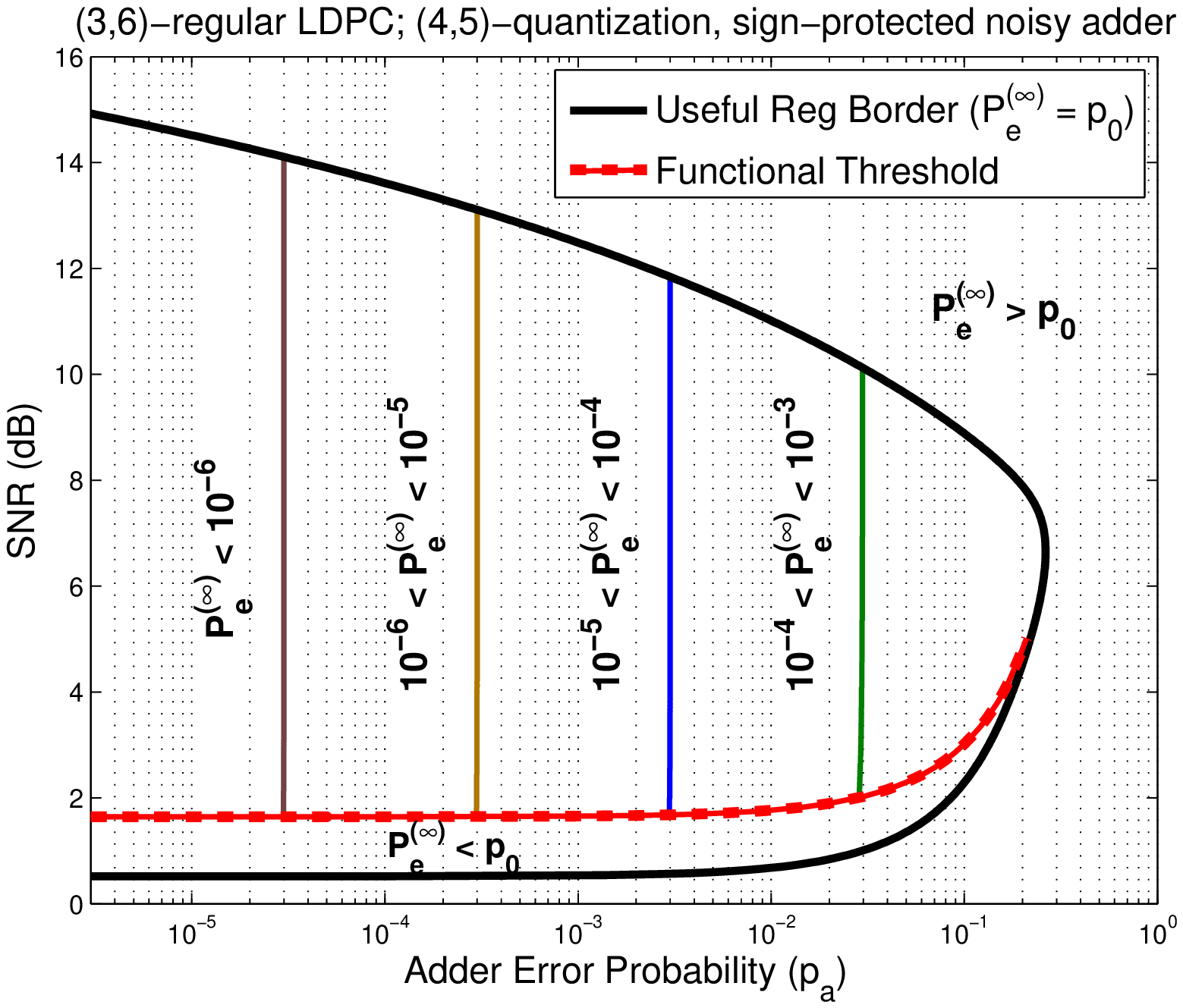}\label{subfig:awgn_45_eta_regions_add_sp}}%
\subfigure[Full-depth noisy adder]{\includegraphics[width=.49\linewidth]{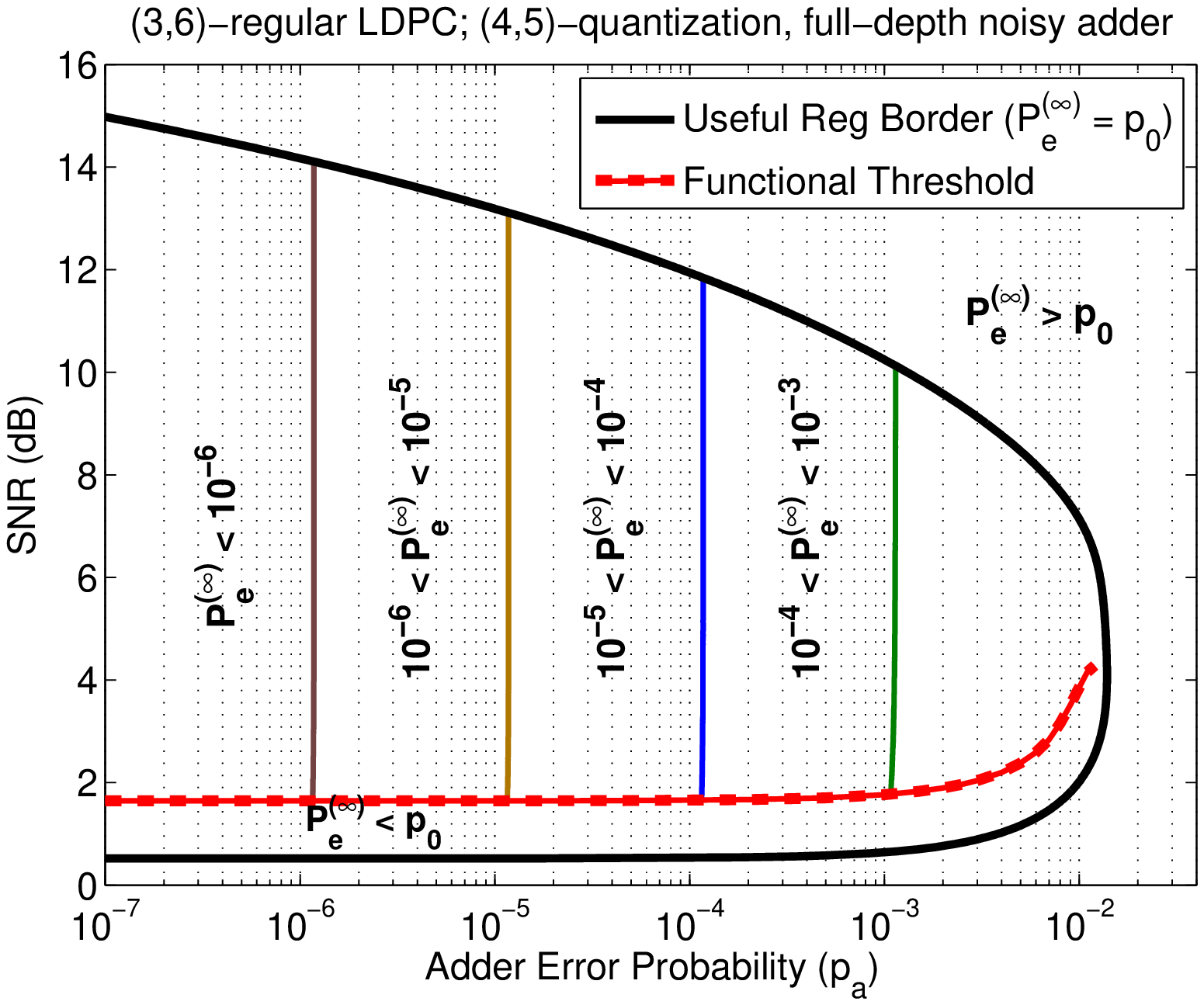}\label{subfig:awgn_45_eta_regions_add_fd}}%
\caption{Useful and $\eta$-threshold regions of the MS decoder with noisy adder ({\sc bi-awgn})}
\label{fig:awgn_45_eta_regions_add}
\end{figure}

Useful and $\eta$-regions of the MS decoder with noisy adders are represented in Figure~\ref{fig:awgn_45_eta_regions_add}, for both sign-preserving and full-depth noisy adder models.
The useful region is delimited by the solid black curve, while vertical lines delimit the $\eta$-threshold regions, for $\eta = 10^{-3}, 10^{-4}, 10^{-5}, 10^{-6}$ (from right to the left). The {\em functional threshold} of the decoder is also displayed by a red dashed curve.

\begin{figure}[!thb]
\centering
\subfigure[sign-preserving noisy adder, $p_a = 10^{-4}$]{\includegraphics[width=.49\linewidth]{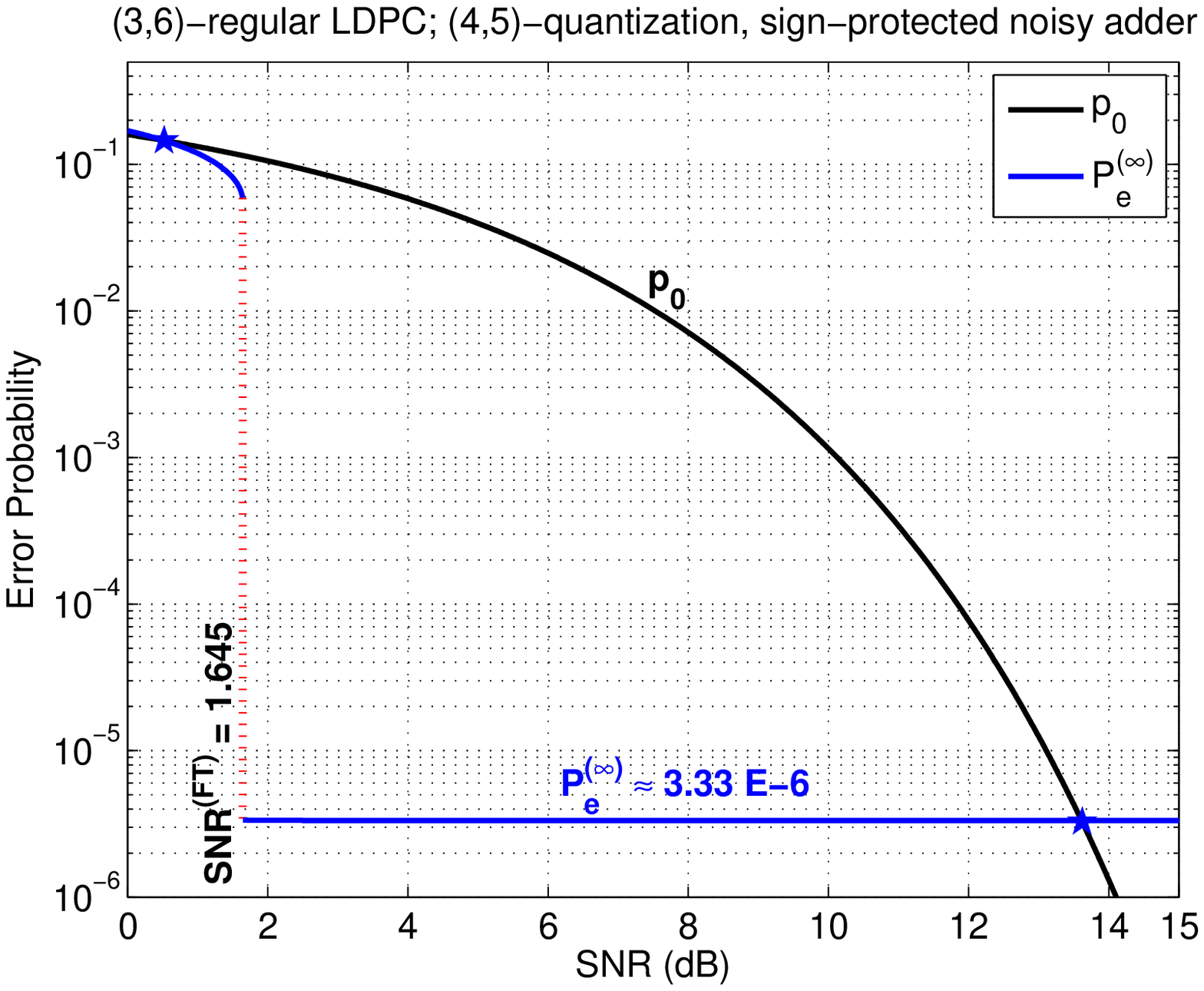}\label{subfig:awgn_pinf_noisy_adder_sp}}%
\subfigure[full-depth noisy adder, $p_a = 10^{-4}$]{\includegraphics[width=.49\linewidth]{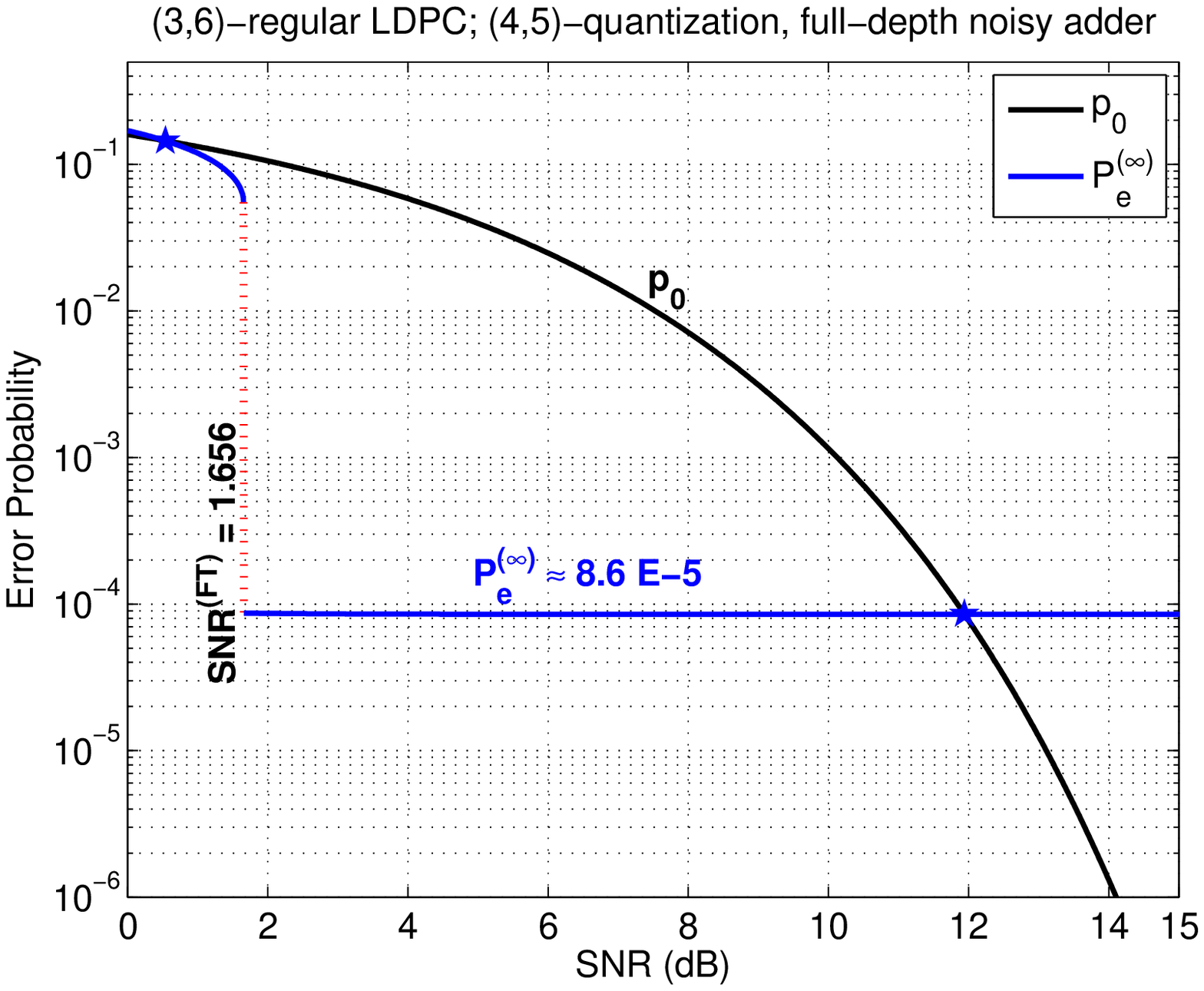}\label{subfig:awgn_pinf_noisy_adder_fd}}%
\caption{Asymptotic error probability $P_e^{(\infty)}$ of the MS decoder with noisy-adder as a function of the SNR}
\label{fig:awgn_pinf_noisy_adder}
\end{figure} 

Figure~\ref{fig:awgn_pinf_noisy_adder} shows the input and output error probabilities of the decoder ($p_0$ and $P_e^{(\infty)}$) as functions of the SNR value, for the sign-preserving and full-depth noisy adder models with $p_a = 10^{-4}$. The two intersection points between the two curves correspond to the points on the lower and upper borders of the useful region in Figure~\ref{fig:awgn_45_eta_regions_add}, for  $p_a = 10^{-4}$. The discontinuity point of the $P_e^{(\infty)}$ curve corresponds to the functional threshold value in Figure~\ref{fig:awgn_45_eta_regions_add}, for  $p_a = 10^{-4}$.

\subsubsection{Study of the impact of the noisy XOR-operator and noisy comparator}

\begin{figure}[!htb]
\centering
\parbox{.49\linewidth}{%
\includegraphics[width=\linewidth]{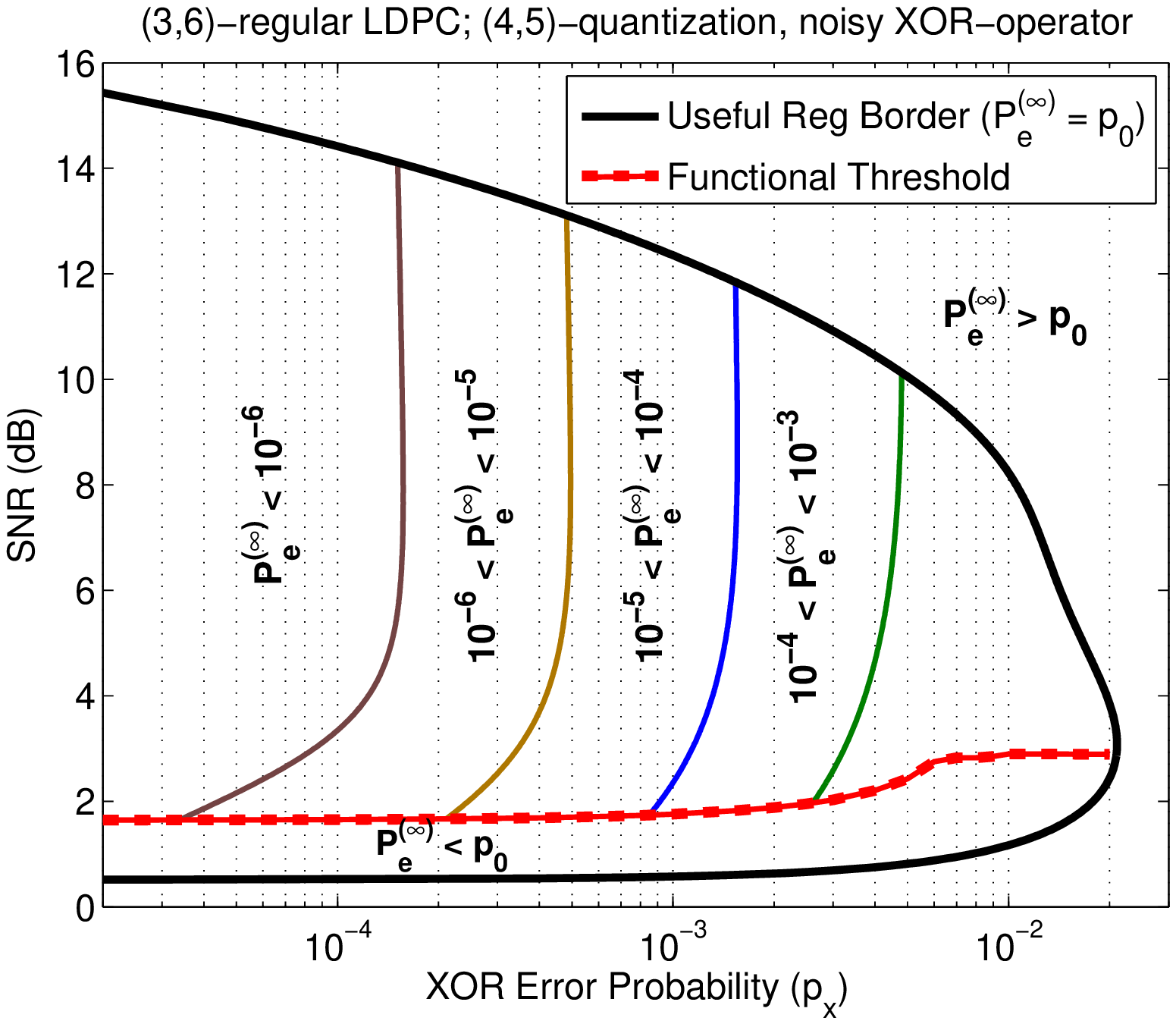}
\caption{Useful and $\eta$-threshold regions of the MS decoder with noisy {\sc xor}-operator ({\sc bi-awgn})}
\label{fig:awgn_45_eta_regions_xor}}\hfill
\parbox{.49\linewidth}{%
\includegraphics[width=\linewidth]{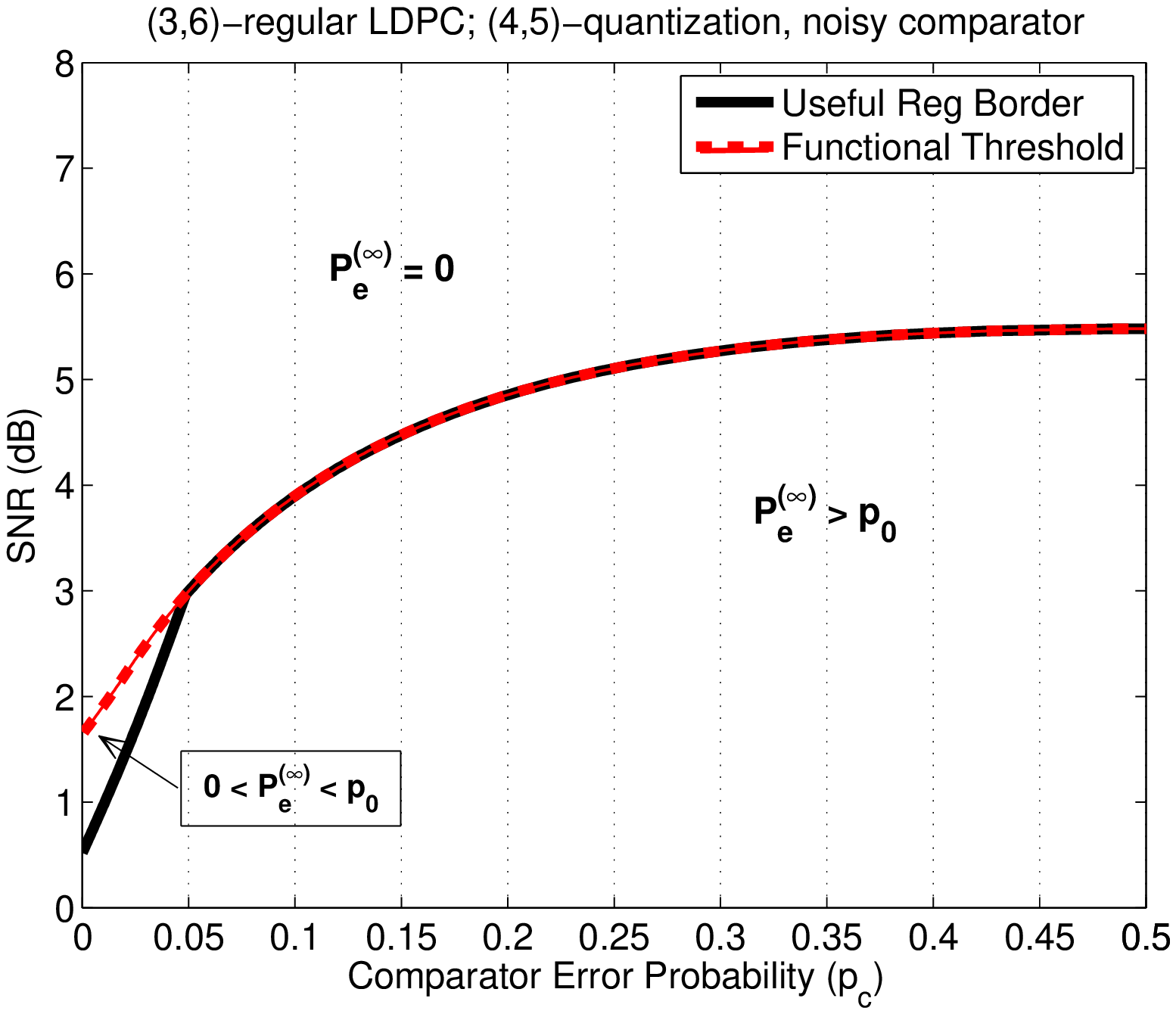}
\caption{Useful region and threshold curve of the MS decoder with noisy comparator ({\sc bi-awgn})}
\label{fig:awgn_45_eta_regions_comp}}%
\end{figure}

The useful region and the $\eta$-threshold regions of the MS decoder, assuming that only the {\sc xor}-operator used within the CN-processing step is noisy, are plotted in Fig. \ref{fig:awgn_45_eta_regions_xor}. The {\em functional threshold} of the decoder is also displayed by a red dashed curve.  

The case of a noisy comparator is illustrated in Figure~\ref{fig:awgn_45_eta_regions_comp}. Similar to the BSC channel, this case exhibits a ``classical'' threshold phenomenon: for any SNR value above the functional threshold curve, one has $P_e^{(\infty)} = 0$. 

\clearpage

\section{Finite Length Performance of Min-Sum based decoders}\label{sec:finite_length_perf}
 The goal of this section is twofold:
\begin{description}
\item[(1)] To corroborate the asymptotic analysis through finite-length simulations;
\item[(2)] To investigate ways of increasing the robustness of the MS decoder to hardware noise.  
\end{description}

\noindent {\bf Assumption:} Unless otherwise stated, the $(3, 6)$-regular LDPC code with length $N = 1008$ bits from \cite{mackaywebsite} will be used for finite length simulations throughout this section.

\subsection{Practical implementation and early stopping criterion}
First of all, we note that the practical implementation of the noisy MS decoder differs slightly from the one presented in Algorithm~\ref{alg:noisy_ms}: 
\begin{itemize}
\item  The order of the {\bf VN-processing} and {\bf AP-update} steps is inverted;
\item  The variable-to-check node messages are computed by subtracting the incoming check-to-variable message from the corresponding a posteriori information value: 

\fbox{\begin{minipage}{.98\linewidth}
{\bf for all} $n=1,\dots,N$ {\bf do} \hfill $\vartriangleright$ {\bf AP-update}

      \hspace*{5mm} $\tilde{\gamma}_n = \addprred\left(\{\gamma_n\} \cup 
      \{\beta_{m,n}\}_{{m} \in{\cal H}(n)}\right)\text{;}$\\

{\bf for all} $n=1,\dots,N$ and $m\in{\cal H}(n)$ {\bf do} \hfill $\vartriangleright$ {\bf VN-processing}

    \hspace*{5mm}$\begin{array}{@{}r@{\ }c@{\ }l}
    \alpha_{m,n} &  = & \addprred\left(\tilde{\gamma}_n, -\beta_{m,n}\right)\text{;} \\
    \alpha_{m,n} & =  & \mathbf{s}_{\cal M}\left(\alpha_{m,n} \right)\text{;}
    \end{array}$ 
\end{minipage}}     
\end{itemize} 

For floating-point noiseless decoders, the two ways of computing the variable-to-check messages are completely equivalent. However, this equivalence does not hold anymore for finite-precision (noisy or noiseless) decoders, because of saturation effects and, in case of noisy decoders, of probabilistic computations.  We note that the practical implementation might result in a degradation of the decoder performance compared to the ``Density-Evolution like'' implementation (Algorithm~\ref{alg:noisy_ms}), since each variable-to-check node message {\em encompasses} $d_v+1$ additions ($d_v$ additions to compute  $\tilde{\gamma}_n $ and one subtraction).

Finally, it is worth noting that the density-evolution analysis cannot be applied to the practical implementation, due to the fact that in the VN-processing step, the computation of variable-to-check messages  $\alpha_{m,n} = \addprred\left(\{\gamma_n\}, -\beta_{m,n}\right)$ involves two correlated variables, namely $\gamma_n$ and $\beta_{m,n}$.

\subsubsection{Early stopping criterion (syndrome check)}

As described in Algorithm~\ref{alg:noisy_ms}, each decoding iteration also comprises a {\em hard decision} step, in which each transmitted bit is estimated according to the sign of the a posteriori information, and a {\em syndrome check} step, in which the syndrome of the estimated word is computed. 

Both steps are assumed to be {\em noiseless}, and the syndrome check step acts as an {\em early stopping criterion}: the decoder stops when whether the syndrome is $+1$ (the estimated word is a codeword) or a maximum number of iterations is reached.
 We note however that the syndrome check step is optional and, if missing, the decoder stops when the maximum number of iterations is reached. 

\medskip \noindent {\bf Remark:} The reason why we stress the difference between the MS decoder with and without the syndrome check step is because, as we will see shortly, the {\em noiseless} early stopping criterion may significantly improve the bit error rate performance of the {\em noisy} decoder in the error floor region. 

\medskip \noindent {\bf Assumptions:} 
\begin{itemize}
\item Unless otherwise stated, the MS decoder is assumed to implement the {\em noiseless} stopping criterion (syndrome check step).
\item The maximum number of decoding iterations is fixed to 100 throughout this section.
\end{itemize}

\subsection{Corroboration of the  asymptotic analysis through finite-length simulations}

\begin{figure}[!thb]
\centering
\includegraphics[width=90mm]{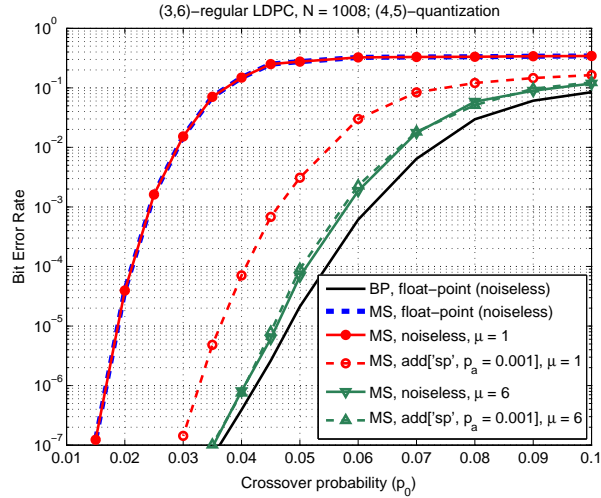}
\caption{BER performance of noiseless and noisy MS decoders with various channel scale factors}
\label{fig:bsc_flen_mu1_mu6}
\end{figure}

We start by analyzing the finite-length decoder performance over the BSC channel. Figure~\ref{fig:bsc_flen_mu1_mu6} shows the bit error rate (BER) performance of the finite-precision MS decoder (both noiseless and noisy) with various channel scale factors. For comparison purposes, we also included the BER performance of the Belief-Propagation decoder (solid black curve, no markers) and of the infinite-precision MS decoder (dashed blue curve, no markers).

It can be observed that the worst performance is achieved by the infinite-precision MS decoder~(!) and the finite-precision noiseless MS decoder with channel scale factor $\mu = 1$ (both curves are virtually indistinguishable). The BER performance of the latter improves significantly when using a sign-preserving noisy adder with error probability $p_a = 0.001$  (dashed red curve with empty circles).

For a channel scale factor $\mu = 6$, both noiseless and noisy decoders have almost the same performance (solid and dashed green curves, with triangular markers). Remarkably, the achieved BER is very close to the one achieved by the Belief-Propagation decoder!

These results corroborate the asymptotic analysis from Section~\ref{subsec:num_results_bsc} concerning the channel scale factor optimization.  

\subsubsection{Error floor performance}

Surprisingly, the BER curves of the noisy decoders from Figure~\ref{fig:bsc_flen_mu1_mu6} do not show any error floor down to $10^{-7}$. However, according to Proposition~\ref{prop:lower-bounds}, the decoding error probability should be lower-bounded by $P_e^{(\ell)} \geq \frac{1}{2\widetilde{Q}}p_a = 3.33 \times 10^{-5}$ (see also 
the $\eta$-threshold regions in Figure~\ref{subfig:bsc_45_eta_regions_add_sp}). 

The fact that the observed decoding error probability may decrease below the above lower-bound is due to the early stopping criterion (syndrome check step) implemented within the MS decoder. Indeed, as we observed in the previous section, the above lower-bound is tight, when $\ell$ (the iteration number) is sufficiently large. Therefore, as the iteration number increases, the expected number of erroneous bits gets closer and closer to $\frac{1}{2\widetilde{Q}}p_aN = 0.034$, and the probability of not having any erroneous bit within one iteration approaches $\left(1 -  \frac{1}{2\widetilde{Q}}p_a\right)^N = 0.967$. As the decoder performs more and more iterations, it will eventually reach an error free iteration. The absence of errors is at once detected by the noiseless syndrome check step, and the decoder stops.

\begin{figure}[!thb]
\centering
\parbox{.49\linewidth}{%
\includegraphics[width=\linewidth]{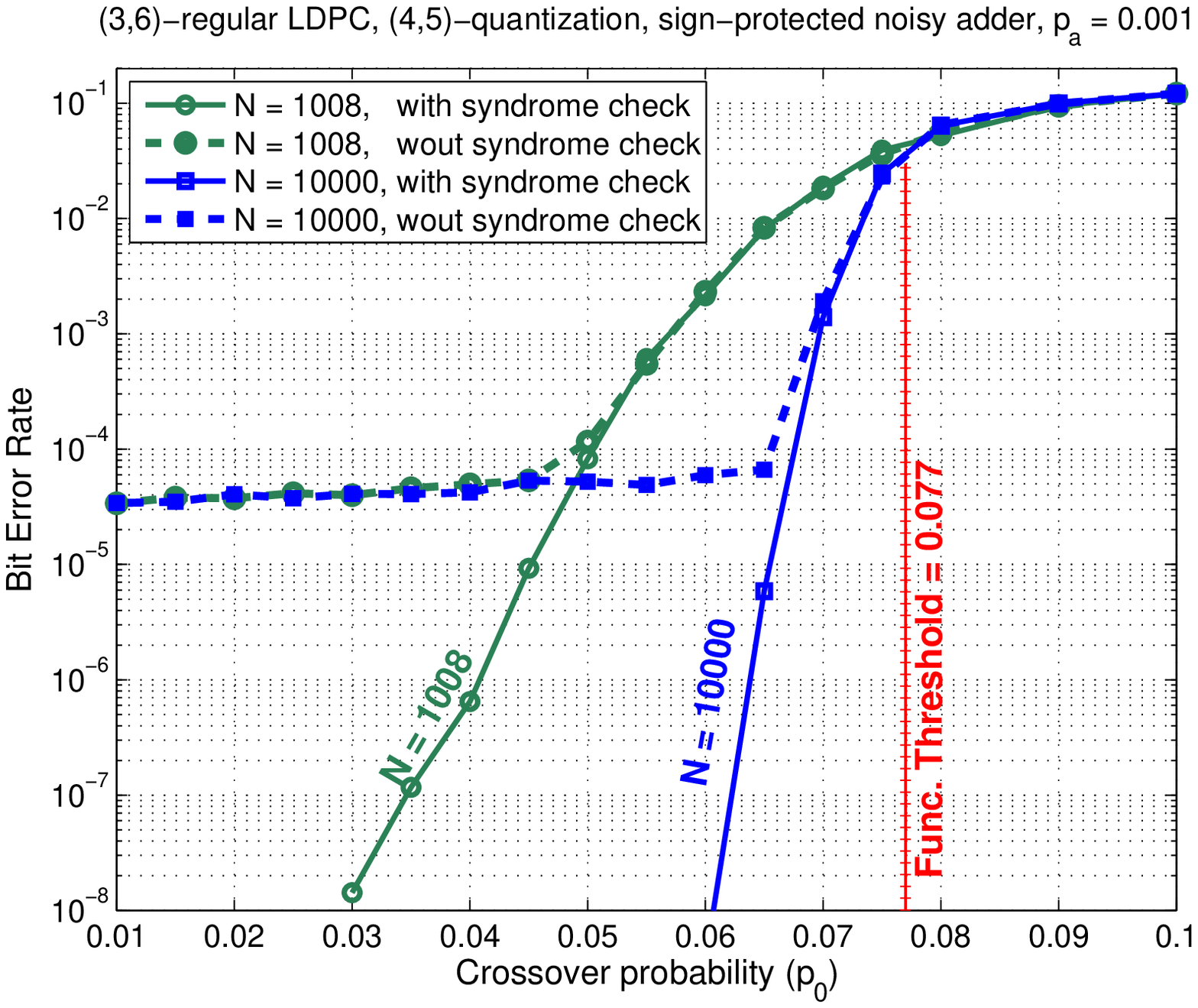}
\caption[BER performance with and without early stopping criterion]{BER performance with and without early stopping criterion (MS decoder with sign-preserving noisy adder, $p_a = 0.001$)}
\label{fig:bsc_ber_variousN}}\hfill
\parbox{.49\linewidth}{%
\includegraphics[width=\linewidth]{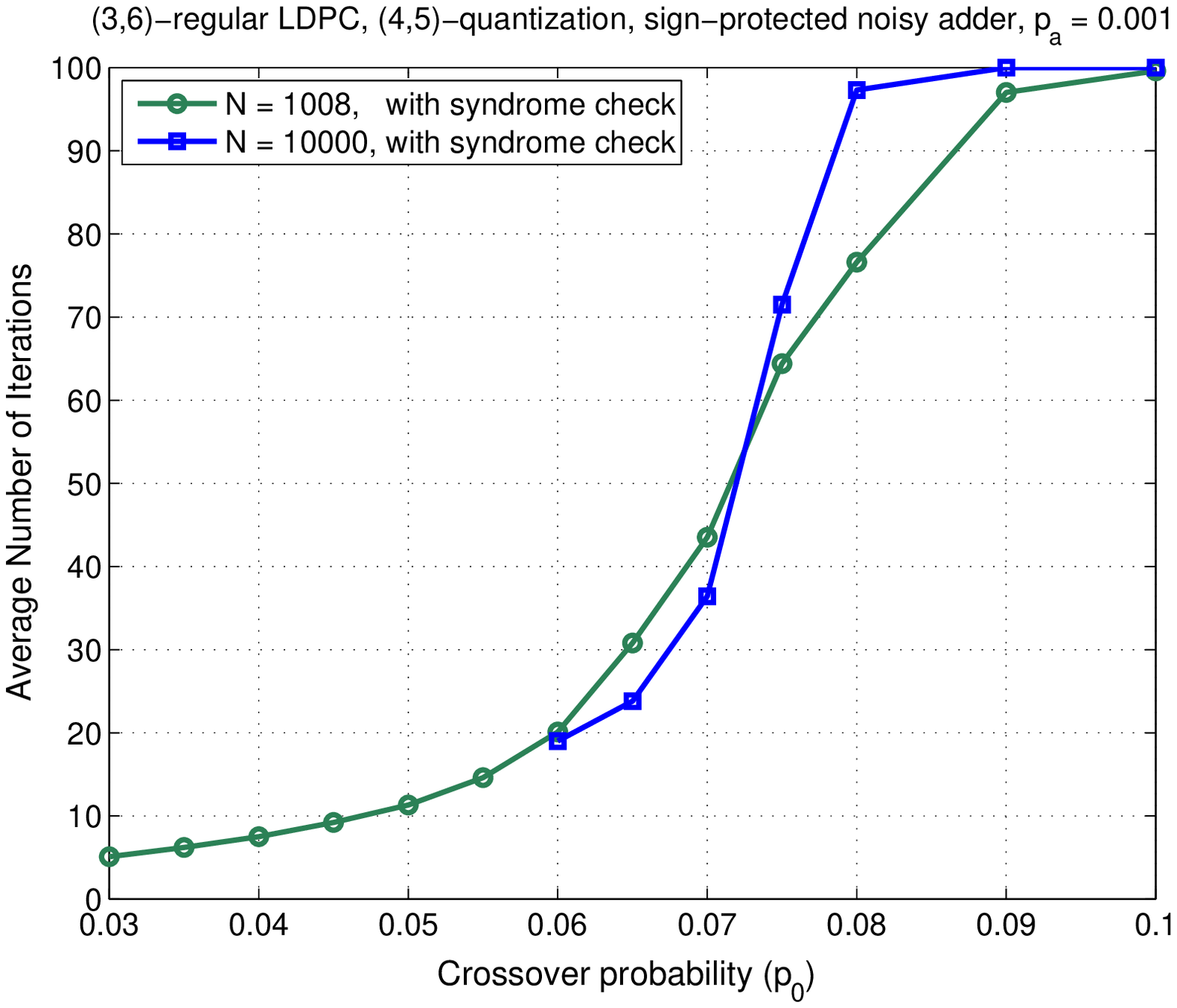}
\caption[Average number of decoding iterations with early stopping criterion]{Average number of decoding iterations with early stopping criterion (MS decoder with sign-preserving noisy adder, $p_a = 0.001$)}
\label{fig:bsc_ave_iter_nb_variousN}}%
\end{figure}

\medskip To illustrate this behavior, we plotted the Figure~\ref{fig:bsc_ber_variousN} the BER performance of the noisy MS decoder, with and without early stopping criterion. The noisy MS decoder comprises a sign-preserving noisy adder with $p_a = 0.001$, while the comparator and the {\sc xor}-operator are assumed to be noiseless ($\pcomp = \pxor = 0$). Two codes are simulated, the first with length $N = 1008$ bits, and the second with length $N = 10000$ bits. 
In case that the noiseless early stopping criterion is implemented (solid curves), it can be seen that none of the BER curves show any error floor down to $10^{-8}$. However, if the early stopping criterion is not implemented (dashed curves), corresponding BER curves exhibit an error floor at $\approx 3.33\times 10^{-5}$, as predicted by Proposition~\ref{prop:lower-bounds}.

In Figure~\ref{fig:bsc_ave_iter_nb_variousN} we plotted the average number of decoding iterations in case that the early stopping criterion is implemented. It can be seen that the average number of decoding iterations decreases with the channel crossover probability $p_0$, or equivalently, with the achieved bit error rate. 
However, for a fixed BER -- say  BER $= 10^{-6}$, achieved either at $p_0 \approx 0.04$ for the code with $N = 1008$, or at $p_0 \approx 0.063$ for the code with $N = 10000$ -- the average number of iterations is about $8$ for the first code and about $21$  for the second. 
Note that in case the early stopping criterion is not implemented, both codes have nearly the same performance 
for the above $p_0$ values. Thus, when the early stopping criterion is implemented, the decoder needs to perform more iterations  to eventually reach an error free iteration when $N = 10000$, which explains the increased average number of decoding iterations.

\subsubsection{Further results on the finite-length performance}

\begin{figure}[!thb]
\centering
\subfigure[sign-preserving noisy adder, $\pcomp = \padd$, $\pxor = 0.0001$]{\includegraphics[width=.49\linewidth]{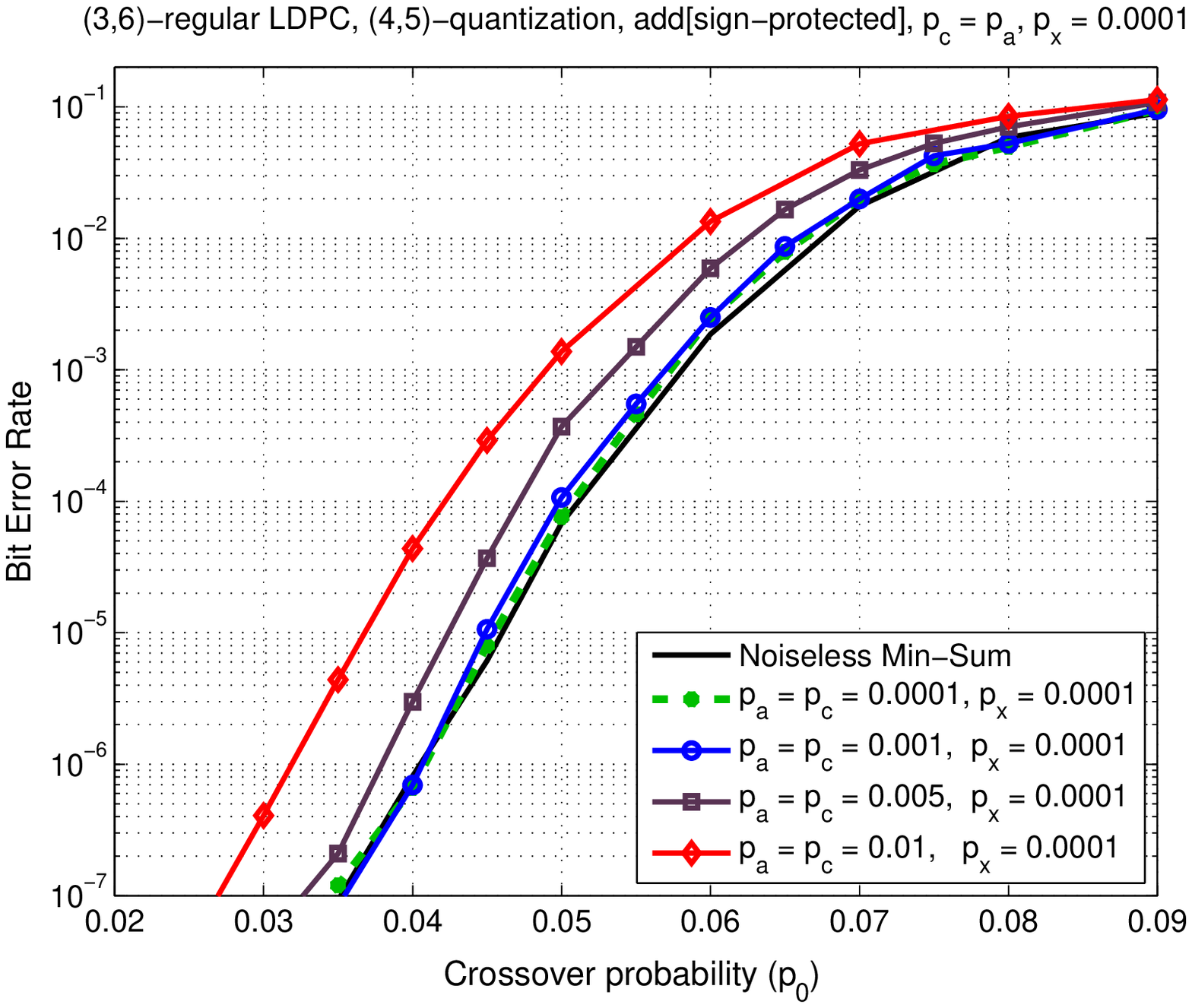}\label{subfig:bsc_ber_add_sp_px0001}}%
\subfigure[full-depth noisy adder, $\pcomp = \padd$, $\pxor = 0.0001$]{\includegraphics[width=.49\linewidth]{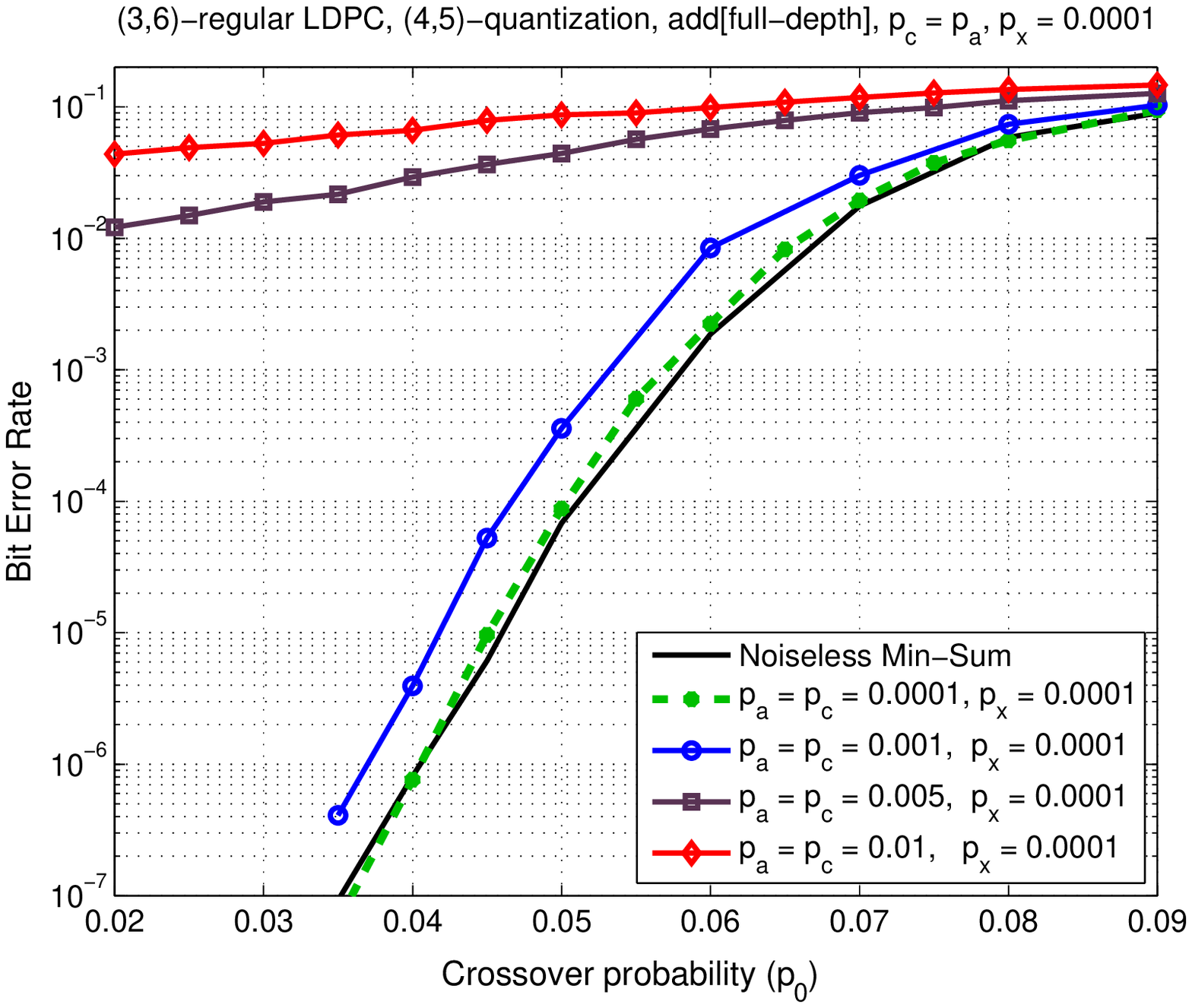}\label{subfig:bsc_ber_add_fd_px0001}}%

\subfigure[sign-preserving noisy adder, $\pcomp = \padd$, $\pxor = 0.001$]{\includegraphics[width=.49\linewidth]{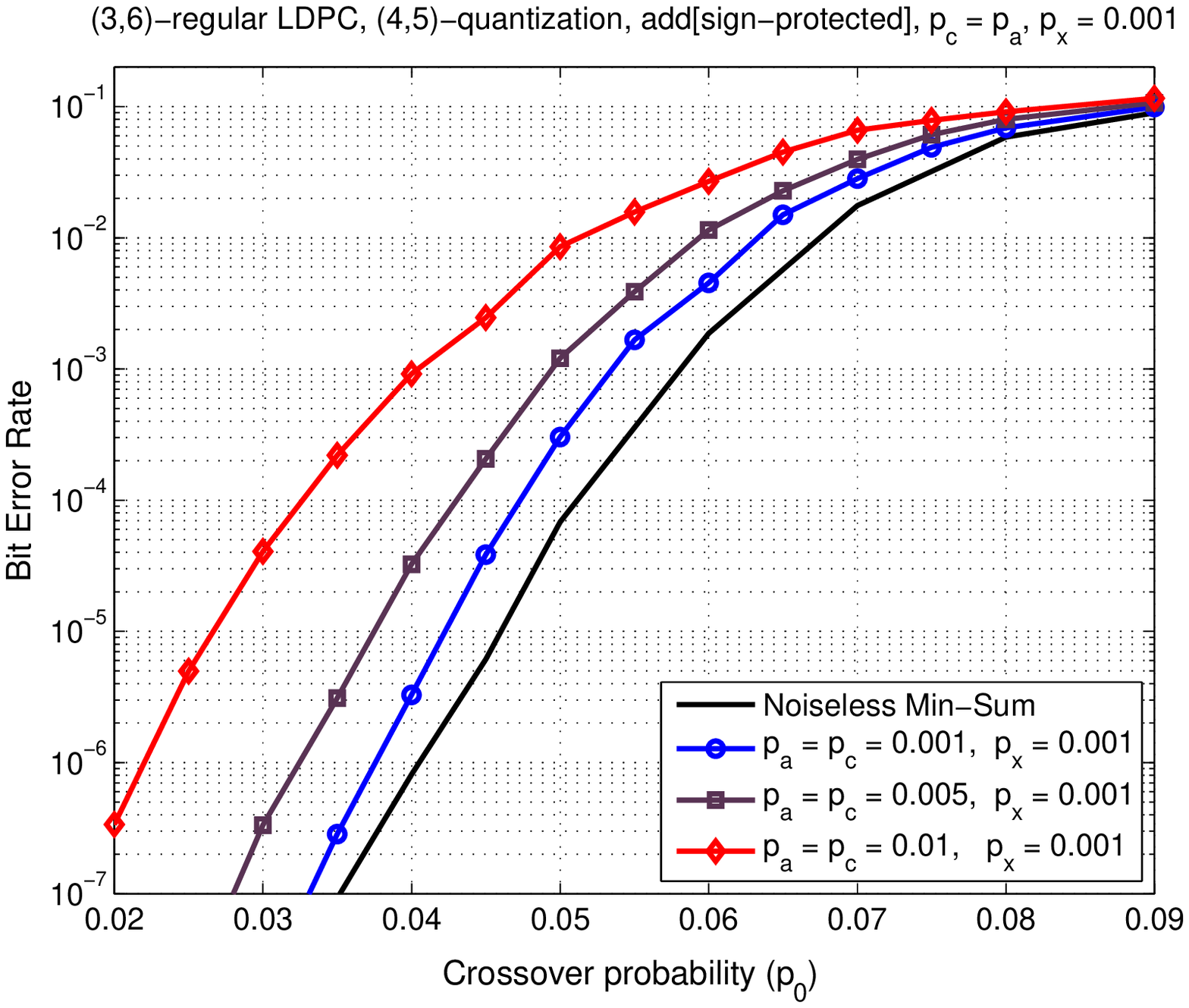}\label{subfig:bsc_ber_add_sp_px001}}%
\subfigure[full-depth noisy adder, $\pcomp = \padd$, $\pxor = 0.001$]{\includegraphics[width=.49\linewidth]{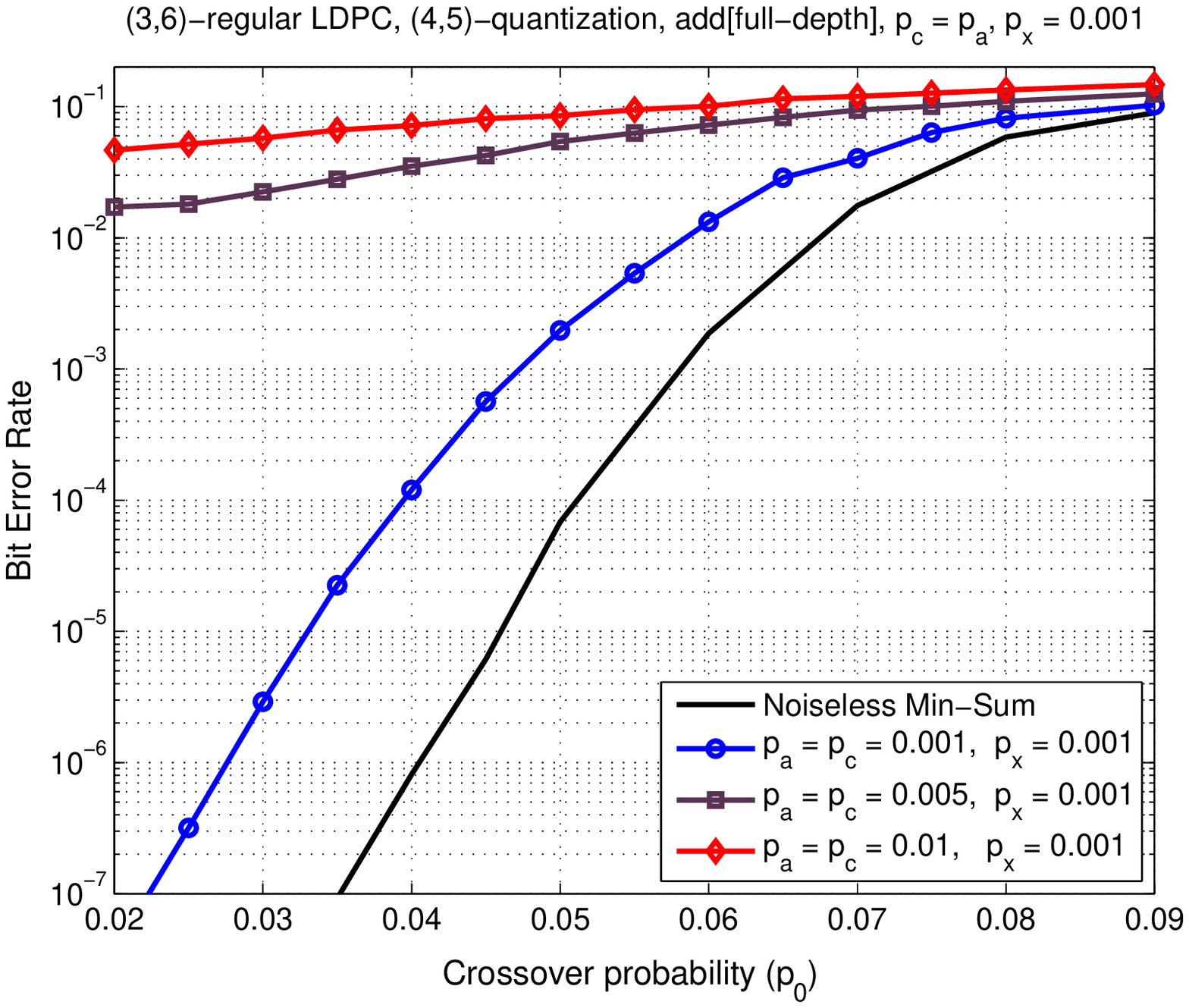}\label{subfig:bsc_ber_add_fd_px001}}%
\caption{BER performance of the noisy MS decoder with various noise parameters}
\label{fig:ber_perf_noisy_adder}
\end{figure}

In this section we investigate the finite-length performance when all the MS components (adder, comparator, and {\sc xor}-operator) are noisy. In order to reduce the number of simulations, we assume that $p_a = p_c \geq p_x$. Concerning the noisy adder, we evaluate the BER performance for both the sign-preserving and the full-depth error models. Simulation results are presented in Figure~\ref{fig:ber_perf_noisy_adder}. The error probability of the {\sc xor}-operator is $\pxor = 0.0001$ in sub-figures~\ref{subfig:bsc_ber_add_sp_px0001} and~\ref{subfig:bsc_ber_add_fd_px0001}, and  $\pxor = 0.001$ in sub-figures~\ref{subfig:bsc_ber_add_sp_px001} and~\ref{subfig:bsc_ber_add_fd_px001}. The noisy adder is sign-preserving in sub-figures~\ref{subfig:bsc_ber_add_sp_px0001} and~\ref{subfig:bsc_ber_add_sp_px001}, and  full-depth in sub-figures~\ref{subfig:bsc_ber_add_fd_px0001} and~\ref{subfig:bsc_ber_add_fd_px001}. 

In case the noisy-adder is sign-preserving, it can be seen that the MS decoder can provide reliable error protection for all the noise parameters that have been simulated. Of course, depending on the error probability parameters of the noisy components, there is a more or less important degradation of the achieved BER with respect to the noiseless case. But in all cases the noisy decoder can achieve a BER less than $10^{-7}$. This is no longer true for the full-depth noisy adder: it can be seen that for $\pcomp = \padd \geq 0.005$, the noisy decoder cannot achieve bit error rates below $10^{-2}$.

\clearpage

\subsection{Noisy Self-Corrected Min-Sum decoder}

In this section we investigate the finite-length performance of the Self-Corrected Min-Sum (SCMS) decoder \cite{savin2008self}. The objective is to determine if a correction circuit ``plugged into'' the noisy MS decoder can improve the robustness of the decoder to hardware noise.

The specificity of the SCMS decoder is to {\em erase} ({\em i.e.} set to zero) any variable-to-check message that 
changes its sign between two consecutive iterations. However, in order to avoid erasures propagation, a message cannot be erased if it has also been erased at the previous iteration.  Hence, the SCMS decoder performs the same computations as the noisy MS, except that the {\bf VN processing} step further includes a {\em correction step}, as follows\footnote{Superscript $(\ell)$ used to denote the iteration number}:

\medskip\noindent\fbox{\begin{minipage}{.98\linewidth}
{\bf for all} $n=1,\dots,N$ and $m\in{\cal H}(n)$ {\bf do} \hfill $\vartriangleright$ {\bf VN-processing}

    \vspace*{2mm}\hspace*{5mm}$\begin{array}{@{}r@{\ }c@{\ }l}
    \alpha_{m,n}^{(\ell)} &  = & \mathbf{s}_{\cal M}\left(\addprred\left(\tilde{\gamma}_n^{(\ell)}, -\beta_{m,n}^{(\ell)}\right)\right)\text{;} 
    \end{array}$ 
    
    \vspace*{2mm}\hspace*{5mm}$\text{\bf if } \ 
       \sgn\left(\alpha_{m,n}^{(\ell)}\right) \neq \sgn\left(\alpha_{m,n}^{(\ell-1)}\right)
       \text{ and }  \alpha_{m,n}^{(\ell-1)} \neq 0 \\
  \hspace*{15mm} \alpha_{m,n}^{(\ell)} = 0\,; \\
  \hspace*{5mm}\text{\bf end}$
\end{minipage}}     

\bigskip The body enclosed between the {\bf if} condition and the matching {\bf end} is referred to as the {\em correction step}. In practical implementations, one needs to store the signs of the variable-to-check node messages and to keep a record of messages that have been erased by the self-correction step. We use the following notation:
\begin{itemize}
\item $s_{m,n}^{(\ell)} = \text{sgn}\left(\alpha_{m,n}^{(\ell)}\right)$, the sign of the message $\alpha_{m,n}^{(\ell)}$;
\item $e_{m,n}^{(\ell)}\in\{0,1\}$, with $e_{m,n}^{(\ell)} = 1$ if and only if the corresponding variable-to-check message has been erased at iteration $\ell$; for $\ell = 0$, these values are all initialized as zero. 
\item $\text{\sc scu}(s_1, s_2, e) \stackrel{\text{def}}{=} (s_1 \oplus s_2) \otimes (1 \oplus e)$, for any $s_1, s_2, e\in\{0, 1\}$, where $\oplus$ denotes the {\sc xor} operation (sum modulo $2$) and $\otimes$ denotes the {\sc and} operation (product). Clearly $\text{\sc scu}(s_1, s_2, e) = 1$ if and only if $s_1\neq s_2$ and $e = 0$.
\end{itemize} 

\noindent Therefore, the VN-processing step of the SCMS decoder can be rewritten as follows:

\medskip\noindent\fbox{\begin{minipage}{.98\linewidth}
{\bf for all} $n=1,\dots,N$ and $m\in{\cal H}(n)$ {\bf do} \hfill $\vartriangleright$ {\bf VN-processing}

    \vspace*{2mm}\hspace*{5mm}$\begin{array}{@{}r@{\ }c@{\ }l}
    \alpha_{m,n}^{(\ell)} &  = & \mathbf{s}_{\cal M}\left(\addprred\left(\tilde{\gamma}_n^{(\ell)}, -\beta_{m,n}^{(\ell)}\right)\right)\text{;} \hbox{\protect\raisebox{0mm}[0mm][4mm]{}}\\
    e_{m,n}^{(\ell)} & = & \text{\sc scu}\left(s_{m,n}^{(\ell)}, s_{m,n}^{(\ell-1)}, 
                               e_{m,n}^{(\ell-1)}\right)\text{;}
    \end{array}$ 
    
    \vspace*{2mm}\hspace*{5mm} $\text{\bf if } \ e_{m,n}^{(\ell)} = 1 \ {\bf then }\ \ 
    \alpha_{m,n}^{(\ell)} = 0\,; \ \ \text{\bf end}$
\end{minipage}}     

\bigskip This reformulation of the VN-processing step allows defining a {\em noisy self-correction} step, by injecting errors in the output of the {\sc scu} operator. The noisy {\sc scu} operator with error probability $\pscu$ is defined by:
\begin{equation}\label{eq:noisy-scu-def}
 \scupr(s_1, s_2, e) = \left\{\begin{array}{cl}
 \text{\sc scu}(s_1, s_2, e), & \mbox{with probability } 1-\pscu\\
 1 - \text{\sc scu}(s_1, s_2, e), & \mbox{with probability } \pscu
 \end{array}\right.
 \end{equation}
This error model captures the effect of the noisy logic or of the noisy storage of $s_{m,n}$ and $e_{m,n}$ values on the {\sc scu} operator. The SCMS decoder with noisy self-correction step is detailed in Algorithm~\ref{alg:noisy_scms}.

\begin{figure}[!b]
\centering
\vspace*{-5mm}
\subfigure[BSC channel ($\mu = 6$)]{\includegraphics[width=.49\linewidth]{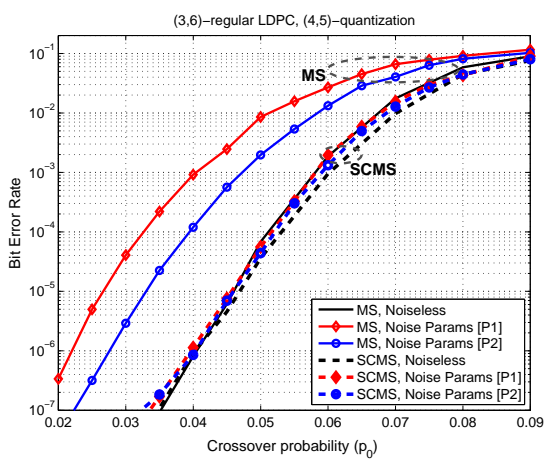} \label{subfig:bsc_ber_ms_scms}}%
\subfigure[BI-AWGN channel ($\mu = 5.5$)]{\includegraphics[width=.49\linewidth]{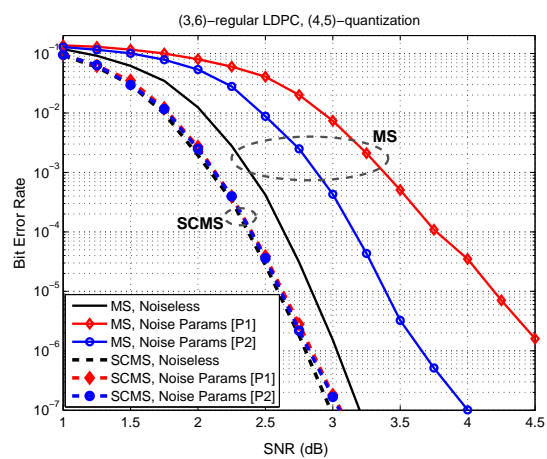}
\label{subfig:awgn_ber_ms_scms}}
\vspace*{-3mm}\caption{BER performance comparison between noisy MS and noisy SCMS decoders}
\label{fig:noisy_ms_vs_scms}
\end{figure}


\begin{algorithm}[!t]
\caption{Noisy Self-Corrected Min-Sum (Noisy-SCMS) decoding }\label{alg:noisy_scms}
\begin{algorithmic}[0]
\State{Input:}  $\vect{y} = (y_1,\dots,y_N)\in{\cal Y}^N$ (${\cal Y}$ is the channel output alphabet)\Comment{received word}
\State{Output:} $\hat{\vect{x}} = (\hat{x}_1,\dots,\hat{x}_N)\in\{-1, +1\}^N$ \Comment{estimated codeword}  

\Init 
   \ForAll{$n=1,\dots,N$} 
          $\gamma_{n} = \mathbf{q}(y_n)$;   
   \EndFor
   \ForAll{$n=1,\dots,N$ and $m\in{\cal H}(n)$} 
          {\bf\{} $\alpha_{m,n} = \gamma_n$; \ 
                  $s_{m,n}      = \sgn(\gamma_n)$; \ 
                  $e_{m,n}      = 0$;\,{\bf\}}
   \EndFor
\EndInit
\Iter 
   \ForAll{$m=1,\dots,M$ and $n\in{\cal H}(m)$} \
       \Comment{{\bf CN-processing}}
       \State $\beta_{m,n}=\displaystyle
		\xorprred\left( \{\sgn(\alpha_{m,n'})\}_{{n'}\in{\cal H}(m)\setminus n} \right) 
		\minprred\left( \{|\alpha_{m,n'}|\}_{{n'}\in{\cal H}(m)\setminus n} \right)$;
   \EndFor
   
   \ForAll{$n=1,\dots,N$}\ 
	    \Comment{{\bf AP-update}}        
        \State $\tilde{\gamma}_n = \addprred\left(\{\gamma_n\} \cup 
        \{\beta_{m,n}\}_{{m} \in{\cal H}(n)}\right)\text{;}$
   \EndFor
	
   \ForAll{$n=1,\dots,N$ and $m\in{\cal H}(n)$} \
       \Comment{{\bf VN-processing}}       
	   \State  $\begin{array}{@{}r@{\ }c@{\ }l}
               \alpha_{m,n} &  = & \mathbf{s}_{\cal M}\left(\addprred\left(\tilde{\gamma}_n, 
                                            -\beta_{m,n}\right)\right)\text{;} 
                                            \hbox{\protect\raisebox{0mm}[0mm][2mm]{}}\\
               e_{m,n} & = & \scuprred\left(\sgn(\alpha_{m,n}), s_{m,n}, e_{m,n}\right)\text{;}
                                             \hbox{\protect\raisebox{0mm}[0mm][2mm]{}}\\
               s_{m,n} & = & \sgn(\alpha_{m,n})\text{;}\hbox{\protect\raisebox{0mm}[0mm][2mm]{}}
               \end{array}$ 
    
    \If{$e_{m,n} = 1$}
         $\alpha_{m,n} = 0\text{;}$
    \EndIf
		
   \EndFor

	\ForAll{$\{v_n\}_{n=1,\dots,N}$} $\hat{x}_n = \sgn(\tilde{\gamma}_{n})$; \ 
	    \Comment{hard decision}
	\EndFor
	
	\If{$\hat{\vect{x}}$ is a codeword} exit the iteration loop \Comment{syndrome check}
	\EndIf
\EndIter
%
%
\end{algorithmic}
\end{algorithm}

\smallskip The finite length performance of the noisy SCMS decoder is presented in Figure~\ref{fig:noisy_ms_vs_scms}, for both BSC and BI-AWGN channels. For comparison purposes, Figure~\ref{fig:noisy_ms_vs_scms} also shows the performance of the noisy MS decoder. The parameters of the different noisy components are as follows:

\medskip \noindent {\bf [P1]} sign-preserving adder with $\padd = 0.01$, $\pcomp = 0.01$, $\pxor = \pscu = 0.001$ (red curves, diamond markers);

\smallskip \noindent {\bf [P2]} full-depth adder with $\padd = 0.001$, $\pcomp = 0.001$, $\pxor = \pscu = 0.001$ (blue curves, circle markers). 


Solid and dashed curves correspond respectively to the MS and SCMS  performance. While the hardware noise alters the performance of the MS decoder, it can be seen that the noisy SCMS decoder exhibits very good performance, very close to that of the noiseless decoder. Therefore, one can think of the self-correction circuit
as a {\em noisy patch} applied to the noisy MS decoder, in order to improve its robustness to hardware noise.  The robustness of the  SCMS decoder to hardware noise is explained by the fact that it has an intrinsic capability to detect unreliable messages, and discards them from the iterative decoding process \cite{savin2008self}.

\section{Conclusion}\label{sec:noisy_ms_conslusion}
This paper investigated the asymptotic and finite length behavior of the noisy MS decoder. We demonstrated the impact of the channel scale factor on the decoder performance, both for the noiseless and for the noisy decoder. We also highlighted the fact that an inappropriate choice the channel scale factor may lead to an {\em unconventional} behavior, in the sense that the noise introduce by the device may actually result in an increased correction capacity with respect to the noiseless decoder. We analyzed the asymptotic performance of the noisy MS decoder in terms of useful regions and target-BER thresholds, and further revealed the existence of a different threshold phenomenon, which was referred to as functional threshold. Finally, we also corroborated the asymptotic analysis through finite-length simulations, and highlighted the excellent performance of the noisy SCMS decoder, which provides virtually the same performance as the noiseless decoder, for a wide range of values of the hardware noise parameters.

\bibliographystyle{IEEEtran}
\bibliography{./biblio/biblio_database}

\begin{thebibliography}{10}
\providecommand{\url}[1]{#1}
\csname url@samestyle\endcsname
\providecommand{\newblock}{\relax}
\providecommand{\bibinfo}[2]{#2}
\providecommand{\BIBentrySTDinterwordspacing}{\spaceskip=0pt\relax}
\providecommand{\BIBentryALTinterwordstretchfactor}{4}
\providecommand{\BIBentryALTinterwordspacing}{\spaceskip=\fontdimen2\font plus
\BIBentryALTinterwordstretchfactor\fontdimen3\font minus
  \fontdimen4\font\relax}
\providecommand{\BIBforeignlanguage}[2]{{%
\expandafter\ifx\csname l@#1\endcsname\relax
\typeout{** WARNING: IEEEtran.bst: No hyphenation pattern has been}%
\typeout{** loaded for the language `#1'. Using the pattern for}%
\typeout{** the default language instead.}%
\else
\language=\csname l@#1\endcsname
\fi
#2}}
\providecommand{\BIBdecl}{\relax}
\BIBdecl

\bibitem{taylor1968reliablestorage}
M.~G. Taylor, ``Reliable information storage in memories designed from
  unreliable components,'' \emph{Bell System Technical Journal}, vol.~47, pp.
  2299--2337, 1968.

\bibitem{taylor1968reliablecomputing}
------, ``Reliable computation in computing systems designed from unreliable
  components,'' \emph{Bell System Technical Journal}, vol.~47, pp. 2339--2366,
  1968.

\bibitem{kuznetsov1973information}
A.~V. Kuznetsov, ``Information storage in a memory assembled from unreliable
  components,'' \emph{Problemy Peredachi Informatsii}, vol.~9, no.~3, pp.
  100--114, 1973.

\bibitem{chilappagari2006analysis}
S.~K. Chilappagari, M.~Ivkovic, and B.~Vasic, ``Analysis of one step majority
  logic decoders constructed from faulty gates,'' in \emph{Proc. of IEEE Int.
  Symp. on Information Theory}, 2006, pp. 469--473.

\bibitem{vasic2007information}
B.~Vasic and S.~K. Chilappagari, ``An information theoretical framework for
  analysis and design of nanoscale fault-tolerant memories based on low-density
  parity-check codes,'' \emph{IEEE Trans. on Circuits and Systems I: Regular
  Papers}, vol.~54, no.~11, pp. 2438--2446, 2007.

\bibitem{winstead2009probabilistic}
C.~Winstead and S.~Howard, ``A probabilistic {LDPC}-coded fault compensation
  technique for reliable nanoscale computing,'' \emph{IEEE Trans. on Circuits
  and Systems II: Express Briefs}, vol.~56, no.~6, pp. 484--488, 2009.

\bibitem{tang2012ldpc}
Y.~Tang, C.~Winstead, E.~Boutillon, C.~Jego, and M.~Jezequel, ``An ldpc
  decoding method for fault-tolerant digital logic,'' in \emph{IEEE Int. Symp.
  on Circuits and Systems (ISCAS)}, 2012, pp. 3025--3028.

\bibitem{hussien2011class}
A.~M. Hussien, M.~S. Khairy, A.~Khajeh, A.~M. Eltawil, and F.~J. Kurdahi, ``A
  class of low power error compensation iterative decoders,'' in \emph{IEEE
  Global Telecom. Conf. (GLOBECOM)}, 2011, pp. 1--6.

\bibitem{varshney2011performance}
L.~R. Varshney, ``Performance of {LDPC} codes under faulty iterative
  decoding,'' \emph{IEEE Trans. Inf. Theory}, vol.~57, no.~7, pp. 4427--4444,
  2011.

\bibitem{yazdi2012probabilistic}
S.~Yazdi, H.~Cho, Y.~Sun, S.~Mitra, and L.~Dolecek, ``Probabilistic analysis of
  {Gallager B} faulty decoder,'' in \emph{IEEE Int. Conf. on Communications
  (ICC)}, 2012, pp. 7019--7023.

\bibitem{yazdi2012optimal}
S.~Yazdi, C.~Huang, and L.~Dolecek, ``Optimal design of a {Gallager B} noisy
  decoder for irregular {LDPC} codes,'' \emph{IEEE Comm. Letters}, vol.~16,
  no.~12, pp. 2052--2055, 2012.

\bibitem{yazdi2013gallager}
S.~Yazdi, H.~Cho, and L.~Dolecek, ``Gallager b decoder on noisy hardware,''
  \emph{IEEE Trans. on Comm.}, vol.~66, no.~5, pp. 1660--1673, 2013.

\bibitem{balatsoukas2014characterization}
A.~Balatsoukas-Stimming, C.~Studer, and A.~Burg, ``Characterization of min-sum
  decoding of {LDPC} codes on unreliable silicon,'' in \emph{Information Theory
  and Applications Workshop (ITA)}, 2014.

\bibitem{balatsoukas2014density}
A.~Balatsoukas-Stimming and A.~Burg, ``Density evolution for min-sum decoding
  of {LDPC} codes under unreliable message storage,'' \emph{IEEE Communications
  Letters}, vol.~PP, no.~99, pp. 1--4, 2014.

\bibitem{gallager1963low}
R.~G. Gallager, ``Low density parity check codes,'' MIT Press, Cambridge, 1963,
  research Monograph series.

\bibitem{tanner1981recursive}
R.~Tanner, ``A recursive approach to low complexity codes,'' \emph{IEEE Trans.
  on Inf. Theory}, vol.~27, no.~5, pp. 533--547, 1981.

\bibitem{pearl1982bp}
J.~Pearl, ``{Reverend Bayes on inference engines: A distributed hierarchical
  approach},'' in \emph{Proc. of the 2nd National Conference on Artificial
  Intelligence (AAAI-82)}, 1982, pp. 133–--136.

\bibitem{pearl1988probabilistic}
------, \emph{Probabilistic reasoning in intelligent systems: networks of
  plausible inference}.\hskip 1em plus 0.5em minus 0.4em\relax Morgan Kaufmann
  Publishers, 1988.

\bibitem{fossorier1999reduced}
M.~Fossorier, M.~Mihaljevic, and H.~Imai, ``Reduced complexity iterative
  decoding of low-density parity check codes based on belief propagation,''
  \emph{IEEE Trans. on Communications}, vol.~47, no.~5, pp. 673--680, 1999.

\bibitem{chung2000construction}
S.~Chung, ``On the construction of some capacity-approaching coding schemes,''
  Ph.D. dissertation, Massachusetts Institute of Technology, 2000.

\bibitem{eleftheriou2001reduced}
E.~Eleftheriou, T.~Mittelholzer, and A.~Dholakia, ``Reduced-complexity decoding
  algorithm for low-density parity-check codes,'' \emph{IET Electronics
  Letters}, vol.~37, no.~2, pp. 102--104, 2001.

\bibitem{dobrushin1977lower}
R.~L. Dobrushin and S.~Ortyukov, ``Lower bound for the redundancy of
  self-correcting arrangements of unreliable functional elements,''
  \emph{Problemy Peredachi Informatsii}, vol.~13, no.~1, pp. 82--89, 1977.

\bibitem{i-RISC-D2.1}
\BIBentryALTinterwordspacing
A.~Amaricai \emph{et~al.}, ``Circuit level fault models for sub-powered {CMOS}
  circuits for uncorrelated and correlated errors,'' FP7\,/\,FET
  OPEN\,/\,309129, i-RISC project, Deliverable D2.1, January 2014. [Online].
  Available: \url{http:www.i-risc.eu}
\BIBentrySTDinterwordspacing

\bibitem{richardson2001capacity}
T.~J. Richardson and R.~L. Urbanke, ``The capacity of low-density parity-check
  codes under message-passing decoding,'' \emph{IEEE Transactions on
  Information Theory}, vol.~47, no.~2, pp. 599--618, 2001.

\bibitem{mackaywebsite}
\BIBentryALTinterwordspacing
D.~J. MacKay. Encyclopedia of sparse graph codes. [Online]. Available:
  \url{http://www.inference.phy.cam.ac.uk/\-mackay/\-codes/data.html}
\BIBentrySTDinterwordspacing

\bibitem{savin2008self}
V.~Savin, ``Self-corrected min-sum decoding of {LDPC} codes,'' in \emph{Proc.
  of IEEE Int. Symp. on Information Theory (ISIT)}, 2008, pp. 146--150.

\end{thebibliography}

  \end{document}